\documentclass[11pt,letterpaper]{article}
\pdfoutput=1

\usepackage[utf8x]{inputenc}%
\usepackage{tcolorbox,fancybox}%
\usepackage{physics}
\usepackage{ntheorem}

\usepackage[english]{babel}
\usepackage{amsmath,amssymb,amsfonts}
\usepackage{graphicx}
\usepackage{booktabs}
\usepackage{color,xcolor}
\usepackage[numbers,sort&compress]{natbib}

\usepackage{mathtools}
\usepackage{mathdots}
\usepackage[mathscr]{euscript}

\usepackage{tikz}
\usepackage[compat=1.1.0]{tikz-feynman}
\usetikzlibrary{positioning}
\usetikzlibrary{decorations.text}
\usetikzlibrary{decorations.pathmorphing}

\usetikzlibrary{calc}
\usetikzlibrary{shapes.misc}
\tikzset{
        >=latex,
    photon/.style={decorate, decoration={snake}, draw=black, thick},
    fermionnoarrow/.style={draw=black, postaction={decorate}, thick},
    scalar/.style={draw=black, postaction={decorate}, decoration={markings,mark=at position .55 with {\arrow{>}}}, thick, dashed},
    scalarnoarrow/.style={draw=black, postaction={decorate},  thick, dashed},
    fermion/.style={draw=black, postaction={decorate},decoration={markings,mark=at position .55 with {\arrow{>}}}, thick},
    gluon/.style={decorate, draw=black, decoration={coil,amplitude=4pt, segment length=5pt}, thick},
    vertex/.style={draw,shape=circle,fill=black,minimum size=3pt,inner sep=0pt},
    fillvertex/.style={draw,shape=circle,fill=black,minimum size=5pt,inner sep=0pt},
    openvertex/.style={draw,shape=circle,minimum size=5pt,inner sep=0pt},
    blob/.style={draw=red,shape=circle,fill=red,minimum size=6pt,inner sep=0pt},
    redvertex/.style={draw=red,shape=circle,fill=red,minimum size=3pt,inner sep=0pt},
    cross/.style={cross out, draw=black,thick, minimum size=5pt, inner sep=0pt, outer sep=0pt}
}

\usepackage{physics}
\usepackage{orcidlink}
\usepackage{slashed}
\usepackage{mathrsfs}
\usepackage{enumerate}
\usepackage{mdframed}
\usepackage{url}
\usepackage[makeroom]{cancel}
\usepackage{geometry}

\usepackage{pifont}
\newcommand{\cmark}{\ding{51}}%
\newcommand{\xmark}{\ding{55}}%

\newtheorem*{thm-non}{Theorem}
\newtheorem*{conj}[thm-non]{Conjecture}
\newtheorem*{law}[thm-non]{Law}

\newtheorem*{define}[thm-non]{Definition}

\usepackage{hyperref} %Automatically links \label and \ref commands; Always load last
\hypersetup{
    colorlinks=true,       % false: boxed links; true: colored links
    linkcolor=red,          % color of internal links
    citecolor=blue,        % color of links to bibliography
    filecolor=magenta,      % color of file links
    urlcolor=purple           % color of external links
}
\usepackage[all]{hypcap} %Link navagates to top of figure instead of caption (below fig)

\def\beqn{\begin{eqnarray}}
\def\eeqn{\end{eqnarray}}
\def\beqs{\begin{subequations}}
\def\eeqs{\end{subequations}}
\def\beq{\begin{equation}}
\def\eeq{\end{equation}}
\def\ba{\begin{array}}
\def\ea{\end{array}}

\def\non{\nonumber\\}

\def\[{\left[}
\def\]{\right]}
\def\({\left(}
\def\){\right)}
\newcommand\para{\paragraph{}}

\def\gSU{\rm SU}

\def\gSO{\rm SO}

\newcommand{\rep}[1]{\mathbf{#1}}
\newcommand{\repb}[1]{\mathbf{\overline{#1}}}

\def\Bc{\mathcal{B}}
\def\Cc{\mathcal{C}}
\def\Dc{\mathcal{D}}
\def\Ec{\mathcal{E}}
\def\Fc{\mathcal{F}}
\def\Gc{\mathcal{G}}
\def\Hc{\mathcal{H}}

\def\Lc{\mathcal{L}}

\def\Nc{\mathcal{N}}
\def\Oc{\mathcal{O}}

\def\Rc{\mathcal{R}}

\def\Tc{\mathcal{T}}
\def\Uc{\mathcal{U}}
\def\Vc{\mathcal{V}}

\def\Xc{\mathcal{X}}
\def\Yc{\mathcal{Y}}

\def\DG{\mathfrak{D}}  \def\dG{\mathfrak{d}}
\def\EG{\mathfrak{E}}  \def\eG{\mathfrak{e}}

  \def\nG{\mathfrak{n}}

\def\UG{\mathfrak{U}}  \def\uG{\mathfrak{u}}

\title{
 {\bf The global $B-L$ symmetry \\ in the flavor-unified ${\rm SU}(N)$ theories} \\
\author{\large Ning Chen$^{\,\heartsuit}$\,\orcidlink{0000-0002-0032-9012}, Ying-nan Mao$^{\,\diamondsuit}$\,\orcidlink{0000-0001-8063-8968}, Zhaolong Teng$^{\,\clubsuit}$\,\orcidlink{0000-0002-7141-2331}}
\date{\small \it
$^\heartsuit\, ^\clubsuit$School of Physics, Nankai University, Tianjin, 300071, China \\
$^\diamondsuit$ Department of Physics, School of Science, Wuhan University of Technology, \\ Wuhan, 430070, Hubei, China \\
}
}

\begin{document}

\maketitle
\setlength{\parskip}{0.2ex}

\begin{abstract}
\bigskip
We study the origin of the global $B-L$ symmetry in a class of flavor-unified theories with gauge groups of ${\rm SU}(N\geq 6)$.
In particular, we focus on the ${\rm SU}(8)$ theory which can minimally embed three-generational SM fermions non-trivially.
A reformulation of the third law for the flavor sector proposed by Georgi is useful to manifest the underlying global symmetries.
The 't Hooft anomaly matching and the generalized neutrality conditions for Higgs fields play the key roles in defining the $B-L$ symmetry.
Based on the global $B-L$ symmetry, we count the Higgs fields that can develop the VEVs and the massless sterile neutrinos in the ${\rm SU}(8)$ theory.
We also prove that a global $B-L$ symmetry can always be defined in any ${\rm SU}(N\geq 6)$ theory when it is spontaneously broken to the SM gauge symmetry.
\end{abstract}

\vspace{9.6cm}
{\emph{Emails:}\\  
$^{\,\heartsuit}$\url{chenning_symmetry@nankai.edu.cn},\\
$^{\,\diamondsuit}$\url{ynmao@whut.edu.cn},\\
$^{\,\clubsuit}$ \url{tengcl@mail.nankai.edu.cn}
 }

\thispagestyle{empty}  
\newpage  
\setcounter{page}{1}  

\vspace{1.0cm}
\tableofcontents

%###################################################################
\section{Introduction}
\label{section:intro}
%###################################################################
%
%

\para
Grand Unified Theories (GUTs), with their original formulations based on the gauge groups of $\gSU(5)$~\cite{Georgi:1974sy} and $\gSO(10)$~\cite{Fritzsch:1974nn}, were proposed to unify all three fundamental symmetries described by the Standard Model (SM) into one fundamental symmetry.
A widely accepted experimental justification of various GUTs is the proton decay processes.
Through the analyses of the corresponding $d=6$ four-fermion operators, it was pointed out by Weinberg, Wilczek, and Zee~\cite{Weinberg:1979sa,Wilczek:1979hc}, that a conserved global $\widetilde{ {\rm U}}(1)_{B-L}$ symmetry~\footnote{Throughout the paper, we use the notion of $\widetilde \Gc$ to denote the groups for global symmetries.} should originate from the GUTs such as the ${\rm SU}(5)$.

\para
Historically, an extension of the unified gauge group beyond the ${\rm SU}(5)$ was motivated by Georgi to achieve the flavor unification~\cite{Georgi:1979md}.
The central idea is to avoid the repetitive generation structure in the flavor sector of the GUT.
Accordingly, Georgi conjectured that the observed three-generational SM fermions are non-trivially embedded in irreducible representations (irreps) of a simple Lie group beyond the ${\rm SU}(5)$.
This insight is profound when confronting with the latest LHC measurements of the SM Higgs boson via different production and decay channels~\cite{CMS:2022dwd,ATLAS:2022vkf}.
The longstanding flavor puzzle in the SM quark and lepton sector can be mapped into a question of how does a SM Higgs boson distinguish fermions in different generations so that hierarchical Yukawa couplings are generated.
At the UV scale, the flavor-unified theories have the intrinsic property that three-generational SM fermions must transform differently~\cite{Georgi:1979md,Frampton:1979cw,Frampton:1979tj,Barr:1979xt,Barbieri:1980tz,Nanopoulos:1980xm,Barr:2008pn,Frampton:2009ce,Chen:2022xge}.
Therefore, one can expect a resolution to the fundamental flavor puzzles within this class of flavor-unified theories.

\para
The purpose of this paper is to study the origin of the global $\widetilde{ {\rm U}}(1)_{B-L}$ symmetry in the context of the flavor-unified theories with gauge groups of ${\rm SU}(N \geq 6)$.
The 't Hooft anomaly matching condition~\cite{tHooft:1979rat} will be applied.
Since the global $\widetilde{ {\rm U}}(1)_{B-L}$ symmetry in the SM is non-anomalous in the sense that $\[ \Gc_{\rm SM} \]^2 \cdot \widetilde{ {\rm U}}(1)_{B-L}=0$, it is natural to expect a global $\widetilde{ {\rm U}}(1)_{T}$ symmetry~\footnote{The subscripts of $_T$ throughout the paper stand for the tensor representations. Collectively, we will use the $\widetilde{ {\rm U}}(1)_{T}$ to denote the generalized global symmetries in different GUTs in our discussion.} originated from the underlying UV theory such that $\[ {\rm SU}(N) \]^2\cdot \widetilde{ {\rm U}}(1)_{T}=0$.
In such flavor-unified theories, the gauge symmetry undergoes several intermediate symmetry-breaking stages before the electroweak symmetry breaking (EWSB).
By performing the 't Hooft anomaly matching condition to the global $\widetilde{ {\rm U}}(1)_{T}$ symmetries defined at each stage, the massless fermions such as the sterile neutrinos in the spectrum, can be counted precisely.
A separate condition that is to require the Higgs fields that can develop the vacuum expectation values (VEVs) for each symmetry-breaking stage are neutral under the corresponding global $\widetilde{ {\rm U}}(1)_{T}$ symmetries.
This is a generalization from the fact that the SM Higgs field in the minimal ${\rm SU}(5)$ theory is $\widetilde{ {\rm U}}(1)_{B-L}$-neutral.

\para
The rest of the paper is organized as follows.
In Sec.~\ref{section:BminusL_SU5}, we revisit the definition of the $B-L$ symmetry in the ${\rm SU}(5)$ theory from the 't Hooft anomaly matching condition.
Generalized definition of the non-anomalous global $\widetilde{ {\rm U}}(1)_{T}$ charge assignments can be given for an ${\rm SU}(N)$ chiral gauge theory with rank-$k$ anti-symmetric fermions.
In Sec.~\ref{section:SU8_pattern}, we describe the minimal flavor-unified ${\rm SU}(8)$ theory with the same fermion contents as were given in Ref.~\cite{Barr:2008pn}.
Since this class of theories were not quite studied since Georgi's early proposal~\cite{Georgi:1979md}, we reformulate what was known as the third law of the flavor unification through the definition of the irreducible anomaly-free fermion set (IAFFS).
The global Dimopoulos-Raby-Susskind (DRS) symmetries~\cite{Dimopoulos:1980hn} will emerge naturally.
We focus on the decomposition of the Higgs fields and fermions under the reasonable symmetry-breaking pattern according to Ref.~\cite{Li:1973mq}.
In Sec.~\ref{section:BminusL_SU8}, we define the global $B-L$ symmetry in the ${\rm SU}(8)$ theory based on the 't Hooft anomaly matching condition.
The consistent global $\widetilde{ {\rm U}}(1)_T$ neutrality conditions to the Higgs fields at different symmetry-breaking stages will be applied so that the global $\widetilde{ {\rm U}}(1)_{B-L}$ symmetry will naturally appear when the theory flows to the electroweak (EW) scale.
In Sec.~\ref{section:BminusL_SUN}, we generalize the discussion to an arbitrary ${\rm SU}(N)$ theory.
It turns out that a global $\widetilde{{\rm U}}(1)_{B-L}$ symmetry can always be defined as long as the ${\rm SU}(N)$ is spontaneously broken into the subgroup of $\Gc_{\rm SM}$.
We summarize the results and discuss the related issues for theories of this class in Sec.~\ref{section:conclusion}.
App.~\ref{section:Br} is given to define the decomposition rules and the charge quantizations for the ${\rm SU}(8)$ theory based on the symmetry-breaking pattern in the current discussion.

%###################################################################
\section{The global $B-L$ symmetry in the ${\rm SU}(5)$ theory}
\label{section:BminusL_SU5}
%###################################################################

\para
Let us start from the minimal ${\rm SU}(5)$ theory~\cite{Georgi:1974sy}, with the chiral fermions of 
\beqn\label{eq:SU5_fermion}
\{  f_L \}_{ {\rm SU}(5)}&=& \repb{5_F} \oplus \rep{10_F}\,,
\eeqn
where the generational indices are suppressed.
The first way to view the global $\widetilde{{\rm U}}(1)_{B-L}$ symmetry is to consider the renormalizable Yukawa couplings of
\beqn
-\Lc_Y &=& Y_\Dc \repb{5_F}  \rep{10_F} \rep{5_H}^\dag +  Y_\Uc \rep{10_F}  \rep{10_F} \rep{5_H} + H.c. \,.
\eeqn
It is straightforward to find the following global $\widetilde{ {\rm U} }(1)_{ T_2}$ charge assignment of
\beqn\label{eq:SU5_T2charge}
&& \Tc_2 ( \repb{5_F}  ) =  -3 t_2\,,\quad \Tc_2 ( \rep{10_F}  ) = +t_2 \,,\quad  \Tc_2 ( \rep{5_H}  ) =  - \Tc_2 ( \rep{5_H}^\dag  ) =  -2 t_2 \,.
\eeqn
The $\widetilde{{\rm U}}(1)_{B-L}$ symmetry can be defined by the linear combination of the $\widetilde{{\rm U}}(1)_{T_2}$ symmetry and the ${\rm U}(1)_{Y}$ symmetry as
\beqn\label{eq:SU5_BminusLcharge}
\Bc - \Lc &\equiv& \frac{1}{5} \Tc_2 + \frac{4}{5} \Yc \,,
\eeqn
with the original convention of $t_2=+1$.

\para
Alternatively, we can view the global $\widetilde{{\rm U}}(1)_{T_2}$ and the $\widetilde{{\rm U}}(1)_{B-L}$ symmetries from the 't Hooft anomaly matching condition~\cite{tHooft:1979rat}.
The ${\rm SU}(5)$ fermions in Eq.~\eqref{eq:SU5_fermion} enjoy the global DRS symmetries~\cite{Dimopoulos:1980hn} of
\beqn\label{eq:SU5_DRS}
\widetilde \Gc_{\rm DRS}&=& \widetilde{ {\rm U} }(1)_{ \overline \Box } \otimes \widetilde{ {\rm U} }(1)_{ 2} \,.
\eeqn
Both $\widetilde{ {\rm U} }(1)_{ \overline \Box }$ and $\widetilde{ {\rm U} }(1)_{ 2}$ are anomalous by the ${\rm SU}(5)$ instanton.
However, one can always find a linear combination $\widetilde{ {\rm U} }(1)_{ T_2}$ of Eq.~\eqref{eq:SU5_DRS} such that the mixed gauge-global anomaly of $\[ {\rm SU}(5) \]^2\cdot \widetilde{ {\rm U} }(1)_{ T_2} $ vanishes.
Apparently, the global $\widetilde{ {\rm U} }(1)_{ T_2}$ charge assignment in Eq.~\eqref{eq:SU5_T2charge} satisfy that $\[ {\rm SU}(5) \]^2\cdot \widetilde{ {\rm U} }(1)_{ T_2} =0$.
Two other non-vanishing global anomalies are
\beqn
&& \[  {\rm grav} \]^2 \cdot \widetilde{ {\rm U} }(1)_{ T_2} = -5 t_2 \,, \quad \[  \widetilde{ {\rm U} }(1)_{ T_2} \]^3 = - 125 t_2^3 \,.
\eeqn

\para
After the ${\rm SU}(5)$ GUT-scale symmetry breaking, we can further define the global $\widetilde{ {\rm U} }(1)_{ B-L }$ symmetry as
\beqn\label{eq:SM_BminusL}
\Bc - \Lc &\equiv& \tilde a_1 \Tc_2 + \tilde a_2 \Yc \,.
\eeqn
Since the mixed gauge-global anomaly of $\[ {\rm SU}(5) \]^2\cdot \widetilde{ {\rm U} }(1)_{ T_2} $ is vanishing, the mixed gauge-global anomaly of $\[ \Gc_{\rm SM} \]^2\cdot \widetilde{ {\rm U} }(1)_{ B-L }$ symmetry must be vanishing as well.
The ${\rm SU}(5)$ fermions and Higgs field are decomposed as
\beqs\label{eqs:SU5_decomp}
\beqn
\repb{5_F}&=& \overbrace{ ( \repb{3}\,, \rep{1} \,, +\frac{1}{3} )_{ \rep{F}} }^{ {d_R}^c } \oplus  \overbrace{ ( \rep{1} \,, \repb{2}\,,  -\frac{1}{2 } )_{ \rep{F}} }^{ \ell_L} \,,\\[1mm]
%%%%%%%%%%%%%%%%%%%%%%%%%%%%%%%%%%%%%%%%
\rep{10_F}&=&  \overbrace{ ( \rep{3}\,, \rep{2} \,, +\frac{1}{6} )_{ \rep{F}} }^{ q_L } \oplus  \overbrace{ ( \repb{3}\,, \rep{1} \,, -\frac{2}{3} )_{ \rep{F}} }^{ {u_R}^c } \oplus   \overbrace{ ( \rep{1}\,, \rep{1} \,, + 1 )_{ \rep{F}} }^{ {e_R}^c } \,,\\[1mm]
%%%%%%%%%%%%%%%%%%%%%%%%%%%%%%%%%%%%%%%%
\rep{5_H}&=& \overbrace{ ( \rep{3}\,, \rep{1} \,, -\frac{1}{3} )_{ \rep{H}} }^{ \Phi_{\rep{3} } } \oplus \overbrace{ ( \rep{1} \,, \rep{2}\,,  + \frac{1}{2} )_{ \rep{H}} }^{ \Phi_{\rm SM} } \,.
\eeqn
\eeqs
Here, the color-triplet scalar of $\Phi_{\rep{3}}$ is expected to obtain masses at the GUT scale, and the $\Phi_{\rm SM}$ is identified as the SM Higgs doublet.
According to the definition in Eq.~\eqref{eq:SM_BminusL}, we have the global $\widetilde{ {\rm U} }(1)_{ B-L }$ charges of
\beqn\label{eq:SM_BminusLcharges}
&& ( \Bc - \Lc ) ( {d_R}^c ) = -3 t_2 \tilde a_1 + \frac{1}{3} \tilde a_2 \,, \quad ( \Bc -\Lc) ( \ell_L ) = -3 t_2 \tilde a_1 - \frac{1}{2} \tilde a_2 \,, \non
&& ( \Bc - \Lc ) ( q_L ) =  t_2 \tilde a_1 + \frac{1}{6} \tilde a_2  \,,\quad  ( \Bc - \Lc ) ( {u_R}^c ) = t_2 \tilde a_1 - \frac{2}{3} \tilde a_2 \,,\quad ( \Bc - \Lc ) ( {e_R}^c ) = t_2 \tilde a_1 + \tilde a_2  \,, \non
&& ( \Bc -  \Lc ) ( \Phi_{ \rep{3}} ) = - 2  t_2 \tilde a_1 - \frac{1}{3} \tilde a_2 \,,\quad  ( \Bc - \Lc  ) ( \Phi_{\rm SM} ) = - 2 t_2 \tilde a_1 + \frac{1}{2} \tilde a_2 \,.
\eeqn
The 't Hooft anomaly matching condition reads
\beqn
&&  \[  {\rm grav} \]^2 \cdot \widetilde{ {\rm U} }(1)_{ T_2} = -5 t_2\,, \quad  \[  {\rm grav} \]^2 \cdot \widetilde{ {\rm U} }(1)_{ B-L } =  -5 t_2 \tilde a_1 \,, \non
&& \[  \widetilde{ {\rm U} }(1)_{ T_2} \]^3= -125 t_2^3 \,, \quad \[  \widetilde{ {\rm U} }(1)_{ B - L } \]^3  = - 125 ( t_2 \tilde a_1 )^3\,.
\eeqn
By matching $\[  {\rm grav} \]^2 \cdot \widetilde{ {\rm U} }(1)_{ T_2} =  \[  {\rm grav} \]^2 \cdot \widetilde{ {\rm U} }(1)_{ B-L }$ and $\[  \widetilde{ {\rm U} }(1)_{ T_2} \]^3 = \[  \widetilde{ {\rm U} }(1)_{ B - L } \]^3$, we arrive at $\tilde a_1=+1$.
A separate condition is to require that the SM Higgs doublet of $\Phi_{\rm SM}$ is $\widetilde{ {\rm U} }(1)_{ B-L }$-neutral~\cite{Georgi:1981pu}, which leads to $\tilde a_2= 4 t_2 \tilde a_1 = 4 t_2$ from Eq.~\eqref{eq:SM_BminusLcharges}.
With both conditions, the $\widetilde{ {\rm U} }(1)_{ B-L }$ charges in Eq.~\eqref{eq:SM_BminusLcharges} become
\beqn
&& ( \Bc -\Lc) ( {d_R}^c ) = - \frac{5 }{3} t_2 \,, \quad ( \Bc -\Lc) ( \ell_L ) = -5 t_2  \,, \non
&& ( \Bc -\Lc ) ( q_L ) =   + \frac{5}{3 } t_2   \,,\quad  ( \Bc -\Lc ) ( {u_R}^c ) =   - \frac{5 }{3} t_2 \,,\quad (  \Bc -\Lc ) ( {e_R}^c ) = 5 t_2    \,, \non
&& ( \Bc -\Lc) ( \Phi_{ \rep{3}} ) = - \frac{10}{3} t_2  \,,\quad  ( \Bc -\Lc) ( \Phi_{\rm SM} ) = 0 \,.
\eeqn
A choice of $t_2=+\frac{1}{5}$ recovers the conventional definition of the $\widetilde{ {\rm U}}(1)_{B-L}$ charges in the SM.

%\para
%\NingC{Let us further consider a hypothetical situation, namely, the $\Phi_{ \rep{3}}$ was $\widetilde{ {\rm U}}(1)_{B-L}$-neutral.
%This leads to the coefficient of $\tilde a_2 = -6  t_2$.
%Thus, we arrive at
%
%
%\beqn
%&& ( \Bc -\Lc) ( {d_R}^c ) = - 5 t_2 \,, \quad ( \Bc -\Lc) ( \ell_L ) = 0  \,, \non
%
%&& ( \Bc -\Lc ) ( q_L ) =   0 \,,\quad  ( \Bc -\Lc ) ( {u_R}^c ) =   5 t_2 \,,\quad (  \Bc -\Lc ) ( {e_R}^c ) = - 5 t_2    \,, \non
%
%&& ( \Bc -\Lc) ( \Phi_{ \rep{3}} ) = 0  \,,\quad  ( \Bc -\Lc) ( \Phi_{\rm SM} ) = -5 t_2  \,.
%\eeqn}

%###################################################################
\section{The flavor-unified ${\rm SU}(8)$ theory and its symmetry-breaking pattern}
\label{section:SU8_pattern}
%###################################################################

%%%%%%%%%%%%%%%%%%%%%%%%%%%%
\subsection{Overview}
%%%%%%%%%%%%%%%%%%%%%%%%%%%%

\para
The flavor-unified theories based on the ${\rm SU}(N)$ groups~\footnote{It was argued that orthogonal groups of ${\rm SO}(N>10)$ cannot have three-generational SM fermions~\cite{Georgi:1979md} at the EW scale, since the spinor representations are always decomposed into pairs of $\rep{16_F} \oplus \repb{16_F}$.} were first proposed by Georgi~\cite{Georgi:1979md}, where he required no repetition of any irrep in his third law, which reads
\begin{law}[Georgi]

no irreducible representation should appear more than once in the representation of the left-handed fermions.

\end{law}
The minimal solution is an ${\rm SU}(11)$ theory with $561$ chiral fermions.
The motivation of the third law in a flavor-unified theory is to avoid the simple repetitive structure of one generational SM fermions.
Some of the examples that can lead to three-generational SM fermions by relaxing Georgi's third law can be found in Refs.~\cite{Frampton:1979cw,Frampton:1979tj,Barr:1979xt,Barr:2008pn,Chen:2021ovs,Chen:2021zwn,Rizzo:2022lpm}.
In this class of theories, only the anti-symmetric representations of the ${\rm SU}(N)$ fermions, denoted as $\[ N\,,k \]_{ \rep{F}}$, will be considered, so that no ${\rm SU}(3)_c/{\rm U}(1)_{\rm EM}$-exotic fermions can be present in the spectrum. 
The current LHC experiments~\cite{CMS:2022dwd,ATLAS:2022vkf} have confirmed the facts that the $125\,{\rm GeV}$ SM Higgs boson (i) only couples to the top quark with the natural $\sim \Oc(1)$ Yukawa coupling, and (ii) can also distinguish lighter SM fermions from different generations so that the suppressed and hierarchical Yukawa couplings are generated.
The flavor-unified theories have the property that both experimental facts of the SM Higgs boson can be naturally explained without extra flavor symmetries introduced by hand.
For example, we have found that the SM Higgs boson from the $\rep{35_H}$ in a two-generational ${\rm SU}(7)$ toy model~\cite{Chen:2022xge} only couples to the top quark at the tree level.

\para
Since the class of the flavor-unified theories were not much studied, we wish to reformulate Georgi's third law~\cite{Georgi:1979md}.
For our purpose, let us generalize the concept of ``generation'' into the concept of the IAFFS as follows.
\begin{define}

An IAFFS is a set of left-handed anti-symmetric fermions of $\sum_\Rc m_\Rc \, \Fc_L(\Rc)$, with $m_\Rc$ being the multiplicities of a particular fermion representation of $\Rc$.
Obviously, the anomaly-free condition reads $\sum_\Rc m_\Rc \, {\rm Anom}(  \Fc_L(\Rc) ) =0$.
We also require the following conditions to be satisfied for an IAFFS.
\begin{itemize}

\item The greatest common divisor (GCD) of the $\{ m_\Rc \}$ should satisfy that ${\rm GCD} \{  m_\Rc \} =1$.

\item The fermions in an IAFFS can no longer be removed, which would otherwise bring non-vanishing gauge anomalies.

\item There should not be any singlet, self-conjugate, or adjoint fermions in an IAFFS.

\end{itemize}

\end{define}
Obviously, one generational SM fermions as well as their unified irreps of $\repb{5_F} \oplus \rep{10_F}$ in the ${\rm SU}(5)$ theory in Eqs.~\eqref{eqs:SU5_decomp} form an IAFFS.
Based on the definition of the IAFFS, we conjecture the third law~\footnote{This third law is a weaker version of the original version by Georgi~\cite{Georgi:1979md}, in the sense that the current setup leads to three distinctive $\rep{10_F}$'s and three identical $\repb{5_F}$'s at the EW scale.
Given the decomposition of $\rep{10_F}= q_L \oplus {u_R}^c \oplus {e_R}^c$, one can find that both left- and right-handed components of up-type quarks transform differently, the left-handed (right-handed) components of down-type quarks (charged leptons) also transform differently.
Altogether, three-generational SM quarks/leptons must be distinguishable.
Georgi's solution of an ${\rm SU}(11)$ theory~\cite{Georgi:1979md} lead to three different $\overline{\mathbf{5_F}}$'s and $\mathbf{10_F}$'s.} of
\begin{conj}[Chen]

only distinctive IAFFSs without repetition are allowed in the GUT.

\end{conj}
The flavor sectors with $3\times \[ \repb{5_F} \oplus \rep{10_F} \]$ in the ${\rm SU}(5)$ theory or $3\times  \rep{16_F} $ in the ${\rm SO}(10)$ theory do not satisfy the third law.
Furthermore, we have also reported in Ref.~\cite{Chen:2022xge} that the early suggestion of the three-generational ${\rm SU}(7)$ models~\cite{Frampton:1979cw} cannot satisfy the third law since the repetitive IAFFSs can be found therein. 
Accordingly, the ${\rm SU}(N)$ chiral fermions can be partitioned into several distinctive IAFFSs as follows~\footnote{We use the notion of $\{ k \}^\prime$ in Eqs.~\eqref{eq:SUN_fermions_realistic} and \eqref{eq:DRS} to indicate distinctive IAFFSs according to the third law.}
\beqn\label{eq:SUN_fermions_realistic}
\{ f_L \}_{ {\rm SU}(N)}^{n_g}&=& \bigoplus_{ \{ k \}^\prime } \Big\{  \underbrace{ \overline{ \[ N\,, 1 \]_{ \rep{F}} }^\lambda \oplus   \[ N\,, k \]_{ \rep{F}} }_{ \textrm{rank-$k$ IAFFS}} \Big\} \,,\quad  \lambda =1 \,,...\,, m_k \,,\quad 2 \leq k\leq \[ \frac{N}{2} \]\,,
\eeqn
with 
\beqn
m_k &=& {\rm Anom} ( \[ N\,,k \]_{\rep{F}}) = \frac{ (N-2k)\, ( N-3)! }{ (N-k-1)!\, (k-1)!}  \,.
\eeqn

\para
Georgi also gave the rules of counting the SM fermion generations in such ${\rm SU}(N)$ theories~\cite{Georgi:1979md}, which read as follows
\begin{itemize}

\item The ${\rm SU}(N)$ fundamental irrep is decomposed under the ${\rm SU}(5)$ as $\[ N\,,1 \]_{\rep{F}}= (N-5) \times \rep{1_F} \oplus \rep{5_F}$. 
The decompositions of other higher-rank irreps can be obtained by tensor products.

\item For an ${\rm SU}(N)$ GUT, one can eventually decompose the set of anomaly-free fermion irreps into the ${\rm SU}(5)$ irreps of $(\rep{1_F}\,, \rep{5_F}\,, \repb{5_F}\,, \rep{10_F}\,, \repb{10_F} )$.

\item Count the multiplicity of each ${\rm SU}(5)$ irrep as $\nu_{\rep{5_F}}$ and so on, and the anomaly-free condition must lead to $\nu_{\rep{5_F}} + \nu_{\rep{10_F}} = \nu_{\repb{5_F}} + \nu_{\repb{10_F}}$.

\item The SM fermion generation is determined by $n_g= \nu_{\repb{5_F}} - \nu_{\rep{5_F}} = \nu_{\rep{10_F}} - \nu_{\repb{10_F}}$.

\end{itemize}
The multiplicity difference between the $\rep{10_F}$ and the $\repb{10_F}$ from a particular rank-$k$ IAFFS was found to be~\cite{Chen:2022xge}
\beqn\label{eq:SUN_multi}
&&  \nu_{ \rep{10_F}} [ N \,, k ]_{ \rep{F}}  - \nu_{ \repb{10_F}} [ N\,, k ]_{ \rep{F}} = \frac{ (N - 2k ) (N-5 )! }{ (k - 2)! ( N-k -2)! } \,, \quad 2 \leq k\leq \[ \frac{N}{2} \]  \,.
\eeqn
The total number of generations at the EW scale can be determined by~\cite{Georgi:1979md}
\beqn\label{eq:SUN_generation}
n_g&=& \sum_{  \{  k\}^\prime } \Big( \nu_{ \rep{10_F}} [ N \,, k ]_{ \rep{F}}  - \nu_{ \repb{10_F}} [ N\,, k ]_{ \rep{F}}  \Big)  \,.
\eeqn
By expressing the fermions in terms of the IAFFSs in Eq.~\eqref{eq:SUN_fermions_realistic}, one can immediately identify the emergent global DRS symmetries~\cite{Dimopoulos:1980hn} of
\beqn\label{eq:DRS}
\widetilde{ \Gc}_{\rm DRS}&=& \bigotimes_{ \{ k \}^\prime } \Big[  \widetilde{ {\rm SU}}(m_{k} ) \otimes \widetilde{ {\rm U}}(1)_{\lambda_k } \otimes \widetilde{ {\rm U}}(1)_{k} \Big]  \,,
\eeqn
when $m_{k } \geq 2$~\footnote{The original work by Dimopoulos, Raby, and Susskind~\cite{Dimopoulos:1980hn} considered the global symmetry in the rank-$2$ anti-symmetric ${\rm SU}(N+4)$ theories.}.
The emergent global DRS symmetries in the general rank-$k$ anti-symmetric chiral gauge theories were previously acknowledged in Refs.~\cite{Appelquist:2013oni,Shi:2015fna}.
With any rank-$k$ IAFFS, the global $\widetilde{ {\rm U}}(1)_{\lambda_k } \otimes \widetilde{ {\rm U}}(1)_{k}$ symmetries are broken by the ${\rm SU}(N)$ instanton.
One can define a non-anomalous global $\widetilde{ {\rm U}}(1)_{T_k }$ symmetry such that the mixed anomaly of $\[ {\rm SU}(N) \]^2 \cdot \widetilde{ {\rm U}}(1)_{T_k}$ vanishes.
According to Refs.~\cite{Appelquist:2013oni,Shi:2015fna}, one finds the general $\widetilde{ {\rm U}}(1)_T$ charge assignments of
\beqn\label{eq:rk_Tcharges_Ferm}
&& \Tc( \overline{ \[ N\,, 1 \]_{ \rep{F}} }^\lambda ) = - \frac{ N-2 }{d_k}  t\,, \quad  \Tc( \[ N\,, k \]_{ \rep{F}} ) = \frac{  N-2k }{ d_k} t \,, \non
&& d_k = {\rm GCD} ( N-2\,, N- 2k ) \,.
\eeqn
The other non-vanishing global anomalies read
\beqs
\beqn
&& \[  {\rm grav} \]^2 \cdot \widetilde{ {\rm U}}(1)_{T_k} = \frac{N ( N-2)! }{ d_k ( N-k)! \, k!} ( N-k -1) (N-2k ) (1-k )t \,, \\[1mm]
&& \[  \widetilde{ {\rm U}}(1)_{T_k} \]^3   = \Big[   \frac{ N! }{ k!\, ( N-k )! } ( N-2k)^3 - N ( N-2)^3 \Big]  \frac{ t^3 }{ d_k^3 }\,.
\eeqn
\eeqs

\para
With the $\widetilde{ {\rm U}}(1)_T$ charge assignments in Eq.~\eqref{eq:rk_Tcharges_Ferm}, we can further determine the $\widetilde{ {\rm U}}(1)_T$ charges for the Higgs field in the rank-$k$ IAFFS through the renormalizable Yukawa couplings of
\beqn\label{eq:rk_Tcharges_Higgs}
-\Lc_Y&=& \overline{ \[ N\,, 1 \]_{ \rep{F}}}^\lambda \otimes  \[ N\,, k \]_{ \rep{F}}  \otimes  \overline{ \[ N\,,  k - 1 \]_{ \rep{H}\,, \lambda } } \non
&+&  \[ N\,, k \]_{ \rep{F}}  \otimes \[ N\,, k \]_{ \rep{F}}  \otimes \overline{ \[ N\,, 2k \]_{ \rep{H}} }  + H.c.  \,, \non
&\Rightarrow& \Tc(   \overline{ \[ N\,,  k - 1 \]_{ \rep{H}\,, \lambda } }   ) = + \frac{ 2k -2 }{d_k}  t\,, \quad  \Tc(   \overline{ \[ N\,, 2k \]_{ \rep{H}} }  ) =  \frac{  4k - 2N }{ d_k} t \,.
\eeqn
The $\widetilde{ {\rm U}}(1)_T$ charges of the Higgs fields are the center of the ${\rm SU}(N)$ group.
Notice that the Yukawa coupling of $ \[ N\,, k \]_{ \rep{F}}  \otimes \[ N\,, k \]_{ \rep{F}}  \otimes \overline{ \[ N\,, 2k \]_{ \rep{H}} } + H.c.$ cannot exist for the odd-$k$ case, since the anti-symmetric property of the chiral fermion $\[ N\,, k \]_{ \rep{F}}$ leads this term to vanish.

\para
The three-generational ${\rm SU}(8)$ theory has the following anomaly-free chiral fermions
\beqn\label{eq:SU8_3gen_fermions}
\{ f_L \}_{ {\rm SU}(8)}^{n_g=3}&=& \Big[  \repb{8_F}^\lambda \oplus \rep{28_F} \Big] \bigoplus \Big[  \repb{8_F}^{ \dot \lambda } \oplus \rep{56_F} \Big] \,,~ {\rm dim}_{ \mathbf{F}}= 156\,, \non
&& \Lambda\equiv ( \lambda \,, \dot \lambda) \,,~ \lambda = ( 3\,, {\rm IV}\,, {\rm V}\,, {\rm VI}) \,, ~  \dot \lambda = (\dot 1\,, \dot 2\,, \dot {\rm VII}\,, \dot {\rm IIX}\,, \dot {\rm IX} )  \,,
\eeqn
with the undotted/dotted indices for the rank-$2$ IAFFS and the rank-$3$ IAFFS, respectively.
The Arabic numbers and the Roman numbers are used for the SM fermions and the heavy partner fermions, respectively.
According to Refs.~\cite{Georgi:1979md,Barr:2008pn,Chen:2022xge,Rizzo:2022lpm}, we find the ${\rm SU}(8)$ theory with the fermions in Eq.~\eqref{eq:SU8_3gen_fermions} is the minimal flavor-unified theory where three-generational SM fermions are embedded non-trivially.
The total number of generations can be determined by Eqs.~\eqref{eq:SUN_multi} and \eqref{eq:SUN_generation}.
The global DRS symmetries~\cite{Dimopoulos:1980hn} from the fermion contents in Eq.~\eqref{eq:SU8_3gen_fermions} are
\beqn\label{eq:SU8_DRS}
\widetilde{ \Gc}_{\rm DRS}&=&  \[ \widetilde{ {\rm SU} }(4)_{ \lambda }  \otimes \widetilde{ {\rm U} }(1)_{ \lambda }  \otimes \widetilde{ {\rm U} }(1)_2 \] \bigotimes   \[ \widetilde{ {\rm SU} }(5)_{\dot \lambda }  \otimes \widetilde{ {\rm U} }(1)_{ \dot \lambda }  \otimes \widetilde{ {\rm U} }(1)_3 \] \,,
\eeqn
at the GUT scale.

\para
The most general gauge-invariant Yukawa couplings are~\footnote{The other possible gauge-invariant Yukawa coupling of $\rep{56_F} \rep{56_F} \rep{28_H} + H.c.$ vanishes due to the rank-$3$ anti-symmetric indices of the $\rep{56_F}$~\cite{Barr:2008pn}.
The Yukawa coupling of $\rep{28_F} \rep{56_F} \rep{56_H} + H.c.$ was recently included in Refs.~\cite{Chen:2021ovs,Rizzo:2022lpm}.
}
\beqn\label{eq:SU8_Yukawa}
-\Lc_Y&=& {(Y_\Bc)_\lambda}^\kappa \repb{8_F}^\lambda \rep{28_F}  \repb{8_{H}}_{\,,\kappa} +  {(Y_\Dc)_{\dot \lambda} }^{\dot \kappa} \repb{8_F}^{\dot \lambda} \rep{56_F}  \repb{28_{H}}_{\,,\dot \kappa}  \non
&+&  Y_{\Tc\Cc}\rep{28_F} \rep{56_F} \rep{56_H}+  Y_\Tc \rep{28_F} \rep{28_F} \rep{70_H} + H.c.\,.
\eeqn
In the $\widetilde{ \Gc}_{\rm DRS}$-invariant limit, the Yukawa couplings are reduced to
\beqn\label{eq:SU8_Yukawa_DRS}
&&  {(Y_{\Bc} )_{\lambda}}^\kappa = Y_{\Bc} \, {\delta_{\lambda}}^\kappa \,,\quad  {(Y_{\Dc} )_{\dot \lambda }}^{\dot \kappa}  = Y_{\Dc}\,  {\delta_{\dot \lambda }}^{\dot \kappa}  \,.
\eeqn
Altogether, we collect the ${\rm SU}(8)$ Higgs fields as follows
\beqn\label{eq:SU8_Higgs}
 \{ H \}_{ {\rm SU}(8)}^{n_g=3} &=&  \repb{8_H}_{\,, \lambda}  \oplus \repb{28_H}_{\,, \dot \lambda }   \oplus \rep{56_H} \oplus \rep{70_H} \oplus \underline{ \rep{63_H} } \,, \non
 &&   \lambda = ( 3\,, {\rm IV}\,, {\rm V}\,, {\rm VI}) \,, ~  \dot \lambda = (\dot 1\,, \dot 2\,, \dot {\rm VII}\,, \dot {\rm IIX}\,, \dot {\rm IX} ) \,.
\eeqn
By partitioning the chiral fermions to the IAFFSs in Eq.~\eqref{eq:SUN_fermions_realistic}, the Higgs fields in the ${\rm SU}(8)$ theory can be determined according to the global DRS symmetries, while this point was not previously observed in Ref.~\cite{Barr:2008pn}.
The adjoint Higgs field of $\rep{63_H}$ is real and only responsible for the GUT scale symmetry breaking, while the others are complex.

%%%%%%%%%%%%%%%%%%%%%%%%%%%%
\subsection{The symmetry-breaking pattern}
\label{section:patterns}
%%%%%%%%%%%%%%%%%%%%%%%%%%%%

\para
The symmetry-breaking pattern defines a realistic GUT from its UV setup to the low-energy effective theories at different scales.
It was known that the Higgs representations are responsible for the gauge symmetry-breaking patterns, as well as the proper choices of the Higgs self couplings~\cite{Li:1973mq}.
For this purpose, we tabulate the symmetry-breaking patterns for the ${\rm SU}(N)$ groups with various Higgs representations in Tab.~\ref{tab:HiggsPatterns}.

\begin{table}[htp]
\begin{center}
\begin{tabular}{c|cc}
\hline\hline
 Higgs irrep&  dimension & pattern   \\
\hline\hline
fundamental  &  $N$  &  $ {\rm SU}(N-1)$  \\[1mm]
rank-$2$ symmetric  &  $\frac{1}{2}N(N+1)$  & ${\rm SO}(N)$   \\
  &  & or ${\rm SU} (N-1)$ \\[1mm]
rank-$2$ anti-symmetric  &  $\frac{1}{2}N(N-1)$  & ${\rm SO}(2k+1)\,, k=[ \frac{N}{2}]$  \\
  &  & or ${\rm SU} (N-2)$ \\[1mm]
adjoint  &  $N^2 -1$  &  ${\rm SU}(N-k) \otimes {\rm SU}(k)\otimes {\rm U}(1)\,, k=[ \frac{N}{2}]$ \\
 &   &  or $ {\rm SU}(N-1)$  \\
\hline\hline
\end{tabular}
\end{center}
\caption{The Higgs representations and the corresponding symmetry-breaking patterns for the $\gSU(N)$ group.}
\label{tab:HiggsPatterns}
\end{table}%

\para
We consider the following symmetry-breaking pattern of the ${\rm SU}(8)$ GUT
\beqn\label{eq:Pattern} 
%%%%%%%%%%%%%%%%%%%%%%%%%%%%%%%%%%%%%%%%%%%%%
&& {\rm SU}(8) \xrightarrow{ v_U } \Gc_{441} \xrightarrow{ v_{441} } \Gc_{341} \xrightarrow{v_{341} } \Gc_{331} \xrightarrow{ v_{331} } \Gc_{\rm SM} \xrightarrow{ v_{\rm EW} } \Gc_{\rm SM}^\prime  \,, \non
&&\Gc_{441} \equiv {\rm SU}(4)_{s} \otimes {\rm SU}(4)_W \otimes  {\rm U}(1)_{X_0 } \,, ~ \Gc_{341} \equiv {\rm SU}(3)_{c} \otimes {\rm SU}(4)_W \otimes  {\rm U}(1)_{X_1 } \,,\non
&&\Gc_{331} \equiv {\rm SU}(3)_{c} \otimes {\rm SU}(3)_W \otimes  {\rm U}(1)_{X_2 } \,,~ \Gc_{\rm SM} \equiv  {\rm SU}(3)_{c} \otimes {\rm SU}(2)_W \otimes  {\rm U}(1)_{Y } \,,\non
&& \Gc_{\rm SM}^\prime \equiv  {\rm SU}(3)_{c}  \otimes  {\rm U}(1)_{\rm EM} \,, ~ \textrm{with}~  v_U\gg v_{441}  \gg v_{341} \gg v_{331} \gg v_{\rm EW} \,.
\eeqn
At the GUT scale, the maximal symmetry-breaking pattern of $ {\rm SU}(8) \xrightarrow{v_U  } \Gc_{441}$ can be achieved~\cite{Li:1973mq} due to the VEV of the ${\rm SU}(8)$ adjoint Higgs field as follows
\beqn
&& \langle \rep{63_H} \rangle = {\rm diag} ( - \frac{1}{4} \mathbb{I}_{4\times 4} \,, +\frac{1}{4} \mathbb{I}_{4 \times 4} )v_U \,.
\eeqn
This symmetry-breaking pattern contains three intermediate scales between the GUT scale of $v_U$ and the EWSB scale of $v_{\rm EW}$.
We denote two extended weak symmetries above the EW scale as ${\rm SU}(4)_W \otimes  {\rm U}(1)_{X_1 }$ and ${\rm SU}(3)_W \otimes  {\rm U}(1)_{X_2 }$, since there can be both the left-handed and the right-handed quark/lepton multiplets under the corresponding extended weak symmetries through the decompositions.

%%%%%%%%%%%%%%%%%%%%%%%%%%%%
\subsection{Decompositions of the ${\rm SU}(8)$ Higgs fields}
\label{section:SU8_Higgs}
%%%%%%%%%%%%%%%%%%%%%%%%%%%%

\para
Below the GUT scale, all other intermediate symmetry-breaking stages will be due to the VEVs of the Higgs fields from the gauge-invariant Yukawa couplings of the ${\rm SU}(8)$ theory.
During the symmetry-breaking stage of $\Gc\to \Hc$, only the Higgs fields that contain the $\Hc$-singlet components are likely to develop the VEV for this symmetry-breaking stage.
We decompose Higgs fields in Eq.~\eqref{eq:SU8_Yukawa} into components that can be responsible for the sequential symmetry-breaking pattern in Eq.~\eqref{eq:Pattern}.
The corresponding decomposition rules are defined in App.~\ref{section:Br}.
For Higgs fields of $\repb{8_H}_{\,,\lambda }$, they read
\beqn\label{eq:SU8_Higgs_Br01}
\repb{8_H}_{\,,\lambda }  &\supset&  \boxed{ ( \repb{4} \,, \rep{1} \,, +\frac{1}{4} )_{\mathbf{H}\,, \lambda }  } \oplus  \underline{ ( \rep{1} \,, \repb{4} \,, -\frac{1}{4} )_{\mathbf{H}\,, \lambda} } \non
&\supset&  \boxed{ ( \rep{1} \,, \repb{4} \,, -\frac{1}{4} )_{\mathbf{H}\,, \lambda }   } \supset  \boxed{ ( \rep{1} \,, \repb{3} \,, -\frac{1}{3} )_{\mathbf{H}\,, \lambda} } \supset  \boxed{ ( \rep{1} \,, \repb{2} \,, -\frac{1}{2} )_{\mathbf{H}\,, \lambda} }   \,.
\eeqn
For Higgs fields of $\repb{28_H}_{\,,\dot \lambda} $, they read
\beqn\label{eq:SU8_Higgs_Br02}
%%%%%%%%%%%%%%%%%%%%%%%%%%%%%%%%%%%%%%%%%%%%%
\repb{28_H}_{\,,\dot \lambda} &\supset& ( \rep{6} \,, \rep{1} \,,  +\frac{1}{2} )_{\mathbf{H}\,, \dot\lambda }  \oplus  \underline{ ( \rep{1} \,, \rep{6} \,, -\frac{1}{2} )_{\mathbf{H}\,, \dot\lambda } } \oplus \underline{ ( \repb{4} \,, \repb{4} \,, 0 )_{\mathbf{H}\,, \dot\lambda } }  \non
&\supset & \underline{ ( \rep{1} \,, \rep{6} \,, -\frac{1}{2} )_{\mathbf{H}\,, \dot\lambda} } \oplus \boxed{ ( \rep{1} \,, \repb{4} \,, -\frac{1}{4} )_{\mathbf{H}\,, \dot\lambda }  }  \non
&\supset& \[ \boxed{ ( \rep{1} \,, \repb{3} \,, -\frac{1}{3} )_{\mathbf{H}\,, \dot\lambda }^\prime }  \oplus \underline{ ( \rep{1} \,, \rep{3} \,, -\frac{2}{3} )_{\mathbf{H}\,, \dot\lambda }}  \] \oplus \boxed{ ( \rep{1} \,, \repb{3} \,, -\frac{1}{3} )_{\mathbf{H}\,, \dot\lambda } }   \non
&\supset& \[ \boxed{ ( \rep{1} \,, \repb{2} \,, -\frac{1}{2} )_{\mathbf{H}\,, \dot\lambda }^\prime} \oplus \boxed{ (  \rep{1} \,, \rep{2} \,, -\frac{1}{2}  )_{\mathbf{H}\,, \dot\lambda }  }  \] \oplus \boxed{ ( \rep{1} \,, \repb{2} \,, -\frac{1}{2} )_{\mathbf{H}\,, \dot\lambda }}  \,.
\eeqn
%
%
%For Higgs field of $\rep{28_H}$, they read
%
%
%\beqn\label{eq:SU8_Higgs_Br03}
%%%%%%%%%%%%%%%%%%%%%%%%%%%%%%%%%%%%%%%%%%%%%
%\rep{28_H} &\supset& ( \rep{6} \,, \rep{1} \,,  -\frac{1}{2} )_{\mathbf{H} }  \oplus  \underline{ ( \rep{1} \,, \rep{6} \,, +\frac{1}{2} )_{\mathbf{H}} } \oplus \underline{ ( \rep{4} \,, \rep{4} \,, 0 )_{\mathbf{H}}  } \non
%
%& \supset& \underline{ \overbrace{  ( \rep{1} \,, \rep{6} \,, +\frac{1}{2} )_{\mathbf{H} } }^{ \rep{\Phi}_{ \rep{6}} }  } \oplus \boxed{  \overbrace{ ( \rep{1} \,, \rep{4} \,, +\frac{1}{4} )_{\mathbf{H}} }^{ \rep{\Phi}_{\rep{4}} } }   \non
%
%&\supset&  \[ \boxed{ ( \rep{1} \,, \rep{3} \,, +\frac{1}{3} )_{\mathbf{H} }^\prime }  \oplus ( \rep{1} \,, \repb{3} \,, +\frac{2}{3} )_{\mathbf{H} }^\prime  \] \oplus  \boxed{  ( \rep{1} \,, \rep{3} \,, +\frac{1}{3} )_{\mathbf{H}}}  \non
%&\supset&  \[ \boxed{( \rep{1} \,, \rep{2} \,, +\frac{1}{2} )_{\mathbf{H} }^\prime}  \oplus \boxed{ ( \rep{1} \,, \repb{2} \,, +\frac{1}{2} )_{\mathbf{H} }^\prime }  \] \oplus \boxed{ ( \rep{1} \,, \rep{2} \,, +\frac{1}{2} )_{\mathbf{H}} }  \,.
%\eeqn
%
%
For Higgs field of $\rep{56_H}$, they read
\beqn\label{eq:SU8_Higgs_Br04}
%%%%%%%%%%%%%%%%%%%%%%%%%%%%%%%%%%%%%%%%%%%%%
\rep{56_H} &\supset&  \underline{ ( \rep{1} \,, \repb{4} \,, +\frac{3}{4} )_{\mathbf{H}} } \oplus  ( \repb{4} \,, \rep{1} \,,  -\frac{3}{4} )_{\mathbf{H}} \oplus \underline{ ( \rep{4} \,, \rep{6} \,, +\frac{1}{4} )_{\mathbf{H}} } \oplus ( \rep{6} \,, \rep{4} \,,  -\frac{1}{4} )_{\mathbf{H}} \non
&\supset&  \underline{ ( \rep{1} \,, \repb{4} \,, +\frac{3}{4} )_{\mathbf{H}}  } \oplus  \underline{ ( \rep{1} \,, \rep{6} \,, +\frac{1}{2} )_{\mathbf{H}}^\prime   } \non
&\supset& \underline{ ( \rep{1} \,, \repb{3} \,, +\frac{2}{3} )_{\mathbf{H}} } \oplus \[  \boxed{ ( \rep{1} \,, \rep{3} \,, +\frac{1}{3} )_{\mathbf{H} }^{\prime\prime }  } \oplus ( \rep{1} \,, \repb{3} \,, +\frac{2}{3} )_{\mathbf{H} }^{\prime\prime}  \]   \non
&\supset&  \boxed{ ( \rep{1} \,, \repb{2} \,, +\frac{1}{2} )_{\mathbf{H}} } \oplus \[  \boxed{ ( \rep{1} \,, \rep{2} \,, +\frac{1}{2} )_{\mathbf{H} }^{\prime\prime } } \oplus \boxed{ ( \rep{1} \,, \repb{2} \,, +\frac{1}{2} )_{\mathbf{H} }^{\prime\prime} }  \]   \,.
\eeqn
For Higgs field of $\rep{70_H}$, they read
\beqn\label{eq:SU8_Higgs_Br05} 
%%%%%%%%%%%%%%%%%%%%%%%%%%%%%%%%%%%%%%%%%%%%%
\rep{70_H} &\supset& ( \rep{1} \,, \rep{1 } \,, -1 )_{\mathbf{H}}^{ \prime \prime } \oplus ( \rep{1} \,, \rep{1 } \,, +1 )_{\mathbf{H}}^{ \prime \prime \prime \prime } \oplus \underline{ ( \rep{4} \,, \repb{4} \,, +\frac{1}{2} )_{\mathbf{H}} } \oplus \underline{ ( \repb{4} \,, \rep{4} \,, -\frac{1}{2} )_{\mathbf{H}}  } \oplus ( \rep{6 } \,, \rep{6 } \,, 0 )_{\mathbf{H}}  \non
&\supset& \underline{ ( \rep{1} \,, \repb{4} \,, +\frac{3}{4} )_{\mathbf{H}}^\prime  } \oplus \underline{  ( \rep{1} \,, \rep{4} \,, -\frac{3}{4} )_{\mathbf{H}}^\prime   }  \non
&\supset&  \underline{ ( \rep{1} \,, \repb{3} \,, +\frac{2}{3} )_{\mathbf{H}}^{\prime\prime\prime} } \oplus  \underline{ ( \rep{1} \,, \rep{3} \,, -\frac{2}{3} )_{\mathbf{H}}^{\prime\prime\prime} } \non
&\supset & \boxed{ ( \rep{1} \,, \repb{2} \,, +\frac{1}{2} )_{\mathbf{H}}^{\prime \prime\prime } } \oplus \boxed{  ( \rep{1} \,, \rep{2} \,, -\frac{1}{2} )_{\mathbf{H}}^{\prime \prime\prime} }  \,. 
\eeqn

%%%%%%%%%%%%%%%%%%%%%%%%%%%%%%%%%%%%%%%%%%%%%%%
\begin{table}[htp]
\begin{center}
\begin{tabular}{c|cccc}
\hline \hline
Higgs & $\Gc_{441} \to \Gc_{341}$ & $\Gc_{341} \to \Gc_{331}$  &  $\Gc_{331} \to \Gc_{\rm SM} $ &  $\Gc_{\rm SM} \to \Gc_{\rm SM}^\prime$  \\
\hline\hline
$\repb{8_H}_{\,,\lambda}$ & \cmark  & \cmark  &  \cmark  & \cmark  \\
$\repb{28_H}_{\,,\dot \lambda}$ &  \xmark  & \cmark  &  \cmark  & \cmark  \\
%$\rep{28_H}$ & \xmark  & \cmark  &  \cmark  &  \cmark  \\
$\rep{56_H}$ & \xmark  & \xmark  & \cmark  &  \cmark \\
$\rep{70_H}$ & \xmark  & \xmark  & \xmark  & \cmark \\
\hline\hline
\end{tabular}
\end{center}
\caption{The Higgs fields and their symmetry-breaking directions in the ${\rm SU}(8)$ GUT.
The \cmark and \xmark represent possible and impossible symmetry breaking directions for a given Higgs field.
%The flavor indices for the Higgs fields that develop VEVs at different stages are also specified in the parentheses.
}
\label{tab:SU8Higgs_directions}
\end{table}%
%%%%%%%%%%%%%%%%%%%%%%%%%%%%%%%%%%%%%%%%%%%%%%%

\para
All components that are likely to develop VEVs for the sequential symmetry-breaking stages are framed with boxes, with their original components in the UV stages marked with underlines.
Other ${\rm SU}(3)_c$-exact and/or ${\rm U}(1)_{\rm em}$-exact components are neglected after the symmetry-breaking stage of $\Gc_{441} \to \Gc_{341}$.
Accordingly, we mark the allowed/disallowed symmetry-breaking directions for all Higgs fields in Tab.~\ref{tab:SU8Higgs_directions}.
At this point, these components are only determined by whether they contain the singlets of the unbroken subgroups.
From the decomposition in Eq.~\eqref{eq:SU8_Higgs_Br05}, we found that the $\rep{70_H}$ only contains the EWSB components of $( \rep{1} \,, \repb{2} \,, +\frac{1}{2} )_{\mathbf{H}}^{\prime \prime\prime }$ and $( \rep{1} \,, \rep{2} \,, -\frac{1}{2} )_{\mathbf{H}}^{\prime \prime\prime}$, while all other Higgs fields contain components for the intermediate symmetry-breaking stages.
Naively, one may expect both components of $( \rep{1} \,, \repb{2} \,, +\frac{1}{2} )_{\mathbf{H}}^{\prime \prime\prime }$ and $( \rep{1} \,, \rep{2} \,, -\frac{1}{2} )_{\mathbf{H}}^{\prime \prime\prime}$ to develop the VEVs of the same order.
This is distinctive from the two-generational ${\rm SU}(7)$ model~\cite{Chen:2022xge}, where the Higgs field of $\rep{35_H}$ from the rank-$2$ IAFFS therein only contains one EWSB component.
As we shall show by analyzing their generalized $\widetilde{ {\rm U}}(1)_T$ symmetries, only the $( \rep{1} \,, \repb{2} \,, +\frac{1}{2} )_{\mathbf{H}}^{\prime \prime\prime }$ can develop the EWSB VEV.
Thus, we expect only the component of $( \rep{1} \,, \repb{2} \,, +\frac{1}{2} )_{\mathbf{H}}^{\prime \prime\prime }\subset \rep{70_H}$ that corresponds to the SM Higgs doublet at the EW scale.
Also, the Higgs field of $\rep{56_H}$ contains both the $\Gc_{331} \to \Gc_{\rm SM}$ breaking component and the EWSB components.
Through the definition of the generalized $\widetilde{ {\rm U}}(1)_T$ symmetries, none of these components can actually develop any VEV since they are always charged.

%{\color{red}
\para
For our later discussions of the symmetry-breaking pattern in the ${\rm SU}(8)$ theory, we denote the minimal set of Higgs VEVs as follows
\beqs\label{eqs:SU8_Higgs_VEVs_mini}
\beqn
\Gc_{441} \to \Gc_{341} ~&:&~ \langle ( \repb{4} \,, \rep{1} \,, +\frac{1}{4} )_{\mathbf{H}\,, {\rm IV}} \rangle \equiv W_{ \repb{4}\,, {\rm IV}}\,, \label{eq:SU8_Higgs_VEVs_mini01}\\[1mm]
%%%%%%%%%%%%%%%%%%%%%%%%%%%%%%%%%%%%%%%%%%%%%
\Gc_{341} \to \Gc_{331} ~&:&~ \langle ( \rep{1} \,, \repb{4} \,, -\frac{1}{4} )_{\mathbf{H}\,, {\rm V} } \rangle \equiv  w_{\repb{4}\,, {\rm V} }\,,~  \langle ( \rep{1} \,, \repb{4} \,, -\frac{1}{4} )_{\mathbf{H}\,, \dot {\rm VII} } \rangle \equiv  w_{\repb{4}\,, \dot {\rm VII} } \,, \label{eq:SU8_Higgs_VEVs_mini02}\\[1mm]
%%%%%%%%%%%%%%%%%%%%%%%%%%%%%%%%%%%%%%%%%%%%%
\Gc_{331} \to \Gc_{\rm SM} ~&:&~  \langle ( \rep{1} \,, \repb{3} \,, -\frac{1}{3} )_{\mathbf{H}\,, {\rm VI}} \rangle \equiv V_{ \repb{3}\,, {\rm VI} } \,, \non
&& \langle ( \rep{1} \,, \repb{3} \,, -\frac{1}{3} )_{\mathbf{H}\,, \dot {\rm IIX}  }^\prime \rangle \equiv V_{ \repb{3}\,, \dot {\rm IIX} }^\prime \,,~  \langle ( \rep{1} \,, \repb{3} \,, -\frac{1}{3} )_{\mathbf{H}\,, \dot {\rm IX} } \rangle \equiv V_{ \repb{3}\,, \dot {\rm IX} }  \,,\label{eq:SU8_Higgs_VEVs_mini03}\\[1mm]
%&& \langle ( \rep{1} \,, \rep{3} \,, +\frac{1}{3} )_{\mathbf{H} } \rangle \equiv V_{ \rep{3}} \,, ~ \langle ( \rep{1} \,, \rep{3} \,, +\frac{1}{3} )_{\mathbf{H} }^{\prime\prime} \rangle \equiv V_{ \rep{3}}^{\prime\prime} \,, \\[1mm]
%%%%%%%%%%%%%%%%%%%%%%%%%%%%%%%%%%%%%%%%%%%%%
 {\rm EWSB} ~&:&~   \langle ( \rep{1} \,, \repb{2} \,, +\frac{1}{2} )_{\mathbf{H} }^{ \prime\prime \prime} \rangle \equiv v_t   \,,\label{eq:SU8_Higgs_VEVs_mini04}
\eeqn
\eeqs
according to the Higgs decompositions in Eqs.~\eqref{eq:SU8_Higgs_Br01}, \eqref{eq:SU8_Higgs_Br02}, and~\eqref{eq:SU8_Higgs_Br05}.
%}

%%%%%%%%%%%%%%%%%%%%%%%%%%%%
\subsection{Decompositions of the ${\rm SU}(8)$ fermions}
\label{section:SU8_fermions}
%%%%%%%%%%%%%%%%%%%%%%%%%%%%

%###############################################################################
\begin{table}[htp] {\small
\begin{center}
\begin{tabular}{c|c|c|c|c}
\hline \hline
   $\gSU(8)$   &  $\Gc_{441}$  & $\Gc_{341}$  & $\Gc_{331}$  &  $\Gc_{\rm SM}$  \\
\hline\hline
 $\repb{ 8_F}^\Lambda$   & $( \repb{4} \,, \rep{1}\,,  +\frac{1}{4} )_{ \mathbf{F} }^\Lambda$  & $(\repb{3} \,, \rep{1} \,, +\frac{1}{3} )_{ \mathbf{F} }^\Lambda$  & $(\repb{3} \,, \rep{1} \,, +\frac{1}{3} )_{ \mathbf{F} }^\Lambda$  &  $( \repb{3} \,, \rep{ 1}  \,, +\frac{1}{3} )_{ \mathbf{F} }^{\Lambda}~:~ { \Dc_R^\Lambda}^c$  \\[1mm]
 &  &  $( \rep{1} \,, \rep{1} \,, 0)_{ \mathbf{F} }^{\Lambda}$  &  $( \rep{1} \,, \rep{1} \,, 0)_{ \mathbf{F} }^{\Lambda}$ &  $( \rep{1} \,, \rep{1} \,, 0)_{ \mathbf{F} }^{\Lambda} ~:~ \check \Nc_L^{\Lambda }$  \\[1.5mm]
%%%%%%%%%%%%%%%%%%%%%%%%%%%%%%%%%%%%%%%%%%%%%
 & $(\rep{1}\,, \repb{4}  \,,  -\frac{1}{4})_{ \mathbf{F} }^\Lambda$  &  $(\rep{1}\,, \repb{4}  \,,  -\frac{1}{4})_{ \mathbf{F} }^\Lambda$  &  $( \rep{1} \,, \repb{3} \,,  -\frac{1}{3})_{ \mathbf{F} }^{\Lambda}$  &  $( \rep{1} \,, \repb{2} \,,  -\frac{1}{2})_{ \mathbf{F} }^{\Lambda } ~:~\Lc_L^\Lambda =( \Ec_L^\Lambda \,, - \Nc_L^\Lambda )^T$   \\[1mm]
 &   &   &   &  $( \rep{1} \,, \rep{1} \,,  0)_{ \mathbf{F} }^{\Lambda^\prime} ~:~ \check \Nc_L^{\Lambda^\prime }$  \\[1mm]
  &   &  &   $( \rep{1} \,, \rep{1} \,, 0)_{ \mathbf{F} }^{\Lambda^{\prime\prime} }$ &   $( \rep{1} \,, \rep{1} \,, 0)_{ \mathbf{F} }^{\Lambda^{\prime\prime} } ~:~ \check \Nc_L^{\Lambda^{\prime \prime} }$   \\[1.5mm]   
 %%%%%%%%%%%%%%%%%%%%%%%%%%%%%%%%%%%%%%%%%%%%%
\hline\hline
\end{tabular}
\caption{
The $\gSU(8)$ fermion representation of $\repb{8_F}^\Lambda$ under the $\Gc_{441}\,,\Gc_{341}\,, \Gc_{331}\,, \Gc_{\rm SM}$ subgroups for the three-generational ${\rm SU}(8)$ theory, with $\Lambda=(\lambda \,, \dot \lambda)$.
Here, we denote $\underline{ {\Dc_R^\Lambda}^c={d_R^\Lambda}^c}$ for the SM right-handed down-type quarks, and ${\Dc_R^\Lambda}^c={\DG_R^\Lambda}^c$ for the right-handed down-type heavy partner quarks.
Similarly, we denote $\underline{ \Lc_L^\Lambda = ( e_L^\Lambda \,, - \nu_L^\Lambda)^T}$ for the left-handed SM lepton doublets, and $\Lc_L^\Lambda =( \eG_L^\Lambda \,, - \nG_L^\Lambda )^T$ for the left-handed heavy lepton doublets.
All left-handed neutrinos of $\check \Nc_L$ are sterile neutrinos that do not couple to the EW gauge bosons.}
\label{tab:SU8_8barferm}
\end{center} 
}
\end{table}%
%###############################################################################

%###############################################################################
\begin{table}[htp] {\small
\begin{center}
\begin{tabular}{c|c|c|c|c}
\hline \hline
   $\gSU(8)$   &  $\Gc_{441}$  & $\Gc_{341}$  & $\Gc_{331}$  &  $\Gc_{\rm SM}$  \\
\hline \hline
 $\rep{28_F}$   & $( \rep{6}\,, \rep{ 1} \,, - \frac{1}{2})_{ \mathbf{F}}$ &  $( \rep{3}\,, \rep{ 1} \,, - \frac{1}{3})_{ \mathbf{F}}$   & $( \rep{3}\,, \rep{ 1} \,, - \frac{1}{3})_{ \mathbf{F}}$  & $( \rep{3}\,, \rep{ 1} \,, - \frac{1}{3})_{ \mathbf{F}} ~:~\DG_L$  \\[1mm]
                        &   & $( \repb{3}\,, \rep{ 1} \,, - \frac{2}{3})_{ \mathbf{F}}$  & $( \repb{3}\,, \rep{ 1} \,, - \frac{2}{3})_{ \mathbf{F}}$  & $\underline{( \repb{3}\,, \rep{ 1} \,, - \frac{2}{3})_{ \mathbf{F}}~:~ {t_R }^c }$   \\[1.5mm]
%%%%%%%%%%%%%%%%%%%%%%%%%%%%%%%%%%%%%%%%%%%%%
                        & $( \rep{1}\,, \rep{ 6} \,, +\frac{1}{2})_{ \mathbf{F}}$ & $( \rep{1}\,, \rep{ 6} \,, +\frac{1}{2})_{ \mathbf{F}}$   &  $( \rep{1}\,, \rep{ 3} \,, +\frac{1}{3})_{ \mathbf{F}}$ & $( \rep{1}\,, \rep{2} \,, +\frac{1}{2})_{ \mathbf{F}} ~:~( {\eG_R }^c \,, { \nG_R }^c)^T$  \\[1mm]
                       &   &   &   & $( \rep{1}\,, \rep{1} \,, 0 )_{ \mathbf{F}} ~:~ \check \nG_R^c $ \\[1mm]
                       &   &   & $( \rep{1}\,, \repb{ 3} \,, +\frac{2}{3})_{ \mathbf{F}}$  & $( \rep{1}\,, \repb{2} \,, +\frac{1}{2})_{ \mathbf{F}}^\prime ~:~( { \nG_R^{\prime} }^c\,, - {\eG_R^{\prime} }^c  )^T$   \\[1mm]
                       &   &   &   & $\underline{ ( \rep{1}\,, \rep{1} \,, +1 )_{ \mathbf{F}} ~:~ {\tau_R}^c}$ \\[1.5mm]
%%%%%%%%%%%%%%%%%%%%%%%%%%%%%%%%%%%%%%%%%%%%%
                        & $( \rep{4}\,, \rep{4} \,,  0)_{ \mathbf{F}}$ &  $( \rep{3}\,, \rep{4} \,,  -\frac{1}{12})_{ \mathbf{F}}$   & $( \rep{3}\,, \rep{3} \,,  0)_{ \mathbf{F}}$  & $\underline{ ( \rep{3}\,, \rep{2} \,,  +\frac{1}{6})_{ \mathbf{F}}~:~ (t_L\,, b_L)^T}$  \\[1mm]
                        &   &   &   & $( \rep{3}\,, \rep{1} \,,  -\frac{1}{3})_{ \mathbf{F}}^{\prime} ~:~\DG_L^\prime$  \\[1mm]
                        &   &   & $( \rep{3}\,, \rep{1} \,,  -\frac{1}{3})_{ \mathbf{F}}^{\prime\prime}$  & $( \rep{3}\,, \rep{1} \,,  -\frac{1}{3})_{ \mathbf{F}}^{\prime\prime} ~:~\DG_L^{\prime \prime}$ \\[1mm]
                        &   & $( \rep{1}\,, \rep{4} \,,  +\frac{1}{4} )_{ \mathbf{F}}$  & $( \rep{1}\,, \rep{3} \,,  +\frac{1}{3} )_{ \mathbf{F}}^{\prime\prime}$  & $( \rep{1}\,, \rep{2} \,,  +\frac{1}{2} )_{ \mathbf{F}}^{\prime\prime} ~:~( {\eG_R^{\prime\prime} }^c \,, { \nG_R^{\prime\prime}}^c )^T$  \\[1mm] 
                        &   &   &   & $( \rep{1}\,, \rep{1}\,, 0)_{ \mathbf{F}}^{\prime} ~:~ \check \nG_R^{\prime\,c}$ \\[1mm]  
                        &   &   & $( \rep{1}\,, \rep{1}\,, 0)_{ \mathbf{F}}^{\prime\prime}$ & $( \rep{1}\,, \rep{1}\,, 0)_{ \mathbf{F}}^{\prime\prime} ~:~\check \nG_R^{\prime \prime \,c}$ \\[1.5mm]  
 %%%%%%%%%%%%%%%%%%%%%%%%%%%%%%%%%%%%%%%%%%%%%
\hline\hline
\end{tabular}
\caption{
The $\gSU(8)$ fermion representation of $\rep{28_F}$ under the $\Gc_{441}\,,\Gc_{341}\,, \Gc_{331}\,, \Gc_{\rm SM}$ subgroups for the three-generational ${\rm SU}(8)$ theory.
All irreps for SM fermions are marked with underlines.}
\label{tab:SU8_28ferm}
\end{center}
}
\end{table}%
%###############################################################################

%###############################################################################
\begin{table}[htp] {\small
\begin{center}
\begin{tabular}{c|c|c|c|c}
\hline \hline
   $\gSU(8)$   &  $\Gc_{441}$  & $\Gc_{341}$  & $\Gc_{331}$  &  $\Gc_{\rm SM}$  \\
\hline \hline
     $\rep{56_F}$   & $( \rep{ 1}\,, \repb{4} \,, +\frac{3}{4})_{ \mathbf{F}}$  &  $( \rep{ 1}\,, \repb{4} \,, +\frac{3}{4})_{ \mathbf{F}}$ & $( \rep{ 1}\,, \repb{3} \,, +\frac{2}{3})_{ \mathbf{F}}^\prime$   &  $( \rep{ 1}\,, \repb{2} \,, +\frac{1}{2})_{ \mathbf{F}}^{\prime\prime\prime} ~:~( {\nG_R^{\prime\prime\prime }}^c \,, -{\eG_R^{\prime\prime\prime } }^c )^T$  \\[1mm]
                            &   &   &   & $\underline{( \rep{ 1}\,, \rep{1} \,, +1)_{ \mathbf{F}}^{\prime} ~:~ {e_R}^c~{\rm or}~ {\mu_R}^c }$ \\[1mm]
                            &   &   & $( \rep{ 1}\,, \rep{1} \,, +1)_{ \mathbf{F}}^{\prime\prime}$  & $( \rep{ 1}\,, \rep{1} \,, +1)_{ \mathbf{F}}^{\prime \prime} ~:~{\EG_R}^c$   \\[1.5mm]
%%%%%%%%%%%%%%%%%%%%%%%%%%%%%%%%%%%%%%%%%%%%%
                       & $( \repb{ 4}\,, \rep{1} \,, -\frac{3}{4})_{ \mathbf{F}}$  &  $( \repb{3}\,, \rep{1} \,, -\frac{2}{3})_{ \mathbf{F}}^{\prime}$ & $( \repb{3}\,, \rep{1} \,, -\frac{2}{3})_{ \mathbf{F}}^\prime$  & $\underline{ ( \repb{3}\,, \rep{1} \,, -\frac{2}{3})_{ \mathbf{F}}^{\prime} ~:~ {u_R}^c~{\rm or}~ {c_R}^c }$ \\[1mm]
                       &   &  $( \rep{1}\,, \rep{1} \,, -1)_{ \mathbf{F}}$ & $( \rep{1}\,, \rep{1} \,, -1)_{ \mathbf{F}}$  &  $( \rep{1}\,, \rep{1} \,, -1)_{ \mathbf{F}} ~:~\EG_L$  \\[1.5mm]
%%%%%%%%%%%%%%%%%%%%%%%%%%%%%%%%%%%%%%%%%%%%%
                       & $( \rep{ 4}\,, \rep{6} \,, +\frac{1}{4})_{ \mathbf{F}}$  &  $( \rep{3}\,, \rep{6} \,, +\frac{1}{6})_{ \mathbf{F}}$ & $( \rep{3}\,, \rep{3} \,, 0 )_{ \mathbf{F}}^\prime$ & $\underline{ ( \rep{3}\,, \rep{2} \,, +\frac{1}{6} )_{ \mathbf{F}}^{\prime} ~:~ (u_L\,,d_L)^T }$  \\[1mm]
                       &   &   &   &  ${\rm or}~\underline{ (c_L\,, s_L)^T}$ \\[1mm]
                       &   &   &   & $( \rep{3}\,, \rep{1} \,, -\frac{1}{3})_{ \mathbf{F}}^{\prime\prime \prime } ~:~\DG_L^{\prime \prime \prime}$ \\[1mm]
                       &   &   & $( \rep{3}\,, \repb{3} \,, +\frac{1}{3})_{ \mathbf{F}}$ & $( \rep{3}\,, \repb{2} \,, +\frac{1}{6})_{ \mathbf{F}}^{\prime\prime} ~:~ (\dG_L \,, - \uG_L )^T$   \\[1mm]
                       &   &   &   & $( \rep{3}\,, \rep{1} \,, +\frac{2}{3})_{ \mathbf{F}} ~:~\UG_L$  \\[1mm]
                       &   & $( \rep{1}\,, \rep{6} \,, +\frac{1}{2})_{ \mathbf{F}}^\prime$ & $( \rep{1}\,, \rep{3} \,, +\frac{1}{3})_{ \mathbf{F}}^\prime $ & $( \rep{1}\,, \rep{2} \,, +\frac{1}{2})_{ \mathbf{F}}^{\prime\prime \prime \prime} ~:~ ( {\eG_R^{\prime\prime\prime\prime }}^c \,, {\nG_R^{\prime\prime\prime\prime } }^c )^T$ \\[1mm]
                       &   &   &   & $( \rep{1}\,, \rep{1} \,, 0 )_{ \mathbf{F}}^{\prime\prime \prime} ~:~ {\check \nG_R}^{\prime \prime\prime \,c}$ \\[1mm]
                       &   &   & $( \rep{1}\,, \repb{3} \,, +\frac{2}{3})_{ \mathbf{F}}^{\prime\prime}$  & $( \rep{1}\,, \repb{2} \,, +\frac{1}{2})_{ \mathbf{F}}^{\prime\prime \prime \prime \prime} ~:~( {\nG_R^{\prime\prime\prime\prime\prime }}^c \,, -{\eG_R^{\prime\prime\prime\prime\prime } }^c )^T$  \\[1mm]
                       &   &   &   & $\underline{ ( \rep{1}\,, \rep{1} \,, +1 )_{ \mathbf{F}}^{\prime\prime \prime } ~:~ {\mu_R}^c~{\rm or}~ {e_R}^c  }$ \\[1.5mm]
%%%%%%%%%%%%%%%%%%%%%%%%%%%%%%%%%%%%%%%%%%%%%
                       & $( \rep{ 6}\,, \rep{4} \,, -\frac{1}{4})_{ \mathbf{F}}$  & $( \rep{3}\,, \rep{4} \,, -\frac{1}{12})_{ \mathbf{F}}^\prime$ & $( \rep{3}\,, \rep{3} \,, 0)_{ \mathbf{F}}^{\prime\prime}$ & $\underline{ ( \rep{3}\,, \rep{2} \,, +\frac{1}{6})_{ \mathbf{F}}^{\prime\prime \prime } ~:~ (c_L\,, s_L)^T} $ \\[1mm]
                        &   &   &   & or~$\underline{ (u_L\,, d_L)^T}$  \\[1mm]
                       &   &   &   &  $( \rep{3}\,, \rep{1} \,, -\frac{1}{3})_{ \mathbf{F}}^{\prime \prime \prime \prime} ~:~\DG_L^{\prime \prime \prime\prime}$ \\[1mm]
                       &   &   &  $( \rep{3}\,, \rep{1} \,, -\frac{1}{3})_{ \mathbf{F}}^{\prime \prime \prime\prime \prime}$ & $( \rep{3}\,, \rep{1} \,, -\frac{1}{3})_{ \mathbf{F}}^{\prime \prime \prime \prime \prime} ~:~ \DG_L^{\prime \prime \prime\prime \prime}$ \\[1mm]
                       &   & $( \repb{3}\,, \rep{4} \,, -\frac{5}{12})_{ \mathbf{F}}$ & $( \repb{3}\,, \rep{3} \,, -\frac{1}{3})_{ \mathbf{F}}$ & $( \repb{3}\,, \rep{2} \,, -\frac{1}{6})_{ \mathbf{F}} ~:~ ( {\dG_R}^c \,,{\uG_R}^c )^T$  \\[1mm]
                       &   &   &   & $( \repb{3}\,, \rep{1} \,, -\frac{2}{3})_{ \mathbf{F}}^{\prime \prime} ~:~{\UG_R}^c$  \\[1mm]
                       &   &   & $( \repb{3}\,, \rep{1} \,, -\frac{2}{3})_{ \mathbf{F}}^{\prime \prime \prime}$ & $\underline{ ( \repb{3}\,, \rep{1} \,, -\frac{2}{3})_{ \mathbf{F}}^{\prime \prime \prime} ~:~ {c_R}^c~{\rm or}~ {u_R}^c  }$  \\[1.5mm]
 %%%%%%%%%%%%%%%%%%%%%%%%%%%%%%%%%%%%%%%%%%%%%
\hline\hline
\end{tabular}
\caption{
The $\gSU(8)$ fermion representation of $\rep{56_F}$ under the $\Gc_{441}\,,\Gc_{341}\,, \Gc_{331}\,, \Gc_{\rm SM}$ subgroups for the three-generational ${\rm SU}(8)$ theory. 
All irreps for SM fermions are marked with underlines.
}
\label{tab:SU8_56ferm}
\end{center}
}
\end{table}%
%###############################################################################

\para
By following the symmetry-breaking pattern in Eq.~\eqref{eq:Pattern}, we tabulate the fermion representations at various stages of the ${\rm SU}(8)$ theory in Tabs.~\ref{tab:SU8_8barferm}, \ref{tab:SU8_28ferm}, and \ref{tab:SU8_56ferm}. 
All charges are obtained according to Eqs.~\eqref{eq:X1charge_4sfund}, \eqref{eq:X2charge_4Wfund}, \eqref{eq:Ycharge_4Wfund}, and~\eqref{eq:Qcharge_4Wfund}.
According to the counting rule by Georgi~\cite{Georgi:1979md}, it is straightforward to find $n_g=3$ in the current setup.
Notice that the original counting rule was performed by decomposing all ${\rm SU}(8)$ irreps into the ${\rm SU}(5)$ irreps, and one may ask if the rule is valid when the symmetry-breaking pattern considered in Eq.~\eqref{eq:Pattern} has no subgroup of ${\rm SU}(5)$.
Explicitly, our decompositions in Tabs.~\ref{tab:SU8_8barferm}, \ref{tab:SU8_28ferm}, and \ref{tab:SU8_56ferm} confirm that the counting rule is independent of the symmetry-breaking pattern.
All chiral fermions are named by their irreps of the $\Gc_{\rm SM}$.
For the right-handed quarks of ${\Dc_R^\Lambda}^c$, they are named as follows
\beqn\label{eq:DR_names}
&& {\Dc_R^{ \dot 1} }^c \equiv {d_R}^c \,, ~ {\Dc_R^{  \dot 2} }^c \equiv {s_R}^c \,,~ {\Dc_R^{\dot {\rm VII} } }^c \equiv {\DG_R^{\prime\prime\prime\prime \prime }}^c \,, ~  {\Dc_R^{\dot {\rm IIX} } }^c \equiv {\DG_R^{\prime\prime \prime }}^c  \,,~ {\Dc_R^{  \dot {\rm IX} } }^c \equiv {\DG_R^{\prime\prime\prime \prime }}^c \,, \non
&&{\Dc_R^{3} }^c \equiv {b_R}^c \,,~  {\Dc_R^{\rm IV } }^c \equiv {\DG_R}^c \,, ~  {\Dc_R^{\rm V } }^c \equiv {\DG_R^{\prime\prime }}^c  \,,~ {\Dc_R^{\rm VI } }^c \equiv {\DG_R^{\prime }}^c  \,.
\eeqn
For the left-handed ${\rm SU}(2)_W$ lepton doublets of $(\Ec_L^\Lambda \,, - \Nc_L^\Lambda )^T$, they are named as follows
\beqn\label{eq:ELNL_names}
&&   ( \Ec_L^{ \dot 1} \,,  - \Nc_L^{\dot 1})  \equiv (e_L\,, - \nu_{e\,L} ) \,, ~( \Ec_L^{  \dot 2} \,,   - \Nc_L^{\dot 2})  \equiv( \mu_L \,, - \nu_{\mu\,L} )  \,, \non
&&  ( \Ec_L^{ \dot {\rm VII} } \,,  - \Nc_L^{ \dot {\rm VII} })  \equiv ( \eG_L^{ \prime\prime \prime \prime} \,, - \nG_L^{\prime\prime \prime \prime } )  \,, ~ ( \Ec_L^{ \dot {\rm IIX} } \,, - \Nc_L^{ \dot {\rm IIX} }  )  \equiv ( \eG_L^{ \prime\prime  \prime} \,, - \nG_L^{\prime\prime \prime} ) \,,~  ( \Ec_L^{\dot {\rm IX} } \,,  - \Nc_L^{ \dot {\rm IX} } ) \equiv ( \eG_L^{ \prime\prime \prime \prime \prime } \,,  - \nG_L^{\prime\prime \prime\prime \prime} )  \,,\non
&&  ( \Ec_L^{ 3} \,, - \Nc_L^{3}) \equiv ( \tau_L \,, - \nu_{\tau\,L})\,,~  ( \Ec_L^{\rm IV } \,, - \Nc_L^{\rm IV }) \equiv ( \eG_L^{\prime\prime} \,, - \nG_L^{\prime\prime} ) \,, \non
&& ( \Ec_L^{\rm V } \,, -  \Nc_L^{\rm V }) \equiv ( \eG_L \,, - \nG_L )  \,,~ ( \Ec_L^{\rm VI } \,, - \Nc_L^{\rm VI } ) \equiv  ( \eG_L^\prime\,, - \nG_L^\prime ) \,.
\eeqn
Through the analyses below, we shall see that all heavy $(\Dc^\Lambda\,, \Ec^\Lambda\,, \Nc^\Lambda)$ (with $\Lambda={\rm IV}\,, \ldots \,,\dot {\rm IX}$ and named by Gothic fonts in Tabs.~\ref{tab:SU8_28ferm} and \ref{tab:SU8_56ferm}) acquire vectorlike masses during the intermediate symmetry-breaking stages.
For the left-handed sterile neutrinos of $(  \check \Nc_L^\Lambda \,, \check \Nc_L^{\Lambda^\prime } \,, \check \Nc_L^{\Lambda^{ \prime\prime} } )$, four of them become massive with their right-handed mates during the intermediate symmetry-breaking stages in Sec.~\ref{section:BminusL_SU8} and they are named as follows
\beqn\label{eq:stNL_names}
&& \check \Nc_L^{{\rm IV}^\prime }  \equiv \check \nG_L^\prime \,,~   \check \Nc_L^{{\rm IV}^{\prime \prime} }  \equiv \check \nG_L^{\prime \prime} \,, ~ \check \Nc_L^{{\rm V}^{\prime } }  \equiv \check \nG_L \,,~  \check \Nc_L^{\dot {\rm VII}^{\prime} }  \equiv \check \nG_L^{\prime \prime \prime } \,.
\eeqn
As we shall show in Sec.~\ref{section:BminusL_SU8}, the $\rep{28_F}$ contains the third-generational fermions of $\rep{10_F}$, while the first- and second-generational fermionic components of $\rep{10_F}$'s are from the $\rep{56_F}$.
The same result was also observed in Ref.~\cite{Barr:2008pn}.
Since we are not analyzing the mass hierarchies of the SM quarks and leptons in the current paper, we only listed the possible names for the first- and second-generational fermionic components in Tab.~\ref{tab:SU8_56ferm}.

%###################################################################
\section{The global $B-L$ symmetry in the ${\rm SU}(8)$ theory}
\label{section:BminusL_SU8}
%###################################################################

\para
By following the 't Hooft anomaly matching approach described in Sec.~\ref{section:BminusL_SU5}, we generalize the approach to the three-generational ${\rm SU}(8)$ theory.
The generalization means two following requirements:
\begin{enumerate}

\item the generalized $\widetilde{ {\rm U}}(1)_T$ symmetries should satisfy the 't Hooft anomaly matching condition,

\item the Higgs fields that can develop the VEVs for the specific symmetry-breaking stage should be neutral under the generalized $\widetilde{ {\rm U}}(1)_T$ symmetries defined at that stage.

\end{enumerate}

\begin{table}[htp]
\begin{center}
\begin{tabular}{c|ccccc}
\hline\hline
 & $\repb{8_F}^\lambda$  &  $\rep{28_F}$  &  $\repb{8_H}_{\,,\lambda}$ & $\rep{56_H}$  &  $\rep{70_H}$  \\
\hline
$\Tc_2$  &  $-3t_2$  &  $+2t_2$  & $+t_2$  &  $-2t_2$  &  $-4t_2$  \\[1mm]
\hline\hline
 & $\repb{8_F}^{\dot \lambda}$  &  $\rep{56_F}$  &  $\repb{28_H}_{\,,\dot \lambda}$ &  $\rep{56_H}$ &    \\
 \hline
 $\Tc_3$  &  $-3t_3$  &  $+t_3$  & $+2t_3$  &  $- t_3$ &  \\[1mm]
 \hline\hline
\end{tabular}
\end{center}
\caption{The $\widetilde{ {\rm U}}(1)_{T_2} \otimes \widetilde{ {\rm U}}(1)_{T_3}$ charges for all massless fermions and Higgs fields in the $\gSU(8)$ theory.}
\label{tab:SU8_Tcharges}
\end{table}%

\para
The non-anomalous global symmetries of $\widetilde{ {\rm U}}(1)_{T_2} \otimes \widetilde{ {\rm U}}(1)_{T_3}$ can be constructed through the linear combinations of the $\widetilde{ {\rm U} }(1)_{ \lambda }  \otimes \widetilde{ {\rm U} }(1)_2 \otimes \widetilde{ {\rm U} }(1)_{ \dot \lambda }  \otimes \widetilde{ {\rm U} }(1)_3$ symmetries in Eq.~\eqref{eq:SU8_DRS}.
According to Eq.~\eqref{eq:rk_Tcharges_Ferm}, the corresponding $\widetilde{ {\rm U}}(1)_{T_2} \otimes \widetilde{ {\rm U}}(1)_{T_3}$ charge assignments are given in Tab.~\ref{tab:SU8_Tcharges} such that both mixed $[ {\rm SU}(8) ]^2\cdot \widetilde{ {\rm U} }(1)_{ T_2}$ and $[ {\rm SU}(8) ]^2 \cdot \widetilde{ {\rm U} }(1)_{ T_3}$ anomalies are vanishing.
The $\widetilde{ {\rm U}}(1)_{T_2} \otimes \widetilde{ {\rm U}}(1)_{T_3}$ charges for Higgs fields are assigned so that the Yukawa couplings in Eq.~\eqref{eq:SU8_Yukawa} are $\widetilde{ {\rm U}}(1)_{T_2} \otimes \widetilde{ {\rm U}}(1)_{T_3}$-invariant.
From Tab.~\ref{tab:SU8_Tcharges}, we find the following global anomalies of
\beqn\label{eq:SU8_U1anoms}
&& \[ {\rm grav} \]^2 \cdot \widetilde{ {\rm U} }(1)_{ T_2 } = -40 t_2 \,,\quad \[ \widetilde{ {\rm U} }(1)_{ T_2 } \]^3 = -640 t_2^3 \,,\non
&& \[ {\rm grav} \]^2 \cdot \widetilde{ {\rm U} }(1)_{ T_3 } = -64 t_3 \,,\quad \[ \widetilde{ {\rm U} }(1)_{ T_3 } \]^3 = -1024 t_3^3 \,.
\eeqn

%%%%%%%%%%%%%%%%%%%%%%%%%%%%
\subsection{The zeroth stage}
%%%%%%%%%%%%%%%%%%%%%%%%%%%%

\begin{table}[htp]
\begin{center}
\begin{tabular}{c|cccc}
\hline\hline
$\repb{8_F}^\lambda$ & $( \repb{4}\,, \rep{1}\,, +\frac{1}{4} )_{\rep{F}}^\lambda$  &  $( \rep{1}\,, \repb{4}\,,  -\frac{1}{4} )_{\rep{F}}^\lambda$  &   &     \\[1mm]
\hline
$\Tc_2^\prime$  &  $-3t_2 \tilde c_1 + \frac{1}{4} \tilde c_2~(-4t_2)$  &  $-3t_2 \tilde c_1 - \frac{1}{4} \tilde c_2~(-2t_2)$  &   &     \\[1mm]
\hline
$\rep{28_F}$ &  $( \rep{6}\,, \rep{1}\,,  -\frac{1}{2} )_{\rep{F}}$ &  $( \rep{1}\,, \rep{ 6}\,,  + \frac{1}{2} )_{\rep{F}}$ &  $( \rep{4 }\,, \rep{4}\,,  0 )_{\rep{F}}$   &    \\[1mm]
\hline
$\Tc_2^\prime$  &  $2t_2 \tilde c_1 - \frac{1}{2} \tilde c_2~(4t_2)$  & $2t_2 \tilde c_1 + \frac{1}{2} \tilde c_2~(0)$ & $2t_2 \tilde c_1 ~(2t_2)$   &    \\[1mm]
\hline
$\repb{8_H}_{\,, \lambda }$  & $( \repb{4}\,, \rep{1}\,, +\frac{1}{4} )_{\rep{H}\,,\lambda}$  &  $( \rep{1}\,, \repb{4}\,,  -\frac{1}{4} )_{\rep{H}\,,\lambda}$  &   &      \\[1mm]
\hline
$\Tc_2^\prime$  &  $t_2 \tilde c_1 + \frac{1}{4} \tilde c_2~(0 )$  &  $t_2 \tilde c_1 - \frac{1}{4} \tilde c_2~(2t_2)$  &   &     \\[1mm]
\hline
$\rep{70_H}$  & $( \rep{4}\,, \repb{4}\,,  +\frac{1}{2} )_{\rep{H}}$   &  $( \repb{4}\,, \rep{4}\,,  - \frac{1}{2} )_{\rep{H}}$  &   &       \\[1mm]
\hline
$\Tc_2^\prime$  & $-4 t_2 \tilde c_1 + \frac{1}{2} \tilde c_2~(-6t_2)$  & $-4 t_2 \tilde c_1 - \frac{1}{2} \tilde c_2~(-2t_2)$ &   &      \\[1mm]
\hline
 $\rep{56_H}$  & $( \rep{1 }\,, \repb{4 }\,,  +\frac{3 }{ 4} )_{\rep{H}}$ &  $( \rep{4 }\,, \rep{6 }\,,  +\frac{1 }{ 4} )_{\rep{H}}$ & $$   &  $$   \\[1mm]
\hline
$\Tc_2^\prime$   & $-2 t_2 \tilde c_1 + \frac{3}{4} \tilde c_2~(-5t_2)$   &  $-2 t_2 \tilde c_1 + \frac{1}{4} \tilde c_2~(-3t_2)$   &  $$  &  $$  \\[1mm]
\hline\hline
$\repb{8_F}^{\dot \lambda}$ & $( \repb{4}\,, \rep{1}\,, +\frac{1}{4} )_{\rep{F}}^{\dot \lambda}$  &  $( \rep{1}\,, \repb{4}\,,  -\frac{1}{4} )_{\rep{F}}^{\dot \lambda}$  &  $$ &  $$     \\[1mm]
 \hline
 $\Tc_3^\prime$  &  $-3t_3 \tilde d_1 + \frac{1}{4} \tilde d_2~(-4t_3)$  &  $ -3t_3 \tilde d_1 - \frac{1}{4} \tilde d_2~(-2t_3)$   &  $$    & $$ \\[1mm]
 \hline
 $\rep{56_F}$  &  $( \rep{1}\,, \repb{4}\,,  +\frac{3}{4} )_{\rep{F}}$ &  $( \repb{4}\,,  \rep{1}\,,  -\frac{3}{4} )_{\rep{F}}$ & $( \rep{4}\,,  \rep{6}\,,  +\frac{1}{4} )_{\rep{F}}$  & $( \rep{6}\,,  \rep{4}\,,  -\frac{1}{4} )_{\rep{F}}$   \\[1mm]
 \hline
 $\Tc_3^\prime$  & $ t_3 \tilde d_1 +  \frac{3}{4} \tilde d_2~(-2t_3)$  &  $t_3 \tilde d_1 -  \frac{3}{4} \tilde d_2~(4t_3)$  & $t_3 \tilde d_1 +  \frac{1}{4} \tilde d_2~( 0)$  & $t_3 \tilde d_1 - \frac{1}{4} \tilde d_2~(2 t_3)$  \\[1mm]
 \hline
 $\repb{28_H}_{\,,\dot \lambda}$ & $( \rep{1}\,, \rep{6}\,,  -\frac{1}{2} )_{\rep{H}\,, \dot \lambda}$  &  $( \repb{4}\,, \repb{4}\,,  0 )_{\rep{H}\,, \dot \lambda}$  &   &    \\[1mm]
  \hline
  $\Tc_3^\prime$  &  $2t_3 \tilde d_1 - \frac{1}{2} \tilde d_2~(4t_3)$  &  $2t_3 \tilde d_1 ~(2t_3)$  &      & $$  \\[1mm]
  \hline
  $\rep{56_H}$ & $( \rep{1 }\,, \repb{4 }\,,  +\frac{3 }{ 4} )_{\rep{H}}$  & $( \rep{4 }\,, \rep{6 }\,,  +\frac{1 }{ 4} )_{\rep{H}}$ & $$  &  $$  \\[1mm]
  \hline
  $\Tc_3^\prime$   & $- t_3 \tilde d_1 + \frac{3}{4 } \tilde d_2~(-4t_3)$  & $- t_3 \tilde d_1 + \frac{1}{4 } \tilde d_2~(-2 t_3)$ &   $$ &  $$ \\[1mm]
 \hline\hline
\end{tabular}
\end{center}
\caption{The $\widetilde{ {\rm U}}(1)_{T_2^\prime } \otimes \widetilde{ {\rm U}}(1)_{T_3^\prime }$ charges for massless fermions and possible symmetry-breaking Higgs components in the $\Gc_{441}$ theory.
With the 't Hooft anomaly matching condition in Eq.~\eqref{eq:G441_match} and the neutrality conditions in Eqs.~\eqref{eq:T2p_neutral} and \eqref{eq:T3p_neutral}, the $\widetilde{ {\rm U}}(1)_{T_2^\prime } \otimes \widetilde{ {\rm U}}(1)_{T_3^\prime }$ charges are displayed in the parentheses.}
\label{tab:G441_Tcharges}
\end{table}%

\para
At the zeroth stage, no fermion in the spectrum obtain its mass.
We define the $\widetilde{ {\rm U}}(1)_{T_2^\prime} \otimes \widetilde{ {\rm U}}(1)_{T_3^\prime}$ charges at this stage as
\beqn\label{eq:G441_U1charges}
&& \Tc_2^\prime \equiv \tilde c_1 \Tc_2 + \tilde c_2 \Xc_0 \,,\quad  \Tc_3^\prime \equiv \tilde d_1 \Tc_3 + \tilde d_2 \Xc_0 \,.
\eeqn
The charge assignments are explicitly listed in Tab.~\ref{tab:G441_Tcharges}.
Accordingly, we find the following global anomalies of
\beqn\label{eq:G441_U1anoms}
&& \[ {\rm grav} \]^2 \cdot \widetilde{ {\rm U} }(1)_{ T_2^\prime } = -40 t_2 \tilde c_1 \,,\quad \[ \widetilde{ {\rm U} }(1)_{ T_2^\prime } \]^3 = -640 ( t_2 \tilde c_1 )^3 \,,\non
&& \[ {\rm grav} \]^2 \cdot \widetilde{ {\rm U} }(1)_{ T_3^\prime } = -64 t_3 \tilde d_1 \,,\quad \[ \widetilde{ {\rm U} }(1)_{ T_3^\prime } \]^3 = -1024 ( t_3 \tilde d_1)^3 \,.
\eeqn
By matching with the global anomalies in Eq.~\eqref{eq:SU8_U1anoms}, we find that
\beqn\label{eq:G441_match}
&& \tilde c_1 = \tilde d_1 = +1\,.
\eeqn
At this stage, the $\tilde c_2$ and $\tilde d_2$ are not determined yet.

%%%%%%%%%%%%%%%%%%%%%%%%%%%%
\subsection{The first stage}
%%%%%%%%%%%%%%%%%%%%%%%%%%%%

\para
The first symmetry-breaking stage of $\Gc_{441} \to \Gc_{341}$ can only be achieved by $( \repb{4} \,, \rep{1} \,, +\frac{1}{4} )_{\mathbf{H}\,,\lambda} \subset \repb{8_H}_{\,,\lambda}$ according to Eq.~\eqref{eq:SU8_Higgs_Br01}.
By requiring its generalized $\widetilde{{\rm U}}(1)_{T_2^\prime}$-neutral condition, we find that 
\beqn\label{eq:T2p_neutral}
&& \tilde c_2= -4 t_2 \,, %\quad \tilde d_2=  \,,
\eeqn
in Eq.~\eqref{eq:G441_U1charges}.
Still, the coefficient of $\tilde d_2$ is not determined at this point.

\para
The Yukawa coupling between $\repb{8_F}^\lambda$ and $\rep{28_F}$ can be expressed in terms of the $\Gc_{341}$ irreps as follows
\beqn\label{eq:Yukawa_441_01}
&&  Y_\Bc \repb{8_F}^\lambda \rep{28_F} \repb{8_H}_{\,,\lambda} + H.c. \non
&\supset& Y_\Bc \Big[ ( \rep{1}\,, \repb{4}\,, -\frac{1}{4})_{ \mathbf{F}}^\lambda \otimes ( \rep{4}\,, \rep{4}\,, 0 )_{ \mathbf{F}} \oplus ( \repb{4}\,, \rep{1}\,, +\frac{1}{4} )_{ \mathbf{F}}^\lambda \otimes ( \rep{6}\,, \rep{1}\,, -\frac{1}{2} )_{ \mathbf{F}}   \Big] \otimes \langle ( \repb{4}\,, \rep{1}\,, +\frac{1}{4})_{ \mathbf{H}\,,\lambda }  \rangle + H.c. \non
&\supset& Y_\Bc \Big[ ( \rep{1}\,, \repb{4}\,, -\frac{1}{4} )_{ \mathbf{F}}^\lambda \otimes ( \rep{1}\,, \rep{4}\,, +\frac{1}{4} )_{ \mathbf{F}}  \oplus ( \repb{3}\,, \rep{1}\,, +\frac{1}{3} )_{ \mathbf{F}}^\lambda \otimes  ( \rep{3}\,, \rep{1}\,, -\frac{1}{3} )_{ \mathbf{F}} \Big]  \non 
&& \otimes ( \rep{1}\,, \rep{1}\,, 0 )_{ \mathbf{H}\,,\lambda } + H.c. \,.
\eeqn
They lead to the following fermion masses
\beqs\label{eqs:masses_441_01}
\beqn
&&  Y_\Bc   ( \rep{1}\,, \repb{4}\,, -\frac{1}{4} )_{ \mathbf{F}}^{\rm IV } \otimes ( \rep{1}\,, \rep{4}\,, +\frac{1}{4} )_{ \mathbf{F}} \otimes ( \rep{1}\,, \rep{1}\,, 0 )_{ \mathbf{H}\,, {\rm IV} } + H.c. \non
& \supset& Y_\Bc \Big[  ( \rep{1}\,, \repb{2}\,, -\frac{1}{2} )_{ \mathbf{F}}^{\rm IV } \otimes ( \rep{1}\,, \rep{2}\,, +\frac{1}{2} )_{ \mathbf{F}}^{\prime\prime} \oplus  ( \rep{1}\,, \rep{1}\,, 0 )_{ \mathbf{F}}^{{\rm IV }^\prime} \otimes ( \rep{1}\,, \rep{1}\,, 0 )_{ \mathbf{F}}^{\prime} \oplus  ( \rep{1}\,, \rep{1}\,, 0 )_{ \mathbf{F}}^{{\rm IV }^{\prime\prime}} \otimes ( \rep{1}\,, \rep{1}\,, 0 )_{ \mathbf{F}}^{\prime\prime}  \Big] \non
&& \otimes ( \rep{1}\,, \rep{1}\,, 0 )_{ \mathbf{H}\,, {\rm IV} } + H.c. \non
&=& \frac{1}{\sqrt{2} }  Y_\Bc \Big( \eG_L^{\prime\prime }  {\eG_R^{\prime\prime}}^c - \nG_L^{\prime\prime}  {\nG_R^{\prime\prime}}^c +  \check \Nc_L^{ {\rm IV}^\prime}  \check \nG_R^{\prime\, c} + \check \Nc_L^{ {\rm IV}^{\prime\prime} }  \check \nG_R^{\prime\prime\, c} \Big) W_{ \repb{4}\,,{\rm IV} }  + H.c. \,, \label{eq:masses_441_01a} \\[1mm]
%%%%%%%%%%%%%%%%%%%%%%%%%%%%%%%%%%%%%%%%%%%%%
&&  Y_\Bc ( \repb{3}\,, \rep{1}\,, +\frac{1}{3} )_{ \mathbf{F}}^{\rm IV} \otimes  ( \rep{3}\,, \rep{1}\,, -\frac{1}{3} )_{ \mathbf{F}}  \otimes ( \rep{1}\,, \rep{1}\,, 0 )_{ \mathbf{H}\,, {\rm IV} }+ H.c. \non
&=&  \frac{1}{\sqrt{2} }  Y_\Bc  \DG_L  {\DG_R}^c W_{ \repb{4}\,,{\rm IV} } + H.c.\,,\label{eq:masses_441_01b}
\eeqn
\eeqs
with the DRS limit of the Yukawa couplings.
Without loss of generality, we choose the massive anti-fundamental fermion to be $\lambda={\rm IV}$ at this stage.
Thus, we can identify that $ ( \rep{1}\,, \repb{2}\,, -\frac{1}{2} )_{ \mathbf{F}}^{\rm IV } \equiv ( \eG_L^{\prime\prime} \,, - \nG_L^{\prime \prime})^T$, $(\check \Nc_L^{ {\rm IV}^\prime}\,, \check \Nc_L^{ {\rm IV}^{\prime\prime} })\equiv ( \check \nG_L^\prime\,, \check \nG_L^{ \prime \prime} )$, and $( \repb{3}\,, \rep{1}\,, +\frac{1}{3} )_{ \mathbf{F}}^{\rm IV} \equiv {\DG_R}^c$.

\para
After this stage, the remaining massless fermions expressed in terms of the $\Gc_{341}$ irreps are the following
\beqn\label{eq:341_fermions}
&& \Big[ ( \repb{3}\,, \rep{1}\,, +\frac{1}{3} )_{ \mathbf{F}}^\Lambda \oplus ( \rep{1}\,, \rep{1}\,, 0 )_{ \mathbf{F}}^\Lambda \Big]  \oplus ( \rep{1}\,, \repb{4}\,, -\frac{1}{4} )_{ \mathbf{F}}^\Lambda  \subset \repb{8_F}^\Lambda \,, \non
&& \Lambda = ( \lambda\,, \dot \lambda) \,,\quad \lambda = (3\,, {\rm V}\,, {\rm VI} ) \,,\quad \dot \lambda = ( \dot 1\,, \dot 2\,, \dot {\rm VII} \,,\dot {\rm IIX} \,, \dot {\rm IX}) \,,\non
&& ( \rep{1}\,, \rep{1}\,, 0 )_{ \mathbf{F}}^{{\rm IV} } \subset \repb{8_F}^{ {\rm IV}} \,,\non
%%%%%%%%%%%%%%%%%%%%%%%%%%%%%%%%%%%%%%%%%%%%%
&& \Big[ \cancel{ ( \rep{3}\,, \rep{1}\,, -\frac{1}{3} )_{ \mathbf{F}} } \oplus ( \repb{3}\,, \rep{1}\,, -\frac{2}{3} )_{ \mathbf{F}} \Big] \oplus ( \rep{1}\,, \rep{6}\,, +\frac{1}{2} )_{ \mathbf{F}}  \oplus \Big[ ( \rep{3}\,, \rep{4}\,, -\frac{1}{12} )_{ \mathbf{F}} \oplus \cancel{ ( \rep{1}\,, \rep{4}\,, +\frac{1}{4} )_{ \mathbf{F}} } \Big] \subset \rep{28_F}\,, \non
%%%%%%%%%%%%%%%%%%%%%%%%%%%%%%%%%%%%%%%%%%%%%
&& ( \rep{1}\,, \repb{4}\,, +\frac{3}{4} )_{ \mathbf{F}} \oplus \Big[ ( \repb{3}\,, \rep{1}\,, -\frac{2}{3} )_{ \mathbf{F}}^\prime \oplus ( \rep{1}\,, \rep{1}\,, -1 )_{ \mathbf{F}} \Big] \oplus  \Big[ ( \rep{3}\,, \rep{6}\,, +\frac{1}{6} )_{ \mathbf{F}} \oplus ( \rep{1}\,, \rep{6}\,, +\frac{1}{2})_{ \mathbf{F}}^\prime \Big] \non
&\oplus& \Big[ ( \rep{3}\,, \rep{4}\,, -\frac{1}{12} )_{ \mathbf{F}}^\prime \oplus ( \repb{3}\,, \rep{4}\,, -\frac{5}{12})_{ \mathbf{F}} \Big]   \subset \rep{56_F}\,.
\eeqn
Fermions that become massive at this stage are crossed out by slashes.
From the Yukawa couplings in Eq.~\eqref{eq:Yukawa_441_01}, no components from the $\rep{56_F}$ obtain their masses at this stage.
Loosely speaking, we find that only one of the massive $\repb{8_F}^\Lambda$ is integrated out from the anomaly-free conditions of $[ {\rm SU}(3)_c]^2 \cdot {\rm U}(1)_{X_1}$, $\[ {\rm SU}(4)_W \]^2 \cdot {\rm U}(1)_{X_1}$, and $\[ {\rm U}(1)_{X_1} \]^3$, except for one left-handed sterile neutrino of $\check \Nc_L^{\rm IV}\equiv ( \rep{1}\,, \rep{1}\,, 0 )_{ \mathbf{F}}^{\rm IV} \subset \repb{8_F}^{\rm IV}$.
One can further find that the remaining massless fermions from three $\repb{8_F}^\lambda$ and the $\rep{28_F}$ form an IAFFS, and the remaining massless fermions from five $\repb{8_F}^{\dot \lambda}$ and the $\rep{56_F}$ form another IAFFS.

\begin{table}[htp]
\begin{center}
\begin{tabular}{c|ccc}
\hline\hline
$\repb{8_F}^\lambda$ & $( \repb{3}\,, \rep{1}\,, +\frac{1}{3} )_{\rep{F}}^\lambda$ & $( \rep{1}\,, \rep{1}\,,  0 )_{\rep{F}}^\lambda$  &  $( \rep{1}\,, \repb{4}\,,  -\frac{1}{4} )_{\rep{F}}^\lambda$     \\[1mm]
\hline
$\Tc_2^{\prime\prime}$  &  $-4t_2 \tilde c_1^\prime + \frac{1}{3} \tilde c_2^\prime~( - \frac{4}{3} t_2 )$ & $-4t_2 \tilde c_1^\prime~( -4t_2)$  &  $-2t_2 \tilde c_1^\prime - \frac{1}{4} \tilde c_2^\prime~( - 4 t_2 )$     \\[1mm]
\hline
 $\rep{28_F}$  &  $( \repb{3}\,, \rep{1}\,,  -\frac{2}{3 } )_{\rep{F}}$ &  $( \rep{1}\,, \rep{ 6}\,,  + \frac{1}{2} )_{\rep{F}}$ & $( \rep{3  }\,, \rep{4}\,, - \frac{1 }{12 } )_{\rep{F}}$    \\[1mm]
\hline
$\Tc_2^{\prime\prime}$   & $4t_2 \tilde c_1^\prime - \frac{2}{3 } \tilde c_2^\prime~(- \frac{4}{3} t_2 )$  &  $ \frac{1}{2} \tilde c_2^\prime~(4 t_2 )$  &  $2t_2 \tilde c_1^\prime - \frac{1 }{12 } \tilde c_2^\prime ~(\frac{4 }{3 } t_2 )$    \\[1mm]
\hline
$\repb{8_H}_{\,, \lambda}$  &  $( \rep{1}\,, \repb{4}\,,  -\frac{1}{4} )_{\rep{H}\,,\lambda}$  &   &    \\[1mm]
\hline
$\Tc_2^{\prime\prime}$  &  $2 t_2 \tilde c_1^\prime - \frac{1}{4 } \tilde c_2^\prime ~( 0 )$  &   &      \\[1mm]
\hline
$\rep{70_H}$ &  $( \rep{1 }\,, \repb{4}\,,  +\frac{3}{4} )_{\rep{H}}^\prime$  & $( \rep{1}\,, \rep{4}\,,  - \frac{3 }{4} )_{\rep{H}}^\prime$   &    \\[1mm]
\hline
$\Tc_2^{\prime\prime}$  & $-6t_2 \tilde c_1^\prime + \frac{3}{4 } \tilde c_2^\prime~( 0)$  &  $-2t_2 \tilde c_1^\prime - \frac{3}{4 } \tilde c_2^\prime~( -8 t_2 )$  &      \\[1mm]
\hline
 $\rep{56_H}$   & $( \rep{1}\,, \repb{4 }\,,  +\frac{3}{4 } )_{\rep{H}}$  & $( \rep{1}\,, \rep{6}\,,  +\frac{1}{2} )_{\rep{H}}^\prime$  &   $$    \\[1mm]
\hline
$\Tc_2^{\prime\prime}$  &      $-5 t_2 \tilde c_1^\prime + \frac{3}{4 } \tilde c_2^\prime ~(t_2 )$ & $-3 t_2 \tilde c_1^\prime + \frac{1}{2 } \tilde c_2^\prime ~( t_2 )$   &    \\[1mm]
\hline\hline
 $\repb{8_F}^{\dot \lambda} $ & $( \repb{3}\,, \rep{1}\,, +\frac{1}{3} )_{\rep{F}}^{\dot \lambda}$ & $( \rep{1}\,, \rep{1}\,,  0 )_{\rep{F}}^{\dot \lambda}$ &  $( \rep{1}\,, \repb{4}\,,  -\frac{1}{4} )_{\rep{F}}^{\dot \lambda}$      \\[1mm]
 \hline
 $\Tc_3^{\prime\prime}$  &  $( -3 t_3 + \frac{1}{4} \tilde d_2) \tilde d_1^\prime + \frac{1}{3} \tilde d_2^\prime~(- \frac{4 }{3} t_3 )$  &  $( -3 t_3 + \frac{1}{4} \tilde d_2) \tilde d_1^\prime~( - 4 t_3 )$  & $( -3 t_3 - \frac{1}{4} \tilde d_2) \tilde d_1^\prime - \frac{1}{4} \tilde d_2^\prime ~( -4 t_3 )$    \\[1mm]
    \hline
$\rep{56_F}$ & $( \rep{1}\,, \repb{4}\,,  + \frac{3 }{4} )_{\rep{F}}$ & $( \repb{3}\,,  \rep{1}\,,  - \frac{2 }{3} )_{\rep{F}}^\prime$ &  $( \rep{1}\,,  \rep{1}\,,  - 1 )_{\rep{F}}$     \\[1mm]
 \hline
 $\Tc_3^{\prime\prime}$  &  $(  t_3 + \frac{3}{4} \tilde d_2) \tilde d_1^\prime +\frac{3}{4} \tilde d_2^\prime ~(4 t_3 )$  &  $(  t_3 - \frac{3}{4} \tilde d_2) \tilde d_1^\prime-\frac{2}{3} \tilde d_2^\prime ~( - \frac{4 }{3} t_3 )$  & $(  t_3 - \frac{3}{4} \tilde d_2) \tilde d_1^\prime - \tilde d_2^\prime ~(-4 t_3 )$     \\[1mm]
    \hline
   & $( \rep{3}\,,  \rep{6}\,,  +\frac{1}{6} )_{\rep{F}}$ & $( \rep{1}\,,  \rep{6}\,,  +\frac{1}{2} )_{\rep{F}}^\prime$ &       \\[1mm]
   \hline
  $\Tc_3^{\prime\prime}$ & $(  t_3 + \frac{1}{4} \tilde d_2) \tilde d_1^\prime + \frac{1}{6} \tilde d_2^\prime ~( \frac{4}{3 } t_3)$ & $(  t_3 + \frac{1}{4} \tilde d_2) \tilde d_1^\prime + \frac{1}{2 } \tilde d_2^\prime ~( 4 t_3 )$    &   \\[1mm]
    \hline
   &  $( \rep{3}\,,  \rep{4}\,,  -\frac{1}{12 } )_{\rep{F}}^\prime$ & $( \repb{3}\,,  \rep{4}\,,  -\frac{5}{12 } )_{\rep{F}}^\prime$     &      \\[1mm]
   \hline
  $\Tc_3^{\prime\prime}$ & $(  t_3 - \frac{1}{4} \tilde d_2) \tilde d_1^\prime  - \frac{1}{12} \tilde d_2^\prime ~( \frac{4}{3 } t_3 )$  & $(  t_3 - \frac{1}{4} \tilde d_2) \tilde d_1^\prime - \frac{5}{12} \tilde d_2^\prime ~( - \frac{4 }{3} t_3 )$   &   \\[1mm]
    \hline
 $\repb{28_H}_{\,, \dot \lambda }$  & $( \rep{1}\,, \rep{6}\,,  -\frac{1}{2} )_{\rep{H}\,,\dot \lambda}$  &  $( \rep{1}\,, \repb{4}\,,  -\frac{1}{4} )_{\rep{H}\,,\dot \lambda}$  &      \\[1mm]
\hline
$\Tc_3^{\prime\prime}$  &  $(2 t_3 - \frac{1}{2} \tilde d_2 ) \tilde d_1^\prime - \frac{1}{2} \tilde d_2^\prime ~( 0)$  &  $2 t_3  \tilde d_1^\prime - \frac{1}{4} \tilde d_2^\prime ~( 0) $  &     \\[1mm]
\hline
 $\rep{56_H}$  &  $( \rep{1}\,, \repb{4}\,,  +\frac{3}{4 } )_{\rep{H}}$  &  $( \rep{1}\,, \rep{6}\,,  +\frac{1}{2} )_{\rep{H}}^\prime$     &    \\[1mm]
\hline
$\Tc_3^{\prime\prime}$  & $(-  t_3 + \frac{3}{4} \tilde d_2 ) \tilde d_1^\prime + \frac{3}{4 } \tilde d_2^\prime ~( 2 t_3 )$  &  $(-  t_3 + \frac{1}{4} \tilde d_2 ) \tilde d_1^\prime + \frac{1}{2} \tilde d_2^\prime~( 2 t_3 )$  &   \\[1mm]
 \hline\hline
\end{tabular}
\end{center}
\caption{The $\widetilde{ {\rm U}}(1)_{T_2^{\prime \prime} } \otimes \widetilde{ {\rm U}}(1)_{T_3^{\prime \prime} }$ charges for massless fermions and possible symmetry-breaking Higgs components in the $\Gc_{341}$ theory.
With the 't Hooft anomaly matching condition in Eq.~\eqref{eq:G341_match} and the neutrality conditions in Eqs.~\eqref{eq:T3p_neutral} and \eqref{eq:T23pp_neutral}, the $\widetilde{ {\rm U}}(1)_{T_2^{\prime \prime} } \otimes \widetilde{ {\rm U}}(1)_{T_3^{\prime \prime} }$ charges are displayed in the parentheses. }
\label{tab:G341_Tcharges}
\end{table}%

\para
Thus, we define the $\widetilde{ {\rm U}}(1)_{T_2^{\prime \prime} } \otimes \widetilde{ {\rm U}}(1)_{T_3^{ \prime \prime} }$ charges at this stage as
\beqn\label{eq:G341_U1charges}
&& \Tc_2^{\prime \prime} \equiv \tilde c_1^\prime \Tc_2^\prime + \tilde c_2^\prime \Xc_1 \,,\quad \Tc_3^{ \prime \prime }\equiv \tilde d_1^\prime \Tc_3^\prime + \tilde d_2^\prime \Xc_1 \,,
\eeqn
and they are explicitly listed in Tab.~\ref{tab:G341_Tcharges} for massless fermions and possible symmetry-breaking Higgs components.
Accordingly, we find the following global anomalies of
\beqn\label{eq:G341_U1anoms}
&& \[ {\rm grav} \]^2 \cdot \widetilde{ {\rm U} }(1)_{ T_2^{\prime \prime} } = -40 t_2 \tilde c_1^\prime \,,\quad \[ \widetilde{ {\rm U} }(1)_{ T_2^{\prime \prime} } \]^3 = -640 ( t_2 \tilde c_1^\prime )^3 \,,\non
&& \[ {\rm grav} \]^2 \cdot \widetilde{ {\rm U} }(1)_{ T_3^{\prime \prime} } = -64 t_3 \tilde d_1^\prime  \,,\quad \[ \widetilde{ {\rm U} }(1)_{ T_3^{\prime \prime} } \]^3 = -1024 ( t_3 \tilde d_1^\prime  )^3 \,,
\eeqn
from the massless fermions in Eq.~\eqref{eq:341_fermions}.
By matching with the global anomalies in Eq.~\eqref{eq:G441_U1anoms}, we find that
\beqn\label{eq:G341_match}
&& \tilde c_1^\prime = \tilde d_1^\prime = +1\,.
\eeqn
Two other coefficients of $\tilde c_2^\prime$ and $\tilde d_2^\prime$ are not determined yet.

\para
Here, one subtlety can be observed.
Since one has $( \rep{3}\,,  \rep{2}\,,  +\frac{1 }{6 } )_{\rep{F}}^\prime \subset ( \rep{3}\,,  \rep{3}\,,  0 )_{\rep{F}}^\prime \subset ( \rep{3}\,,  \rep{6}\,,  +\frac{1}{6} )_{\rep{F}}$ and $( \rep{3}\,,  \rep{2}\,,  +\frac{1 }{6 } )_{\rep{F}}^{\prime \prime\prime} \subset ( \rep{3}\,,  \rep{3}\,,  0 )_{\rep{F}}^{\prime \prime} \subset ( \rep{3}\,,  \rep{4}\,,  -\frac{1}{12} )_{\rep{F}}^\prime$ according to Tab.~\ref{tab:SU8_56ferm}, one thus expect the identical $\Tc_3^{\prime\prime}$ charges of $( \rep{3}\,,  \rep{6}\,,  +\frac{1}{6} )_{\rep{F}}$ and $( \rep{3}\,,  \rep{4}\,,  -\frac{1}{12} )_{\rep{F}}^\prime$ such that the first- and second-generational left-handed quark doublets carry the same $\Bc-\Lc$ charges.
This condition determines the coefficient of
\beqn\label{eq:T3p_neutral}
&& \tilde d_2 = -4 t_3\,,
\eeqn
in Eq.~\eqref{eq:G441_U1charges}.

%%%%%%%%%%%%%%%%%%%%%%%%%%%%
\subsection{The second stage}
%%%%%%%%%%%%%%%%%%%%%%%%%%%%

\para
The second symmetry-breaking stage of $\Gc_{341} \to \Gc_{331}$ can be achieved by $( \rep{1} \,, \repb{4} \,, -\frac{1}{4} )_{\mathbf{H}\,,\lambda} \subset \repb{8_H}_{\,,\lambda}$ in Eq.~\eqref{eq:SU8_Higgs_Br01} and $( \rep{1} \,, \repb{4} \,, -\frac{1}{4} )_{\mathbf{H}\,, \dot\lambda} \subset \repb{28_H}_{\,,\dot \lambda }$ in Eq.~\eqref{eq:SU8_Higgs_Br02}.
Correspondingly, their generalized $\widetilde{{\rm U}}(1)_{T_2^{\prime \prime} } \otimes \widetilde{{\rm U}}(1)_{T_3^{\prime \prime} }$ charges should be vanishing.
We find that 
\beqn\label{eq:T23pp_neutral}
&& \tilde c_2^\prime = 8t_2 \,, \quad \tilde d_2^\prime = 8 t_3 \,,
\eeqn
in Eq.~\eqref{eq:G341_U1charges} and Tab.~\ref{tab:G341_Tcharges}.

\para
The Yukawa coupling between $\repb{8_F}^\lambda$ and $\rep{28_F}$ can be expressed in terms of the $\Gc_{331}$ irreps as follows
\beqn\label{eq:Yukawa_341_01} 
&& Y_\Bc \repb{8_F}^\lambda \rep{28_F} \repb{8_H}_{\,,\lambda } + H.c. \non
&\supset& Y_\Bc \Big[ ( \rep{1}\,, \repb{4}\,, -\frac{1}{4})_{ \mathbf{F}}^\lambda \otimes ( \rep{1}\,, \rep{6}\,, +\frac{1}{2} )_{ \mathbf{F}} \oplus ( \repb{4}\,, \rep{1}\,, +\frac{1}{4})_{ \mathbf{F}}^\lambda \otimes ( \rep{4}\,, \rep{4}\,, 0 )_{ \mathbf{F}} \Big] \otimes  ( \rep{1}\,, \repb{4}\,, -\frac{1}{4})_{ \mathbf{H}\,,\lambda } + H.c. \non
&\supset& Y_\Bc  \Big[ ( \rep{1}\,, \repb{4}\,, -\frac{1}{4})_{ \mathbf{F}}^\lambda \otimes ( \rep{1}\,, \rep{6}\,, +\frac{1}{2} )_{ \mathbf{F}} \oplus ( \rep{1 }\,, \rep{1}\,, 0 )_{ \mathbf{F}}^\lambda \otimes ( \rep{1}\,, \rep{4}\,, +\frac{1 }{ 4} )_{ \mathbf{F}} \non 
&\oplus& ( \repb{3}\,, \rep{1}\,, +\frac{1}{3})_{ \mathbf{F}}^\lambda \otimes ( \rep{3 }\,, \rep{4}\,, - \frac{1 }{12 } )_{ \mathbf{F}}  \Big] \otimes  \langle ( \rep{1}\,, \repb{4}\,, -\frac{1}{4})_{ \mathbf{H}\,,\lambda }  \rangle + H.c. \non
&\supset& Y_\Bc \Big[  ( \rep{1}\,, \repb{3}\,, -\frac{1}{3} )_{ \mathbf{F}}^\lambda \otimes ( \rep{1}\,, \rep{3}\,, +\frac{1}{3} )_{ \mathbf{F}}  \oplus ( \rep{1}\,, \rep{1}\,, 0 )_{ \mathbf{F}}^\lambda \otimes ( \rep{1}\,, \rep{1}\,, 0 )_{ \mathbf{F}}^{\prime\prime}  \non
&\oplus&  (  \repb{3}\,, \rep{1}\,, +\frac{1}{3} )_{ \mathbf{F}}^\lambda \otimes ( \rep{3}\,, \rep{1}\,, -\frac{1}{3} )_{ \mathbf{F}}^{\prime\prime}   \Big]  \otimes ( \rep{1}\,, \rep{1}\,, 0 )_{ \mathbf{H}\,, \lambda } + H.c.  \,.
%%%%%%%%%%%%%%%%%%%%%%%%%%%%%%%%%%%%%%%%%%%%%
\eeqn
They lead to fermion masses of
\beqs\label{eqs:masses_341_01}
\beqn
&&  Y_\Bc  ( \rep{1}\,, \repb{3}\,, -\frac{1}{3} )_{ \mathbf{F}}^{\rm V} \otimes ( \rep{1}\,, \rep{3}\,, +\frac{1}{3} )_{ \mathbf{F}} \otimes ( \rep{1}\,, \rep{1}\,, 0 )_{ \mathbf{H}\,, {\rm V} } + H.c. \non
&\supset& Y_\Bc \Big[  ( \rep{1}\,, \repb{2}\,, -\frac{1}{2} )_{ \mathbf{F}}^{\rm V} \otimes ( \rep{1}\,, \rep{2}\,, +\frac{1}{2} )_{ \mathbf{F}}  \oplus ( \rep{1}\,, \rep{1}\,, 0 )_{ \mathbf{F}}^{{\rm V}^\prime } \otimes ( \rep{1}\,, \rep{1}\,, 0 )_{ \mathbf{F}}  \Big] \otimes  ( \rep{1}\,, \rep{1}\,, 0 )_{ \mathbf{H}\,, {\rm V} } \non
&=&  \frac{1}{ \sqrt{2} } Y_\Bc  \Big(  \eG_L {\eG_R}^c -   \nG_L {\nG_R }^c +  \check \Nc_L^{{\rm V}^\prime}  \check \nG_R^c \Big) w_{ \repb{4}\,, {\rm V}} + H.c.  \,, \label{eq:masses_341_01a}  \\[1mm]
%%%%%%%%%%%%%%%%%%%%%%%%%%%%%%%%%%%%%%%%%%%%%
&& Y_\Bc  \Big[  ( \rep{1}\,, \rep{1}\,, 0 )_{ \mathbf{F}}^{\rm V} \otimes ( \rep{1}\,, \rep{1}\,, 0 )_{ \mathbf{F}}^{\prime\prime} \oplus  ( \repb{3}\,, \rep{1}\,, +\frac{1}{3} )_{ \mathbf{F}}^{\rm V} \otimes ( \rep{3}\,, \rep{1}\,, -\frac{1}{3} )_{ \mathbf{F}}^{\prime\prime }   \Big]  \otimes ( \rep{1}\,, \rep{1}\,, 0 )_{ \mathbf{H}\,, {\rm V} } + H.c. \non
&=&  \frac{1}{ \sqrt{2} }  Y_\Bc  \Big(  \check \Nc_L^{\rm V}  \check \nG_R^{\prime\prime\,c} + \DG_L^{\prime\prime}  {\DG_R^{\prime\prime}}^c  \Big) w_{ \repb{4}\,, {\rm V}} + H.c.  \,. \label{eq:masses_341_01b}
\eeqn
\eeqs
Without loss of generality, we choose $\lambda= \kappa= {\rm V}$ for the DRS limit in Eq.~\eqref{eq:SU8_Yukawa_DRS} at this stage.
Thus, we can identify that $ ( \rep{1}\,, \repb{2}\,, -\frac{1}{2} )_{ \mathbf{F}}^{\rm V} \equiv ( \eG_L \,, - \nG_L )^T$, $\check \Nc_L^{{\rm V}^\prime} \equiv \check \nG_L$, and $( \repb{3}\,, \rep{1}\,, +\frac{1}{3} )_{ \mathbf{F}}^{\rm V} \equiv {\DG_R^{\prime\prime}}^c$.
The $\check \Nc_L^{\rm V}  \check \nG_R^{\prime\prime\,c}+H.c.$ in Eq.~\eqref{eq:masses_341_01b} is a mass mixing term, since the right-handed $ \check \nG_R^{\prime\prime\,c}$ has already obtained the Dirac mass with its left-handed mate in Eq.~\eqref{eq:masses_441_01a}.

\para
The Yukawa coupling between $\repb{8_F}^{\dot \lambda}$ and $\rep{56_F}$ can be expressed in terms of the $\Gc_{331}$ irreps as follows
\beqn\label{eq:Yukawa_341_02} 
&&  Y_\Dc \repb{8_F}^{\dot \lambda} \rep{56_F} \repb{28_H}_{\,,\dot \lambda } + H.c. \non
&\supset& Y_\Dc \Big[ ( \rep{1}\,, \repb{4}\,, -\frac{1}{4})_{ \mathbf{F}}^{\dot \lambda} \otimes ( \rep{1}\,, \rep{6}\,, +\frac{1}{2} )_{ \mathbf{F}}^\prime  \oplus ( \repb{3}\,, \rep{1}\,, +\frac{1}{3})_{ \mathbf{F}}^{\dot \lambda} \otimes ( \rep{3}\,, \rep{4}\,, -\frac{1}{12} )_{ \mathbf{F}}^\prime \Big]  \otimes \langle ( \rep{1}\,, \repb{4}\,, -\frac{1}{4})_{ \mathbf{H}\,, \dot \lambda } \rangle + H.c. \non
&\supset& Y_\Dc \Big[ ( \rep{1}\,, \repb{3}\,, -\frac{1}{3})_{ \mathbf{F}}^{\dot \lambda} \otimes ( \rep{1}\,, \rep{3}\,, +\frac{1}{3} )_{ \mathbf{F}}^\prime \oplus ( \repb{3}\,, \rep{1}\,, +\frac{1}{3})_{ \mathbf{F}}^{\dot \lambda} \otimes ( \rep{3}\,, \rep{1}\,, -\frac{1}{3} )_{ \mathbf{F}}^{\prime \prime\prime \prime \prime}  \Big]  \non
&& \otimes ( \rep{1}\,, \rep{1}\,, 0)_{ \mathbf{H}\,, \dot \lambda }  + H.c.  \,.
%%%%%%%%%%%%%%%%%%%%%%%%%%%%%%%%%%%%%%%%%%%%%
\eeqn
They lead to fermion masses of
\beqs\label{eqs:masses_341_02}
\beqn
&&  Y_\Dc  ( \rep{1}\,, \repb{3}\,, -\frac{1}{3} )_{ \mathbf{F}}^{\dot {\rm VII} } \otimes ( \rep{1}\,, \rep{3}\,, +\frac{1}{3} )_{ \mathbf{F}}^\prime \otimes ( \rep{1}\,, \rep{1}\,, 0 )_{ \mathbf{H}\,, \dot {\rm VII} } + H.c. \non
&\supset&  Y_\Dc  \Big[  ( \rep{1}\,, \repb{2}\,, -\frac{1}{2} )_{ \mathbf{F}}^{\dot {\rm VII} } \otimes ( \rep{1}\,, \rep{2}\,, +\frac{1}{2} )_{ \mathbf{F}}^{\prime\prime \prime \prime }  \oplus  ( \rep{1}\,, \rep{1}\,, 0 )_{ \mathbf{F}}^{ \dot {{\rm VII}^\prime } }  \otimes ( \rep{1}\,, \rep{1}\,, 0 )_{ \mathbf{F}}^{\prime\prime\prime}  \Big] \otimes ( \rep{1}\,, \rep{1}\,, 0 )_{ \mathbf{H}\,, \dot {\rm VII} } + H.c. \non
&=& \frac{1}{ \sqrt{2}} Y_\Dc  \Big(  \eG_L^{\prime \prime  \prime\prime }  {\eG_R^{\prime \prime  \prime\prime}}^c -   \nG_L^{\prime \prime  \prime\prime }  {\nG_R^{\prime \prime  \prime\prime }}^c +  \check \Nc_L^{ \dot {\rm VII }^\prime }  \check \nG_R^{\prime\prime\prime\,c }  \Big) w_{\repb{4}\,, \dot {\rm VII}} + H.c.  \,, \label{eq:masses_341_02a} \\[1mm]
%%%%%%%%%%%%%%%%%%%%%%%%%%%%%%%%%%%%%%%%%%%%%
&& Y_\Dc  ( \repb{3}\,, \rep{1}\,, +\frac{1}{3} )_{ \mathbf{F} }^{ \dot {\rm VII}  } \otimes  ( \rep{3}\,, \rep{1}\,, -\frac{1}{3} )_{ \mathbf{F} }^{\prime\prime\prime\prime \prime } \otimes ( \rep{1}\,, \rep{1}\,, 0 )_{ \mathbf{H}\,, \dot {\rm VII} }   + H.c. \non
&=& \frac{1}{ \sqrt{2}}  Y_\Dc  \DG_L^{\prime\prime\prime\prime \prime }  {\DG_R^{\prime\prime\prime\prime \prime } }^c w_{\repb{4}\,, \dot {\rm VII}}+ H.c.  \,. \label{eq:masses_341_02b}
\eeqn
\eeqs
Without loss of generality, we choose $\dot \lambda = \dot \kappa= \dot {\rm VII}$ for the DRS limit in Eq.~\eqref{eq:SU8_Yukawa_DRS} at this stage.
Thus, we can identify that $ ( \rep{1}\,, \repb{2}\,, -\frac{1}{2} )_{ \mathbf{F}}^{\dot {\rm VII} } \equiv ( \eG_L^{ \prime\prime \prime \prime } \,, - \nG_L^{ \prime\prime \prime \prime } )^T$, $\check \Nc_L^{ \dot {\rm VII }^\prime } \equiv \check \nG_L^{ \prime \prime \prime }$, and $( \repb{3}\,, \rep{1}\,, +\frac{1}{3} )_{ \mathbf{F}}^{\dot {\rm VII}} \equiv {\DG_R^{\prime\prime\prime \prime \prime }}^c$.

\para
A renormalizable Yukawa coupling between two $\rep{56_F}$'s cannot be present in Eq.~\eqref{eq:SU8_Yukawa}.
If no further Yukawa couplings were possible, the vectorlike fermions of $(\EG\,, \uG\,,\dG\,, \UG)$ from the $\rep{56_F}$ will be massless in the spectrum.
It was suggested in Ref.~\cite{Barr:2008pn} to consider the following $d=5$ operator 
\beqn\label{eq:Yukawa_341_03}
%%%%%%%%%%%%%%%%%%%%%%%%%%%%%%%%%%%%%%%%%%%%%
&& \frac{1}{ M_{\rm pl}} \rep{56_F} \rep{56_F} \cdot \langle \rep{63_H} \rangle \cdot (\repb{28_H}_{\,, \dot \lambda} )^\dag + H.c. \non
&\supset&   \frac{  v_U }{ M_{\rm pl}} \Big[ ( \rep{1 } \,, \repb{4 } \,,  + \frac{3}{4} )_{ \rep{F}} \otimes ( \repb{4 } \,, \rep{1 } \,, - \frac{3}{4} )_{ \rep{F}}  \oplus  ( \rep{4 } \,, \rep{6 } \,, + \frac{1 }{4} )_{ \rep{F}} \otimes  ( \rep{6 } \,, \rep{4 } \,, -  \frac{1 }{4} )_{ \rep{F}} \Big] \otimes ( \repb{4 } \,, \repb{4 } \,, 0 )_{ \rep{H}\,, \dot \lambda}^\dag + H.c. \non
&\supset& \frac{ v_U }{ M_{\rm pl} } \Big[ ( \rep{1 } \,, \repb{4 } \,,  + \frac{3}{4} )_{ \rep{F}} \otimes ( \rep{1 } \,, \rep{1 } \,, - 1 )_{ \rep{F}}  \oplus  ( \rep{3 } \,, \rep{6 } \,, + \frac{1 }{6 } )_{ \rep{F}} \otimes  ( \repb{3 } \,, \rep{4 } \,, -  \frac{5 }{12 } )_{ \rep{F}}  \Big]  \otimes \langle ( \rep{1 } \,, \repb{4 } \,, -\frac{1 }{4 } )_{ \rep{H}\,, \dot \lambda}^\dag \rangle + H.c. \non
%
%&\supset& \frac{ v_U }{ M_{\rm pl} } \Big[ ( \rep{1 } \,, \rep{1 } \,,  + 1 )_{ \rep{F}}^{ \prime \prime } \otimes ( \rep{1 } \,, \rep{1 } \,, - 1 )_{ \rep{F}}  \oplus  ( \rep{3 } \,, \repb{3 } \,, + \frac{1 }{3 } )_{ \rep{F}} \otimes  ( \repb{3 } \,, \rep{3 } \,, -  \frac{1 }{3 } )_{ \rep{F}}  \Big]  \otimes ( \rep{1 } \,, \rep{1 } \,, 0 )_{ \rep{H}\,, \dot \lambda}^\dag + H.c. \non
%
&\Rightarrow& \frac{ v_U }{ M_{\rm pl} } w_{ \repb{4}\,, \dot {\rm VII} } (  \EG_L {\EG_R}^c + \dG_L {\dG_R}^c - \uG_L {\uG_R}^c + \UG_L {\UG_R}^c ) + H.c. \,.
\eeqn
This $d=5$ bi-linear fermion operator obviously breaks the global DRS symmetries in Eq.~\eqref{eq:SU8_DRS}, and is only due to the gravitational effect~\footnote{The gravitational effect to the SM fermion masses was first put forth in the ${\rm SU}(5)$ GUT~\cite{Ellis:1979fg}, and was also discussed in the context of supergravity~\cite{Nanopoulos:1982zm} where the ratio of $v_U/M_{\rm pl}$ was conjectured as a small parameter for light SM fermions.}.

\begin{table}[htp]
\begin{center}
\begin{tabular}{c|cccc}
\hline\hline
 $\repb{8_F}^\lambda$ & $( \repb{3}\,, \rep{1}\,, +\frac{1}{3} )_{\rep{F}}^\lambda$ & $( \rep{1}\,, \rep{1}\,,  0 )_{\rep{F}}^\lambda$  &  $( \rep{1}\,, \repb{3}\,,  -\frac{1}{3} )_{\rep{F}}^\lambda$  &  $( \rep{1}\,, \rep{1}\,,  0 )_{\rep{F}}^{\lambda^{ \prime\prime} }$   \\[1mm]
\hline
$\Tc_2^{ \prime \prime \prime }$  &  $- \frac{4}{3} t_2 \tilde c_1^{ \prime \prime }+ \frac{1}{3} \tilde c_2^{\prime \prime}~(- \frac{4}{3} t_2)$ & $-4t_2 \tilde c_1^{ \prime \prime}~(- 4 t_2)$  &  $-4t_2 \tilde c_1^{ \prime \prime} - \frac{1}{3} \tilde c_2^{\prime \prime}~(- 4 t_2)$  & $-4t_2 \tilde c_1^{ \prime \prime}~ ( -4 t_2 )$     \\[1mm]
\hline
$\rep{28_F}$ & $( \repb{3}\,, \rep{1}\,, - \frac{2}{3} )_{\rep{F}}$  &  $( \rep{1}\,, \repb{3}\,,  + \frac{2}{3} )_{\rep{F}}$  &  $( \rep{3}\,, \rep{3 }\,,  0 )_{\rep{F}}$   & $$   \\[1mm]
 \hline
 $\Tc_2^{ \prime \prime \prime }$  & $- \frac{4}{3} t_2 \tilde c_1^{ \prime \prime } - \frac{2}{3} \tilde c_2^{\prime \prime} ~( - \frac{4}{3} t_2)$  &  $4 t_2 \tilde c_1^{ \prime \prime }+ \frac{2 }{3} \tilde c_2^{\prime \prime}~ ( 4 t_2)$  &  $ \frac{4}{3} t_2 \tilde c_1^{ \prime \prime }~ (  \frac{4}{3} t_2)$ & $$   \\[1mm]
\hline
$\repb{8_H}_{ \,, \lambda }$   &  $( \rep{1}\,, \repb{3}\,,  - \frac{1}{3} )_{\rep{H}\,,\lambda}$  &   &   & $$   \\[1mm]
\hline
$\Tc_2^{ \prime \prime \prime }$  &  $- \frac{1}{3} \tilde c_2^{\prime\prime }~( 0)$  &   &  & $$  \\[1mm]
\hline
 $\rep{70_H}$  &  $( \rep{1}\,, \repb{3}\,,  + \frac{2}{3} )_{\rep{H}}^{ \prime\prime \prime}$  &  $( \rep{1}\,, \rep{3}\,,  - \frac{2}{3} )_{\rep{H}}^{ \prime\prime \prime}$  &  & $$   \\[1mm]
\hline
$\Tc_2^{ \prime \prime \prime }$  &   $ \frac{2}{3} \tilde c_2^{\prime\prime }~(0) $ &   $-8 t_2 \tilde c_1^{ \prime\prime } -\frac{2}{3} \tilde c_2^{\prime\prime }~( -8 t_2)$ &  & $$  \\[1mm]
\hline
 $\rep{56_H}$   & $( \rep{1}\,, \repb{3}\,,  + \frac{2}{3} )_{\rep{H}}$  & $( \rep{1}\,, \rep{3}\,,  + \frac{1}{3} )_{\rep{H}}^{\prime \prime}$ & $( \rep{1}\,, \repb{3}\,,  + \frac{2}{3} )_{\rep{H}}^{\prime \prime}$  &  $$ \\[1mm]
\hline
$\Tc_2^{ \prime \prime \prime }$ & $t_2 \tilde c_1^{\prime\prime} + \frac{2}{3} \tilde c_2^{ \prime\prime}~( t_2)$  & $ t_2 \tilde c_1^{\prime\prime} + \frac{1}{3} \tilde c_2^{ \prime\prime} ~(t_2)$  & $ t_2 \tilde c_1^{\prime\prime} + \frac{2}{3} \tilde c_2^{ \prime\prime} ~( t_2)$ &  $ $    \\[1mm]
\hline\hline
  $\repb{8_F}^{\dot \lambda}$& $( \repb{3}\,, \rep{1}\,, +\frac{1}{3} )_{\rep{F}}^{\dot \lambda}$ & $( \rep{1}\,, \rep{1}\,,  0 )_{\rep{F}}^{\dot \lambda}$ &  $( \rep{1}\,, \repb{3 }\,,  -\frac{1}{3 } )_{\rep{F}}^{\dot \lambda}$  &  $( \rep{1}\,, \rep{1}\,,  0 )_{\rep{F}}^{\dot \lambda^{ \prime\prime } }$   \\[1mm]
 \hline
 $\Tc_3^{ \prime \prime \prime }$  &  $- \frac{4}{3} t_3 \tilde d_1^{\prime \prime} + \frac{1}{3} \tilde d_2^{\prime \prime}~( - \frac{4}{3} t_3 )$  &  $- 4 t_3 \tilde d_1^{\prime \prime} ~(- 4 t_3)$  & $- 4 t_3 \tilde d_1^{\prime \prime} - \frac{1}{3} \tilde d_2^{\prime \prime}~(- 4 t_3)$  &  $- 4 t_3 \tilde d_1^{\prime \prime} ~(- 4 t_3)$  \\[1mm]
 \hline
$\rep{56_F}$  & $( \rep{1}\,,  \repb{3}\,, +\frac{2}{3} )_{\rep{F}}^\prime$ &  $( \repb{3 }\,, \rep{1}\,,  -\frac{2}{3 } )_{\rep{F}}^\prime$  &    &     \\[1mm]
 \hline
 $\Tc_3^{ \prime \prime \prime }$  &  $4 t_3 \tilde d_1^{\prime \prime} + \frac{2}{3} \tilde d_2^{\prime \prime}~( 4 t_3)$  &  $ -\frac{4}{3 } t_3 \tilde d_1^{\prime \prime} - \frac{2}{3} \tilde d_2^{\prime \prime}~( -\frac{4}{3 } t_3)$ &   &   \\[1mm]
 \hline
  $\rep{56_F}$ & $( \rep{3}\,,  \rep{3}\,, 0 )_{\rep{F}}^\prime$ & $(  \rep{1}\,, \repb{3 }\,,  +\frac{2}{3 } )_{\rep{F}}^{\prime\prime} $ &  $( \rep{3}\,,  \rep{3}\,, 0 )_{\rep{F}}^{\prime \prime} $  & $( \repb{3 }\,,   \rep{1}\,,  -\frac{2}{3 } )_{\rep{F}}^{\prime\prime \prime}$  \\[1mm]
 \hline
 $\Tc_3^{ \prime \prime \prime }$  &  $\frac{4}{3} t_3 \tilde d_1^{\prime \prime} ~(\frac{4}{3} t_3)$  &  $ 4 t_3 \tilde d_1^{\prime \prime} + \frac{2}{3} \tilde d_2^{\prime \prime} ~( 4 t_3 )$  &  $\frac{4}{3} t_3 \tilde d_1^{\prime \prime} ~( \frac{4}{3} t_3)$  &  $ -\frac{4}{3} t_3 \tilde d_1^{\prime \prime} - \frac{2}{3} \tilde d_2^{\prime \prime} ~( -\frac{4}{3} t_3)$  \\[1mm]
 \hline
  $\repb{28_H}_{ \,, \dot \lambda }$  &  $( \rep{1}\,,  \repb{3}\,, -\frac{1}{ 3} )_{\rep{H}\,, \dot \lambda}^\prime$  & $( \rep{1}\,,  \rep{3}\,, -\frac{2 }{ 3} )_{\rep{H}\,, \dot \lambda}$   &  $( \rep{1}\,,  \repb{3}\,, -\frac{1 }{ 3} )_{\rep{H}\,, \dot \lambda}$  & $$   \\[1mm]
 \hline
 $\Tc_3^{ \prime \prime \prime }$  & $- \frac{1}{3} \tilde d_2^{\prime \prime}~(0)$  & $- \frac{2}{3} \tilde d_2^{\prime \prime}~( 0)$  & $- \frac{1}{3} \tilde d_2^{\prime \prime}~( 0)$ &     \\[1mm]
 \hline
 $\rep{ 56_H}$  &  $( \rep{1}\,,  \repb{3}\,, +\frac{2}{ 3} )_{\rep{H}}$  & $( \rep{1}\,,  \rep{3}\,, +\frac{1}{ 3} )_{\rep{H}}^{\prime \prime}$   &  $( \rep{1}\,,  \repb{3}\,, +\frac{2}{ 3} )_{\rep{H}}^{\prime \prime}$  & $$    \\[1mm]
    \hline
$\Tc_3^{ \prime \prime \prime }$ &  $2t_3 \tilde d_1^{\prime\prime} + \frac{2}{3} \tilde d_2^{ \prime\prime}~( 2 t_3 )$  & $2t_3 \tilde d_1^{\prime\prime} + \frac{1}{3} \tilde d_2^{ \prime\prime}~( 2 t_3 )$   &  $2t_3 \tilde d_1^{\prime\prime} + \frac{2}{3} \tilde d_2^{ \prime\prime} ~( 2 t_3 )$  & $$    \\[1mm]
 \hline\hline
\end{tabular}
\end{center}
\caption{The $\widetilde{ {\rm U}}(1)_{T_2^{\prime \prime \prime } } \otimes \widetilde{ {\rm U}}(1)_{T_3^{\prime \prime \prime} }$ charges for massless fermions and possible symmetry-breaking Higgs components in the $\Gc_{331}$ theory.
With the 't Hooft anomaly matching condition in Eq.~\eqref{eq:G331_match} and the neutrality conditions in Eq.~\eqref{eq:T23ppp_neutral}, the $\widetilde{ {\rm U}}(1)_{T_2^{\prime \prime \prime} } \otimes \widetilde{ {\rm U}}(1)_{T_3^{\prime \prime \prime} }$ charges are displayed in the parentheses.
}
\label{tab:G331_Tcharges}
\end{table}

\para
After integrating out the massive fermions in Eqs.~\eqref{eqs:masses_341_01}, \eqref{eqs:masses_341_02}, and \eqref{eq:Yukawa_341_03}, the remaining massless fermions expressed in terms of the $\Gc_{331}$ irreps are the following
\beqn\label{eq:331_fermions}
&& \Big[ ( \repb{3}\,, \rep{1}\,, +\frac{1}{3} )_{ \mathbf{F}}^\Lambda \oplus ( \rep{1}\,, \rep{1}\,, 0 )_{ \mathbf{F}}^\Lambda \Big]  \oplus \Big[ ( \rep{1}\,, \repb{3}\,, -\frac{1}{3} )_{ \mathbf{F}}^\Lambda \oplus ( \rep{1}\,, \rep{1}\,, 0 )_{ \mathbf{F}}^{\Lambda^{\prime \prime }}  \Big]  \subset \repb{8_F}^\Lambda \,, \non
&& \Lambda = ( \lambda\,, \dot \lambda ) \,, \quad \lambda = (3\,, {\rm VI})\,,  \quad \dot \lambda = (\dot 1\,, \dot 2\,, \dot {\rm IIX}\,, \dot {\rm IX} )\,,\non
&& ( \rep{1}\,, \rep{1}\,, 0)_{ \mathbf{F}}^{{\rm IV} } \subset \repb{8_F}^{{\rm IV} } \,, \quad ( \rep{1}\,, \rep{1}\,, 0)_{ \mathbf{F}}^{{\rm V} \,, {\rm V}^{\prime\prime} } \subset \repb{8_F}^{{\rm V} } \,, \non
&&  ( \rep{1}\,, \rep{1}\,, 0)_{ \mathbf{F}}^{\dot {\rm VII} } \oplus ( \rep{1}\,, \rep{1}\,, 0)_{ \mathbf{F}}^{\dot {\rm VII}^{ \prime \prime} } \subset \repb{8_F}^{\dot {\rm VII} } \,,  \non
%%%%%%%%%%%%%%%%%%%%%%%%%%%%%%%%%%%%%%%%%%%%%
&&\Big[ \cancel{ ( \rep{3}\,, \rep{1}\,, -\frac{1}{3} )_{ \mathbf{F}} } \oplus ( \repb{3}\,, \rep{1}\,, -\frac{2}{3} )_{ \mathbf{F}} \Big] \oplus \Big[ \bcancel{ ( \rep{1}\,, \rep{3}\,, +\frac{1}{3} )_{ \mathbf{F}} } \oplus ( \rep{1}\,, \repb{3}\,, +\frac{2}{3} )_{ \mathbf{F}} \Big] \non
&\oplus& \Big[ ( \rep{3}\,, \rep{3}\,, 0 )_{ \mathbf{F}} \oplus \bcancel{ ( \rep{3}\,, \rep{1}\,, -\frac{1}{3} )_{ \mathbf{F}}^{\prime\prime} } \Big]  \oplus \cancel{ \Big[ ( \rep{1}\,, \rep{3}\,, +\frac{1}{3} )_{ \mathbf{F}} \oplus  ( \rep{1}\,, \rep{1}\,, 0 )_{ \mathbf{F}}^{\prime\prime}  \Big] }  \subset \rep{28_F}\,, \non
%%%%%%%%%%%%%%%%%%%%%%%%%%%%%%%%%%%%%%%%%%%%%
&&\Big[ ( \rep{1}\,, \repb{3}\,, +\frac{2}{3} )_{ \mathbf{F}}^\prime \oplus \bcancel{ ( \rep{1}\,, \rep{1}\,, +1 )_{ \mathbf{F}}^{\prime\prime} } \Big] \oplus \Big[  ( \repb{3}\,, \rep{1}\,, -\frac{2}{3} )_{ \mathbf{F}}^\prime \oplus  \bcancel{ ( \rep{1}\,, \rep{1}\,, -1 )_{ \mathbf{F}} }  \Big] \non
&\oplus&  \Big[ ( \rep{3}\,, \rep{3}\,, 0 )_{ \mathbf{F}}^\prime \oplus \bcancel{( \rep{3}\,, \repb{3}\,, +\frac{1}{3} )_{ \mathbf{F}} } \oplus \bcancel{  ( \rep{1}\,, \rep{3}\,, +\frac{1}{3} )_{ \mathbf{F}}^\prime } \oplus ( \rep{1}\,, \repb{3}\,, +\frac{2}{3})_{ \mathbf{F}}^{\prime \prime} \Big] \non
&\oplus& \Big[ ( \rep{3}\,, \rep{3}\,, 0)_{ \mathbf{F}}^{\prime\prime} \oplus \bcancel{ ( \rep{3}\,, \rep{1}\,, -\frac{1}{3})_{ \mathbf{F}}^{\prime\prime\prime\prime\prime}  } \oplus \bcancel{ ( \repb{3}\,, \rep{3}\,, -\frac{1}{3})_{ \mathbf{F}} }  \oplus ( \repb{3}\,, \rep{1}\,, -\frac{2}{3})_{ \mathbf{F}}^{\prime\prime\prime} \Big]   \subset \rep{56_F}\,.
\eeqn
We use the slashes and the back slashes to cross out massive fermions at the first and the second stages, respectively.
From the anomaly-free conditions, we find that one of the $\repb{8_F}^\lambda$ and one of the $\repb{8_F}^{\dot \lambda}$ are integrated out.
Without loss of generality, we choose the massive anti-fundamental fermions to be $\lambda={\rm V}$ and $\dot \lambda=\dot {\rm VII}$ at this stage.

\para
By looking at the remaining massless fermions in Eq.~\eqref{eq:331_fermions}, one finds that a three-generational repetitive structure has already emerged in the effective $\Gc_{331}$ theory.

\para
Thus, we define the $\widetilde{ {\rm U}}(1)_{T_2^{\prime \prime \prime} } \otimes \widetilde{ {\rm U}}(1)_{T_3^{ \prime \prime \prime} }$ charges after this symmetry-breaking stage as
\beqn\label{eq:G331_U1charges}
&& \Tc_2^{\prime \prime \prime} \equiv \tilde c_1^{\prime \prime} \Tc_2^{ \prime \prime} + \tilde c_2^{ \prime \prime} \Xc_2 \,,\quad \Tc_3^{ \prime \prime \prime }\equiv \tilde d_1^{ \prime \prime } \Tc_3^{\prime \prime } + \tilde d_2^{ \prime \prime} \Xc_2 \,,
\eeqn
and they are explicitly listed in Tab.~\ref{tab:G331_Tcharges} for massless fermions and possible symmetry-breaking Higgs components.
Accordingly, we find the following global anomalies of
\beqn\label{eq:G331_U1anoms}
&& \[ {\rm grav} \]^2 \cdot \widetilde{ {\rm U} }(1)_{ T_2^{\prime \prime \prime} } = -40 t_2 \tilde c_1^{\prime \prime} \,,\quad \[ \widetilde{ {\rm U} }(1)_{ T_2^{ \prime\prime \prime } } \]^3 = -640 ( t_2 \tilde c_1^{ \prime \prime } )^3 \,,\non
&& \[ {\rm grav} \]^2 \cdot \widetilde{ {\rm U} }(1)_{ T_3^{\prime \prime \prime} } = -64 t_3 \tilde d_1^{ \prime \prime } \,,\quad \[ \widetilde{ {\rm U} }(1)_{ T_3^{\prime \prime \prime } } \]^3 = -1024 ( t_3 \tilde d_1^{ \prime \prime } )^3 \,.
\eeqn
By matching with the global anomalies in Eq.~\eqref{eq:G341_U1anoms}, we find that
\beqn\label{eq:G331_match}
&& \tilde c_1^{\prime \prime} = \tilde d_1^{ \prime \prime } = 1\,.
\eeqn

%%%%%%%%%%%%%%%%%%%%%%%%%%%%
\subsection{The third stage}
%%%%%%%%%%%%%%%%%%%%%%%%%%%%

\para
From the group theoretical consideration, the third symmetry-breaking stage of $\Gc_{331} \to \Gc_{\rm SM}$ can be achieved by Higgs fields of $( \rep{1} \,, \repb{3} \,, -\frac{1}{3} )_{\mathbf{H}\,,\lambda}   \subset \repb{8_H}_{\,,\lambda}$ in Eq.~\eqref{eq:SU8_Higgs_Br01}, $( \rep{1} \,, \repb{3} \,, -\frac{1}{3} )_{\mathbf{H}\,,\dot \lambda}^\prime \oplus ( \rep{1} \,, \repb{3} \,, -\frac{1}{3} )_{\mathbf{H}\,,\dot \lambda}   \subset \repb{28_H}_{\,,\dot\lambda}$ in Eq.~\eqref{eq:SU8_Higgs_Br02}, 
%$\[ ( \rep{1} \,, \repb{3} \,, -\frac{1}{3} )_{\mathbf{H}\,,\dot \lambda}^\prime \oplus ( \rep{1} \,, \repb{3} \,, -\frac{1}{3} )_{\mathbf{H}\,,\dot \lambda}  \]  \subset ( \rep{1} \,, \rep{6} \,, -\frac{1}{2} )_{\mathbf{H}\,,\dot \lambda}  \oplus  ( \rep{1} \,, \repb{4} \,, -\frac{1}{4} )_{\mathbf{H}\,,\dot \lambda}  \subset ( \rep{1} \,, \rep{6} \,, -\frac{1}{2} )_{\mathbf{H}\,,\dot \lambda} \oplus  ( \repb{4} \,, \repb{4} \,, 0 )_{\mathbf{H}\,, \dot \lambda}  \subset \repb{28_H}_{\,,\dot\lambda}$ in Eq.~\eqref{eq:SU8_Higgs_Br02}, 
and $( \rep{1} \,, \rep{3} \,, +\frac{1}{3} )_{\mathbf{H}}^{\prime\prime }\subset \rep{56_H}$ in Eq.~\eqref{eq:SU8_Higgs_Br04}.
However, the Higgs field of $( \rep{1} \,, \rep{3} \,, +\frac{1}{3} )_{\mathbf{H}}^{\prime\prime }\subset \rep{56_H}$ has different $\widetilde{ {\rm U}}(1)_{T_2^{\prime \prime \prime} } \otimes \widetilde{ {\rm U}}(1)_{T_3^{ \prime \prime \prime} }$ charges from the $( \rep{1} \,, \repb{3} \,, -\frac{1}{3} )_{\mathbf{H}\,,\lambda}\subset \repb{8_H}_{\,, \lambda}$ and the $( \rep{1} \,, \repb{3} \,, -\frac{1}{3} )_{\mathbf{H}\,,\dot \lambda}^\prime \oplus ( \rep{1} \,, \repb{3} \,, -\frac{1}{3} )_{\mathbf{H}\,,\dot \lambda} \subset \repb{28_H}_{\,, \dot \lambda}$.
If one assumes that the $( \rep{1} \,, \rep{3} \,, +\frac{1}{3} )_{\mathbf{H}}^{\prime\prime }\subset \rep{56_H}$ is $\widetilde{ {\rm U}}(1)_{T_2^{\prime \prime \prime} } \otimes \widetilde{ {\rm U}}(1)_{T_3^{ \prime \prime \prime} }$-neutral, we find that $\tilde c_2^{ \prime \prime}= -3 t_2$ and $\tilde d_2^{ \prime \prime}= -6 t_3$ together with Eq.~\eqref{eq:G331_match}.
With such a solution, the global $\widetilde{ {\rm U}}(1)_{ T_2^{\prime\prime \prime} }$ charges for fermions in the rank-$2$ sector cannot match with the global $\widetilde{ {\rm U}}(1)_{ T_3^{\prime\prime \prime} }$ charges in the rank-$3$ sector.
Thus it is most natural to require the $\widetilde{ {\rm U}}(1)_{T_2^{\prime \prime \prime} } \otimes \widetilde{ {\rm U}}(1)_{T_3^{ \prime \prime \prime} }$-neutral conditions for Higgs fields of $( \rep{1} \,, \repb{3} \,, -\frac{1}{3} )_{\mathbf{H}\,,\lambda} \subset  \repb{8_H}_{\,,\lambda}$ and $ ( \rep{1} \,, \repb{3} \,, -\frac{1}{3} )_{\mathbf{H}\,,\dot \lambda}^\prime \oplus ( \rep{1} \,, \repb{3} \,, -\frac{1}{3} )_{\mathbf{H}\,,\dot \lambda} \subset \repb{28_H}_{\,,\dot\lambda}$, and we find that
\beqn\label{eq:T23ppp_neutral}
&& \tilde c_2^{ \prime \prime} = 0 \,,\quad  \tilde d_2^{ \prime \prime} = 0\,,
\eeqn
in Eq.~\eqref{eq:G331_U1charges}.
These relations lead to non-vanishing $\widetilde{ {\rm U}}(1)_{ T_2^{\prime\prime \prime} }\otimes \widetilde{ {\rm U}}(1)_{ T_3^{\prime\prime \prime} }$ charges of
\beqn
&& \Tc_2^{ \prime\prime \prime } ( ( \rep{1}\,, \rep{3}\,, + \frac{1}{3} )_{\rep{H}}^{ \prime \prime}) =t_2 \,, \quad \Tc_3^{ \prime\prime \prime } ( ( \rep{1}\,, \rep{3}\,, + \frac{1}{3} )_{\rep{H}}^{ \prime \prime} ) = 2t_3 \,,
\eeqn
according to Tab.~\ref{tab:G331_Tcharges}.
Though the Higgs field of $( \rep{1}\,, \rep{3}\,, + \frac{1}{3} )_{\rep{H}}^{ \prime \prime}$ is likely to develop the VEV for this symmetry-breaking stage since it contains the $\Gc_{\rm SM}$-singlet component, its non-vanishing $\widetilde{ {\rm U}}(1)_{ T_2^{\prime\prime \prime} }\otimes \widetilde{ {\rm U}}(1)_{ T_3^{\prime\prime \prime} }$ charges prohibit this to occur.

\para
The Yukawa couplings between the $\repb{8_F}^\lambda$ and the $\rep{28_F}$ can be expressed as follows
\beqn\label{eq:Yukawa_331_01}
&& Y_\Bc \repb{8_F}^\lambda \rep{28_F} \repb{8_H}_{\,,\lambda } + H.c. \non
&\supset& Y_\Bc \Big[ ( \rep{1}\,, \repb{4}\,, -\frac{1}{4} )_{ \mathbf{F}}^\lambda \otimes ( \rep{1}\,, \rep{6}\,, -\frac{1}{2} )_{ \mathbf{F}}  \oplus ( \repb{4}\,, \rep{1}\,, +\frac{1}{4} )_{ \mathbf{F}}^\lambda \otimes ( \rep{4}\,, \rep{4}\,, 0 )_{ \mathbf{F}}   \Big]  \otimes ( \rep{1}\,, \repb{4}\,, -\frac{1}{4} )_{ \mathbf{H}\,, \lambda } + H.c. \non 
&\supset& Y_\Bc  \Big[ ( \rep{1}\,, \repb{4}\,, -\frac{1}{4} )_{ \mathbf{F}}^\lambda \otimes ( \rep{1}\,, \rep{6}\,, -\frac{1}{2} )_{ \mathbf{F}}  \oplus ( \rep{1}\,, \rep{1}\,, 0 )_{ \mathbf{F}}^\lambda \otimes ( \rep{1}\,, \rep{4}\,, +\frac{1}{4} )_{ \mathbf{F}} \non
&\oplus& ( \repb{3}\,, \rep{1}\,, +\frac{1}{3} )_{ \mathbf{F}}^\lambda \otimes ( \rep{3}\,, \rep{4}\,, -\frac{1}{12} )_{ \mathbf{F}}   \Big] \otimes ( \rep{1}\,, \repb{4}\,, -\frac{1}{4} )_{ \mathbf{H}\,, \lambda } + H.c. \non 
&\supset& Y_\Bc  \Big[  ( \rep{1}\,, \repb{3}\,, -\frac{1}{3} )_{ \mathbf{F}}^\lambda \otimes ( \rep{1}\,, \repb{3}\,, +\frac{2}{3} )_{ \mathbf{F}} \oplus ( \rep{1}\,, \rep{1}\,, 0 )_{ \mathbf{F}}^\lambda \otimes ( \rep{1}\,, \rep{3}\,, +\frac{1}{3} )_{ \mathbf{F}} \non
&\oplus&  (  \repb{3}\,, \rep{1}\,, +\frac{1}{3} )_{ \mathbf{F}}^\lambda \otimes ( \rep{3}\,, \rep{3}\,, 0 )_{ \mathbf{F}}  \Big] \otimes \langle ( \rep{1}\,, \repb{3}\,, -\frac{1}{3} )_{ \mathbf{H}\,, \lambda }  \rangle + H.c. \non
&\supset& Y_\Bc \Big[  ( \rep{1}\,, \repb{2}\,, -\frac{1}{2} )_{ \mathbf{F}}^\lambda \otimes ( \rep{1}\,, \repb{2}\,, +\frac{1}{2} )_{ \mathbf{F}}^\prime \oplus ( \rep{1}\,, \rep{1}\,, 0)_{ \mathbf{F}}^{\lambda } \otimes ( \rep{1}\,, \rep{1}\,, 0 )_{ \mathbf{F}} \non
&\oplus&  (  \repb{3}\,, \rep{1}\,, +\frac{1}{3} )_{ \mathbf{F}}^\lambda \otimes ( \rep{3}\,, \rep{1}\,, -\frac{1}{3} )_{ \mathbf{F}}^\prime  \Big] \otimes ( \rep{1}\,, \rep{1}\,, 0 )_{ \mathbf{H}\,, \lambda }  + H.c.  \,.
\eeqn
They lead to fermion masses of
\beqn\label{eq:masses_331_01}
&& Y_\Bc  \Big[  ( \rep{1}\,, \repb{2}\,, -\frac{1}{2} )_{ \mathbf{F}}^{\rm VI} \otimes ( \rep{1}\,, \repb{2}\,, +\frac{1}{2} )_{ \mathbf{F}}^\prime \oplus ( \rep{1}\,, \rep{1}\,, 0)_{ \mathbf{F}}^{\rm VI } \otimes ( \rep{1}\,, \rep{1}\,, 0 )_{ \mathbf{F}} \oplus  (  \repb{3}\,, \rep{1}\,, +\frac{1}{3} )_{ \mathbf{F}}^{\rm VI} \otimes ( \rep{3}\,, \rep{1}\,, -\frac{1}{3} )_{ \mathbf{F}}^\prime  \Big] \non
&& \otimes ( \rep{1}\,, \rep{1}\,, 0 )_{ \mathbf{H}\,, {\rm VI} }  + H.c. \non
&=&  \frac{1}{ \sqrt{2} }  Y_\Bc   \Big(  \nG_L^\prime \nG_R^{\prime\,c }- \eG_L^\prime  \eG_R^{\prime\,c}  + \check \Nc_L^{\rm VI}  \check \nG_R^c + \DG_L^\prime  \DG_R^{\prime\,c}  \Big)  V_{ \repb{3}\,, {\rm VI} } + H.c. \,,
\eeqn
where we choose $\lambda ={\rm VI}$ at this stage.
Thus, we can identify that $( \rep{1}\,, \repb{2}\,, -\frac{1}{2} )_{ \mathbf{F}}^{\rm VI}  \equiv ( \eG_L^\prime\,, - \nG_L^\prime)$, and $(  \repb{3}\,, \rep{1}\,, +\frac{1}{3} )_{ \mathbf{F}}^{\rm VI} \equiv \DG_R^{\prime\,c}$.
The $\check \Nc_L^{\rm VI}  \check \nG_R^c + H.c.$ in Eq.~\eqref{eq:masses_331_01} is a mass mixing term of sterile neutrinos.

\para
The Yukawa couplings between the $\repb{8_F}^{\dot \lambda}$ and the $\rep{56_F}$ can be expressed as follows
\beqs\label{eqs:Yukawa_331_02}
\beqn
%%%%%%%%%%%%%%%%%%%%%%%%%%%%%%%%%%%%%%%%%%%%%
&& Y_\Dc  \repb{8_F}^{\dot \lambda} \rep{56_F} \repb{28_H}_{\,,\dot \lambda } + H.c. \non
&\supset& Y_\Dc \Big[ ( \rep{1}\,, \repb{4}\,, -\frac{1}{4})_{ \mathbf{F}}^{\dot \lambda} \otimes ( \rep{1}\,, \repb{4}\,, +\frac{3}{4})_{ \mathbf{F}}  \oplus  ( \repb{3}\,, \rep{1}\,, +\frac{1}{3})_{ \mathbf{F}}^{\dot \lambda} \otimes ( \rep{3}\,, \rep{6}\,, +\frac{1}{6})_{ \mathbf{F}}  \non
&\oplus&  ( \rep{1}\,, \rep{1}\,, 0 )_{ \mathbf{F}}^{\dot \lambda} \otimes ( \rep{1}\,, \rep{6}\,, +\frac{1}{2})_{ \mathbf{F}}^\prime \Big] \otimes   ( \rep{1}\,, \rep{6}\,, -\frac{1}{2})_{ \mathbf{H}\,, \dot \lambda } + H.c.  \non
&\supset& Y_\Dc \Big[   ( \rep{1}\,, \repb{3}\,, -\frac{1}{3})_{ \mathbf{F}}^{\dot \lambda} \otimes ( \rep{1}\,, \repb{3}\,, +\frac{2}{3})_{ \mathbf{F}}^\prime  \oplus ( \repb{3}\,, \rep{1}\,, +\frac{1}{3})_{ \mathbf{F}}^{\dot \lambda} \otimes ( \rep{3}\,, \rep{3}\,, 0 )_{ \mathbf{F}}^\prime \non
& \oplus& ( \rep{1}\,, \rep{1}\,,  0 )_{ \mathbf{F}}^{\dot \lambda} \otimes ( \rep{1}\,, \rep{3}\,, +\frac{1}{3} )_{ \mathbf{F}}^\prime  \Big]  \otimes  \langle  ( \rep{1}\,, \repb{3}\,, -\frac{1}{3})_{ \mathbf{H}\,, \dot \lambda }^\prime \rangle + H.c.  \non
&\supset& Y_\Dc \Big[ ( \rep{1}\,, \repb{2}\,, -\frac{1}{2} )_{ \mathbf{F}}^{\dot \lambda} \otimes ( \rep{1}\,, \repb{2}\,, +\frac{1}{2})_{ \mathbf{F}}^{\prime  \prime\prime}   \oplus   ( \repb{3}\,, \rep{1}\,, +\frac{1}{3})_{ \mathbf{F}}^{\dot \lambda} \otimes ( \rep{3}\,, \rep{1}\,, -\frac{1}{3} )_{ \mathbf{F}}^{\prime \prime\prime }  \non
& \oplus& ( \rep{1}\,, \rep{1}\,,  0 )_{ \mathbf{F}}^{\dot \lambda} \otimes ( \rep{1}\,, \rep{1}\,, 0 )_{ \mathbf{F}}^{\prime \prime \prime } \Big]  \otimes ( \rep{1}\,, \rep{1}\,, 0 )_{ \mathbf{H}\,, \dot \lambda }^\prime + H.c.   \,,\label{eq:Yukawa_331_02a} \\[1mm]
%%%%%%%%%%%%%%%%%%%%%%%%%%%%%%%%%%%%%%%%%%%%%
&& Y_\Dc \repb{8_F}^{\dot \lambda} \rep{56_F} \repb{28_H}_{\,,\dot \lambda } + H.c. \non
&\supset& Y_\Dc \Big[ ( \rep{1}\,, \repb{4}\,, -\frac{1}{4})_{ \mathbf{F}}^{\dot \lambda} \otimes ( \rep{4}\,, \rep{6}\,, +\frac{1}{4})_{ \mathbf{F}}  \oplus  ( \repb{4}\,, \rep{1}\,, +\frac{1}{4})_{ \mathbf{F}}^{\dot \lambda} \otimes ( \rep{6}\,, \rep{4}\,, -\frac{1}{4})_{ \mathbf{F}}  \Big]  \otimes ( \repb{4}\,, \repb{4}\,, 0 )_{ \mathbf{H}\,, \dot \lambda } + H.c.  \non
&\supset& Y_\Dc  \Big[   ( \rep{1}\,, \repb{4}\,, -\frac{1}{4})_{ \mathbf{F}}^{\dot \lambda} \otimes ( \rep{1}\,, \rep{6}\,, +\frac{1}{2})_{ \mathbf{F}}^\prime  \oplus  ( \repb{3}\,, \rep{1}\,, +\frac{1}{3})_{ \mathbf{F}}^{\dot \lambda} \otimes ( \rep{3}\,, \rep{4}\,, -\frac{1}{12})_{ \mathbf{F}}^\prime  \Big]  \otimes ( \rep{1}\,, \repb{4}\,, -\frac{1}{4} )_{ \mathbf{H}\,, \dot \lambda } + H.c. \non
&\supset& Y_\Dc \Big[  ( \rep{1}\,, \repb{3}\,, -\frac{1}{3})_{ \mathbf{F}}^{\dot \lambda} \otimes ( \rep{1}\,, \repb{3}\,, +\frac{2}{3})_{ \mathbf{F}}^{ \prime\prime }   \oplus  ( \rep{1}\,, \rep{1}\,, 0 )_{ \mathbf{F}}^{{\dot \lambda}^{\prime \prime}  } \otimes ( \rep{1}\,, \rep{3}\,, +\frac{1}{3})_{ \mathbf{F}}^{ \prime}   \non
& \oplus& ( \repb{3}\,, \rep{1}\,, +\frac{1}{3})_{ \mathbf{F}}^{\dot \lambda} \otimes ( \rep{3}\,, \rep{3}\,, 0 )_{ \mathbf{F}}^{\prime \prime } \Big] \otimes  \langle ( \rep{1}\,, \repb{3}\,, -\frac{1}{3} )_{ \mathbf{H}\,, \dot \lambda }  \rangle + H.c. \non
&\supset& Y_\Dc \Big[  ( \rep{1}\,, \repb{2}\,, -\frac{1}{2})_{ \mathbf{F}}^{\dot \lambda} \otimes ( \rep{1}\,, \repb{2}\,, +\frac{1}{2})_{ \mathbf{F}}^{ \prime\prime\prime \prime \prime }   \oplus  ( \rep{1}\,, \rep{1}\,, 0 )_{ \mathbf{F}}^{{\dot \lambda}^{\prime \prime}  } \otimes ( \rep{1}\,, \rep{1}\,, 0  )_{ \mathbf{F}}^{ \prime\prime \prime } \non
& \oplus&  ( \repb{3}\,, \rep{1}\,, +\frac{1}{3})_{ \mathbf{F}}^{\dot \lambda} \otimes ( \rep{3}\,, \rep{1}\,, -\frac{1}{3} )_{ \mathbf{F}}^{\prime \prime \prime\prime } \Big]  \otimes ( \rep{1}\,, \rep{1}\,, 0 )_{ \mathbf{H}\,, \dot \lambda } + H.c.  \,. \label{eq:Yukawa_331_02b}
\eeqn
\eeqs
%
%
%Without loss of generality, we choose $\dot \lambda= \dot \kappa=\dot {\rm IIX}$ in Eq.~\eqref{eq:Yukawa_331_02a} and $\dot \lambda= \dot \kappa= \dot {\rm IX}$ in Eq.~\eqref{eq:Yukawa_331_02b} for the DRS limit, respectively.
%
%
The fermion masses from Eq.~\eqref{eq:Yukawa_331_02a} are
\beqs
\beqn
&& Y_\Dc   (  \rep{1}\,, \repb{2}\,, -\frac{1}{2} )_{ \mathbf{F}}^{\dot {\rm IIX}} \otimes ( \rep{1}\,, \repb{2}\,, +\frac{1}{2} )_{ \mathbf{F}}^{ \prime\prime \prime } \otimes ( \rep{1}\,, \rep{1}\,, 0 )_{ \mathbf{H}\,, \dot {\rm IIX}} + H.c. \non
&=& \frac{1}{\sqrt{2}}  Y_\Dc  \Big( \nG_L^{\prime \prime \prime }  \nG_R^{\prime \prime \prime\,c}   -   \eG_L^{\prime \prime \prime }  \eG_R^{\prime \prime \prime\,c}   \Big) V_{ \repb{3}\,, {\dot {\rm IIX}} }^\prime + H.c. \,, \label{eq:masses_331_02a}  \\[1mm]
%%%%%%%%%%%%%%%%%%%%%%%%%%%%%%%%%%%%%%%%%%%%%
&& Y_\Dc  (  \rep{1}\,, \rep{1}\,, 0 )_{ \mathbf{F}}^{\dot {\rm IIX} } \otimes ( \rep{1}\,, \rep{1}\,, 0 )_{ \mathbf{F}}^{ \prime \prime \prime  }  \otimes ( \rep{1}\,, \rep{1}\,, 0 )_{ \mathbf{H}\,, \dot {\rm IIX} } + H.c. \non
&=& \frac{1}{\sqrt{2}} Y_\Dc  \check \Nc_L^{ \dot {\rm IIX} }  \check \nG_R^{ \prime\prime \prime\,c } V_{ \repb{3}\,,\dot {\rm IIX} }^\prime + H.c.  \,, \label{eq:masses_331_02b} \\[1mm]
%%%%%%%%%%%%%%%%%%%%%%%%%%%%%%%%%%%%%%%%%%%%%
&& Y_\Dc  (  \repb{3}\,, \rep{1}\,, +\frac{1}{3} )_{ \mathbf{F}}^{\dot {\rm IIX}} \otimes ( \rep{3}\,, \rep{1}\,, -\frac{1}{3} )_{ \mathbf{F}}^{ \prime \prime\prime  } \otimes ( \rep{1}\,, \rep{1}\,, 0 )_{ \mathbf{H}\,, \dot {\rm IIX}} + H.c. \non
&=&  \frac{1}{\sqrt{2}} Y_\Dc \DG_L^{\prime\prime \prime }  \DG_R^{\prime\prime \prime\,c } V_{ \repb{3}\,,\dot  {\rm IIX} }^\prime + H.c.  \,, \label{eq:masses_331_02c}
\eeqn
\eeqs
where we choose $\dot \lambda=  \dot {\rm IIX}$.
Thus, we can identify that $( \rep{1}\,, \repb{2}\,, -\frac{1}{2} )_{ \mathbf{F}}^{ \dot {\rm IIX} }  \equiv ( \eG_L^{\prime \prime\prime} \,, - \nG_L^{\prime\prime\prime} )$, and $(  \repb{3}\,, \rep{1}\,, +\frac{1}{3} )_{ \mathbf{F}}^{\dot {\rm IIX}} \equiv \DG_R^{\prime\prime\prime\,c}$.
The fermion masses from Eq.~\eqref{eq:Yukawa_331_02b} are
\beqs
\beqn
&& Y_\Dc   (  \rep{1}\,, \repb{2}\,, -\frac{1}{2} )_{ \mathbf{F}}^{\dot {\rm IX}} \otimes ( \rep{1}\,, \repb{2}\,, +\frac{1}{2} )_{ \mathbf{F}}^{ \prime\prime\prime \prime \prime } \otimes ( \rep{1}\,, \rep{1}\,, 0 )_{ \mathbf{H}\,, \dot {\rm IX}} + H.c. \non
&=& \frac{1}{\sqrt{2}}  Y_\Dc  \Big( \nG_L^{\prime \prime \prime \prime \prime}  \nG_R^{\prime \prime \prime \prime \prime\,c}   -   \eG_L^{\prime \prime \prime \prime \prime}  \eG_R^{\prime \prime \prime \prime \prime\,c}   \Big) V_{ \repb{3}\,, {\dot {\rm IX}} } + H.c. \,, \label{eq:masses_331_02d} \\[1mm]
%%%%%%%%%%%%%%%%%%%%%%%%%%%%%%%%%%%%%%%%%%%%%
&& Y_\Dc  (  \rep{1}\,, \rep{1}\,, 0 )_{ \mathbf{F}}^{\dot {\rm IX}^{\prime\prime} } \otimes ( \rep{1}\,, \rep{1}\,, 0 )_{ \mathbf{F}}^{ \prime \prime \prime  }  \otimes ( \rep{1}\,, \rep{1}\,, 0 )_{ \mathbf{H}\,, \dot \kappa} + H.c. \non
&=& \frac{1}{\sqrt{2}} Y_\Dc \check \Nc_L^{ \dot {\rm IX}^{\prime\prime} }   \check \nG_R^{ \prime\prime \prime\,c } V_{ \repb{3}\,,\dot {\rm IX} } + H.c.  \,, \label{eq:masses_331_02e} \\[1mm]
%%%%%%%%%%%%%%%%%%%%%%%%%%%%%%%%%%%%%%%%%%%%%
&& Y_\Dc  (  \repb{3}\,, \rep{1}\,, +\frac{1}{3} )_{ \mathbf{F}}^{\dot {\rm IX}} \otimes ( \rep{3}\,, \rep{1}\,, -\frac{1}{3} )_{ \mathbf{F}}^{ \prime\prime \prime\prime  } \otimes ( \rep{1}\,, \rep{1}\,, 0 )_{ \mathbf{H}\,, \dot {\rm IX}} + H.c. \non
&=&  \frac{1}{\sqrt{2}} Y_\Dc   \DG_L^{\prime\prime \prime \prime }  \DG_R^{\prime\prime \prime \prime\,c } V_{ \repb{3}\,,\dot  {\rm IX} } + H.c.  \,, \label{eq:masses_331_02f}
\eeqn
\eeqs
where we choose $\dot \lambda = \dot {\rm IX}$.
Thus, we can identify that $(  \rep{1}\,, \repb{2}\,, -\frac{1}{2} )_{ \mathbf{F}}^{\dot {\rm IX}} \equiv ( \eG_L^{\prime \prime\prime \prime \prime }\,, - \nG_L^{\prime \prime\prime \prime \prime } )$, and $(  \repb{3}\,, \rep{1}\,, +\frac{1}{3} )_{ \mathbf{F}}^{\dot {\rm IX}} \equiv \DG_R^{\prime \prime\prime \prime\,c }$.
The $\check \Nc_L^{ \dot {\rm IIX} }  \check \nG_R^{ \prime\prime \prime\,c } + H.c.$ in Eq.~\eqref{eq:masses_331_02b} and $\check \Nc_L^{ \dot {\rm IX}^{\prime\prime} }   \check \nG_R^{ \prime\prime \prime\,c }+ H.c.$ in Eq.~\eqref{eq:masses_331_02e} are mass mixing terms of sterile neutrinos.

\para
One may wonder why two anti-fundamental fermions of $\repb{8_F}^{\dot \lambda}\, ( \dot \lambda = \dot {\rm IIX} \,, \dot {\rm IX})$ should obtain masses from different components of $( \rep{1}\,, \repb{3}\,, -\frac{1}{3})_{ \mathbf{H}\,, \dot \lambda }^\prime  \subset ( \rep{1}\,, \rep{6}\,, -\frac{1}{2})_{ \mathbf{H}\,, \dot \lambda }  \subset \repb{28_H}_{\,,\dot \lambda }$ and $( \rep{1}\,, \repb{3}\,, -\frac{1}{3})_{ \mathbf{H}\,, \dot \lambda }  \subset ( \repb{4}\,, \repb{4}\,, 0 )_{ \mathbf{H}\,, \dot \lambda }  \subset \repb{28_H}_{\,,\dot \lambda }$ at this stage.
If both $\repb{8_F}^{\dot \lambda}$ coupled to the $( \rep{1}\,, \rep{3}\,, -\frac{1}{3})_{ \mathbf{H}\,, \dot \lambda}^\prime$, the fermions of $( \nG^{ \prime\prime \prime \prime \prime } \,, \eG^{ \prime\prime \prime \prime \prime } \,, \DG^{ \prime \prime \prime \prime } )$ only mixed with the heavy fermions of $( \nG^{ \prime\prime \prime  } \,, \eG^{  \prime \prime \prime } \,, \DG^{ \prime \prime \prime } )$, and they remain massless.
Clearly, the massless fermion spectrum was no longer anomaly-free.
The same inconsistency is also obtained when both $\repb{8_F}^{\dot \lambda}$ coupled to the $( \rep{1}\,, \repb{3}\,, -\frac{1}{3})_{ \mathbf{H}\,, \dot \lambda}$.
Therefore, this situation should be prohibited.

\para
The remaining massless fermions of the $\Gc_{\rm SM}$ are listed as follows
\beqn\label{eq:SM_fermions}
&& \Big[ ( \repb{3}\,, \rep{1}\,, +\frac{1}{3} )_{ \mathbf{F}}^\Lambda \oplus ( \rep{1}\,, \rep{1}\,, 0 )_{ \mathbf{F}}^\Lambda \Big]  \oplus \Big[ ( \rep{1}\,, \repb{2}\,, -\frac{1}{2} )_{ \mathbf{F}}^\Lambda \oplus ( \rep{1}\,, \rep{1}\,, 0 )_{ \mathbf{F}}^{\Lambda^{\prime}} \oplus ( \rep{1}\,, \rep{1}\,, 0 )_{ \mathbf{F}}^{\Lambda^{\prime\prime}}   \Big]  \subset \repb{8_F}^\Lambda \,, \quad \Lambda = ( \dot 1\,, \dot 2\,, 3 ) \,, \non
&& ( \rep{1}\,, \rep{1}\,, 0)_{ \mathbf{F}}^{{\rm IV} } \oplus ... \oplus ( \rep{1}\,, \rep{1}\,, 0)_{ \mathbf{F}}^{\dot {\rm IX} } \subset \repb{8_F}^{\Lambda } \,,  \non
&& ( \rep{1}\,, \rep{1}\,, 0)_{ \mathbf{F}}^{{\rm VI}^\prime } \oplus  ( \rep{1}\,, \rep{1}\,, 0)_{ \mathbf{F}}^{\dot {\rm IIX}^\prime } \oplus ( \rep{1}\,, \rep{1}\,, 0)_{ \mathbf{F}}^{\dot {\rm IX}^\prime } \subset \repb{8_F}^{\Lambda^\prime } \,,  \non
&& ( \rep{1}\,, \rep{1}\,, 0)_{ \mathbf{F}}^{{\rm V}^{ \prime \prime} } \oplus ... \oplus ( \rep{1}\,, \rep{1}\,, 0)_{ \mathbf{F}}^{\dot {\rm IX}^{ \prime \prime } } \subset \repb{8_F}^{\Lambda^{ \prime\prime} } \,,  \non
%%%%%%%%%%%%%%%%%%%%%%%%%%%%%%%%%%%%%%%%%%%%%
&& \Big[ \cancel{ ( \rep{3}\,, \rep{1}\,, -\frac{1}{3} )_{ \mathbf{F}} } \oplus ( \repb{3}\,, \rep{1}\,, -\frac{2}{3} )_{ \mathbf{F}} \Big]  \oplus \Big[ \bcancel{ ( \rep{1}\,, \rep{2}\,, +\frac{1}{2} )_{ \mathbf{F}} \oplus ( \rep{1}\,, \rep{1}\,, 0 )_{ \mathbf{F}} } \oplus \xcancel{ ( \rep{1}\,, \repb{2}\,, +\frac{1}{2} )_{ \mathbf{F}}^\prime }  \oplus ( \rep{1}\,, \rep{1}\,, +1 )_{ \mathbf{F}} \Big] \non
&\oplus& \Big[ ( \rep{3}\,, \rep{2}\,, +\frac{1}{6} )_{ \mathbf{F}} \oplus \xcancel{ ( \rep{3}\,, \rep{1}\,, -\frac{1}{3} )_{ \mathbf{F}}^\prime }\oplus \bcancel{ ( \rep{3}\,, \rep{1}\,, -\frac{1}{3} )_{ \mathbf{F}}^{\prime \prime } } \non
&\oplus&\cancel{ ( \rep{1}\,, \rep{2}\,, +\frac{1}{2} )_{ \mathbf{F}}^{ \prime \prime} \oplus ( \rep{1}\,, \rep{1}\,, 0 )_{ \mathbf{F}}^{ \prime } \oplus ( \rep{1}\,, \rep{1}\,, 0 )_{ \mathbf{F}}^{ \prime \prime } } \Big]  \subset \rep{28_F}\,, \non
%%%%%%%%%%%%%%%%%%%%%%%%%%%%%%%%%%%%%%%%%%%%%
&&\Big[ \xcancel{ ( \rep{1}\,, \repb{2}\,, +\frac{1}{2} )_{ \mathbf{F}}^{\prime\prime\prime }} \oplus  ( \rep{1}\,, \rep{1}\,, +1 )_{ \mathbf{F}}^\prime \oplus  \bcancel{ ( \rep{1}\,, \rep{1}\,, +1 )_{ \mathbf{F}}^{\prime\prime} } \Big] \oplus  \Big[ ( \repb{3}\,, \rep{1}\,, -\frac{2}{3} )_{ \mathbf{F}}^\prime \oplus \bcancel{ ( \rep{1}\,, \rep{1}\,, -1 )_{ \mathbf{F}}  } \Big] \non
&\oplus&  \Big[ ( \rep{3}\,, \rep{2}\,, +\frac{1}{6} )_{ \mathbf{F}}^\prime \oplus \xcancel{ ( \rep{3}\,, \rep{1}\,, -\frac{1}{3} )_{ \mathbf{F}}^{\prime\prime \prime} } \oplus  \bcancel{  ( \rep{3}\,, \repb{2}\,, +\frac{1}{6} )_{ \mathbf{F}}^{\prime\prime} \oplus ( \rep{3}\,, \rep{1}\,, +\frac{2}{3} )_{ \mathbf{F}} } \non 
&\oplus& \bcancel{ ( \rep{1}\,, \rep{2}\,, +\frac{1}{2} )_{ \mathbf{F}}^{\prime\prime\prime\prime} \oplus  ( \rep{1}\,, \rep{1}\,, 0 )_{ \mathbf{F}}^{\prime\prime \prime}  } \oplus \xcancel{ ( \rep{1}\,, \repb{2}\,, +\frac{1}{2})_{ \mathbf{F}}^{\prime\prime\prime\prime \prime} } \oplus ( \rep{1}\,, \rep{1}\,, +1)_{ \mathbf{F}}^{\prime\prime \prime} \Big] \non
&\oplus& \Big[ ( \rep{3}\,, \rep{2}\,, +\frac{1}{6} )_{ \mathbf{F}}^{\prime\prime \prime} \oplus \xcancel{ ( \rep{3}\,, \rep{1}\,, -\frac{1}{3})_{ \mathbf{F}}^{\prime\prime\prime \prime} }  \oplus \bcancel{ ( \rep{3}\,, \rep{1}\,, -\frac{1}{3})_{ \mathbf{F}}^{\prime\prime\prime \prime \prime }  } \non
&\oplus& \bcancel{ ( \repb{3}\,, \rep{2}\,, -\frac{1}{6})_{ \mathbf{F}}   \oplus ( \repb{3}\,, \rep{1}\,, -\frac{2}{3})_{ \mathbf{F}}^{\prime\prime} } \oplus ( \repb{3}\,, \rep{1}\,, -\frac{2}{3})_{ \mathbf{F}}^{\prime\prime\prime} \Big]   \subset \rep{56_F} \,.
\eeqn
The fermions that become massive at this stage are further crossed outs.
After this stage of symmetry breaking, there are three generations of the SM fermion irreps, together with twenty-three left-handed massless sterile neutrinos.
The third-generational SM fermions are from the rank-$2$ IAFFS of $\[ \repb{8_F}^\lambda \oplus \rep{28_F} \]$, while the first- and second-generational SM fermions are from the rank-$3$ IAFFS of $\[ \repb{8_F}^{\dot \lambda }\oplus \rep{56_F}\]$.
The flavor indices for the massive anti-fundamental fermions at this stage are chosen to be $\lambda={\rm VI}$ and $\dot \lambda=( \dot {\rm IIX}\,, \dot {\rm IX})$, respectively.

\begin{table}[htp]
\begin{center}
\begin{tabular}{c|cccc}
\hline\hline
$\repb{8_F}^\lambda$  & $( \repb{3}\,, \rep{1}\,, +\frac{1}{3} )_{\rep{F}}^\lambda$ & $( \rep{1}\,, \rep{1}\,,  0 )_{\rep{F}}^\lambda$  &  $$  &  $$    \\[1mm]
\hline
$\Tc_2^{ \prime \prime \prime \prime }$  & $-\frac{4}{3} t_2 \tilde c_1^{ \prime\prime \prime} + \frac{1}{3} \tilde c_2^{ \prime\prime \prime} ~ (-\frac{4}{3} t_2)$  &  $-4 t_2 \tilde c_1^{ \prime\prime \prime} ~ ( -4 t_2)$  &  $$ &  $$   \\[1mm]
\hline
$\repb{8_F}^\lambda$  &  $( \rep{1}\,, \repb{2}\,,  -\frac{1}{2 } )_{\rep{F}}^\lambda$  &  $( \rep{1}\,, \rep{1}\,,  0 )_{\rep{F}}^{\lambda^{ \prime } }$ &  $( \rep{1}\,, \rep{1}\,,  0 )_{\rep{F}}^{\lambda^{ \prime\prime} }$ & $$   \\[1mm]
\hline
$\Tc_2^{ \prime \prime \prime \prime }$  &  $-4 t_2 \tilde c_1^{ \prime\prime \prime} - \frac{1}{2} \tilde c_2^{ \prime\prime \prime}~ (-4 t_2)$ &  $-4 t_2 \tilde c_1^{ \prime\prime \prime} ~ (- 4 t_2)$ & $-4 t_2 \tilde c_1^{ \prime\prime \prime} ~ (- 4 t_2)$  & $$    \\[1mm]
\hline
$\rep{28_F}$  & $( \repb{3}\,, \rep{1}\,, - \frac{2}{3} )_{\rep{F}}$  &  $( \rep{1}\,, \rep{1}\,,  + 1 )_{\rep{F}}$  &  $( \rep{3}\,, \rep{2 }\,,  +\frac{1}{6} )_{\rep{F}}$ &  $$    \\[1mm]
 \hline
$\Tc_2^{ \prime \prime \prime \prime }$  &  $-\frac{4}{3} t_2 \tilde c_1^{ \prime\prime \prime} - \frac{2}{3} \tilde c_2^{ \prime\prime \prime}~ (-\frac{4}{3} t_2)$ & $4 t_2 \tilde c_1^{ \prime\prime \prime} + \tilde c_2^{ \prime\prime \prime}~ (4 t_2)$  &  $\frac{4}{3} t_2 \tilde c_1^{ \prime\prime \prime} + \frac{1}{6} \tilde c_2^{ \prime\prime \prime}~ (\frac{4}{3} t_2)$  & $$     \\[1mm]
\hline
$\repb{8_H}_{\,, \lambda}$ &  $( \rep{1}\,, \repb{2}\,,  - \frac{1}{2} )_{\rep{H}\,,\lambda}$  &   &   & $$   \\[1mm]
\hline
$\Tc_2^{ \prime \prime \prime \prime }$  &  $- \frac{1}{2} \tilde c_2^{ \prime\prime \prime}~ (0)$  &   &   & $$   \\[1mm]
\hline
$\rep{70_H}$  &  $( \rep{1}\,, \repb{2}\,,  + \frac{1}{2} )_{\rep{H}}^{ \prime\prime \prime}$  &  $( \rep{1}\,, \rep{2}\,,  - \frac{1}{2 } )_{\rep{H}}^{ \prime\prime \prime}$ &   & $$   \\[1mm]
\hline
$\Tc_2^{ \prime \prime \prime \prime }$  &   $ \frac{1}{2} \tilde c_2^{ \prime\prime \prime}~ ( 0)$ &  $-8 t_2 \tilde c_1^{ \prime\prime \prime} - \frac{1}{2} \tilde c_2^{ \prime\prime \prime}~ (- 8 t_2)$ &   & $$   \\[1mm]
\hline
$\rep{56_H}$ &  $( \rep{1}\,, \repb{2}\,,  +\frac{1}{2} )_{\rep{H} }$  &  $( \rep{1}\,, \rep{2}\,,  +\frac{1}{2} )_{\rep{H} }^{ \prime \prime}$ &  $( \rep{1}\,, \repb{2}\,,  +\frac{1}{2} )_{\rep{H} }^{ \prime \prime}$  &    \\[1mm]
\hline
$\Tc_2^{ \prime \prime \prime \prime }$  &  $t_2 \tilde c_1^{ \prime\prime \prime} + \frac{1 }{2} \tilde c_2^{ \prime\prime \prime} ~(t_2)$  & $t_2 \tilde c_1^{ \prime\prime \prime} + \frac{1 }{2} \tilde c_2^{ \prime\prime \prime}~(t_2)$  & $t_2 \tilde c_1^{ \prime\prime \prime} + \frac{1 }{2} \tilde c_2^{ \prime\prime \prime}~(t_2)$  & $$   \\[1mm]
\hline\hline
$ \repb{8_F}^{ \dot \lambda}$  & $( \repb{3}\,, \rep{1}\,, +\frac{1}{3} )_{\rep{F}}^{\dot \lambda}$ & $( \rep{1}\,, \rep{1}\,,  0 )_{\rep{F}}^{\dot \lambda}$ &  $$  &  $$    \\[1mm]
 \hline
$\Tc_3^{ \prime \prime \prime \prime}$  & $-\frac{4}{3} t_3 \tilde d_1^{ \prime\prime \prime} + \frac{1}{3} \tilde d_2^{ \prime\prime \prime}~ (-\frac{4}{3} t_3) $  &  $-4 t_3 \tilde d_1^{ \prime\prime \prime} ~ (-4 t_3)$  &  $$ &  $$    \\[1mm]
 \hline
 $ \repb{8_F}^{ \dot \lambda}$  &  $( \rep{1}\,, \repb{2 }\,,  -\frac{1}{2 } )_{\rep{F}}^{\dot \lambda}$  &  $( \rep{1}\,, \rep{1}\,,  0 )_{\rep{F}}^{\dot \lambda^{ \prime } }$ &  $( \rep{1}\,, \rep{1}\,,  0 )_{\rep{F}}^{\dot \lambda^{ \prime\prime } }$ & $$   \\[1mm]
 \hline
$\Tc_3^{ \prime \prime \prime \prime}$   &  $ -4 t_3 \tilde d_1^{ \prime\prime \prime} - \frac{1}{2} \tilde d_2^{ \prime\prime \prime}~ (-4 t_3)$ &  $-4 t_3 \tilde d_1^{ \prime\prime \prime} ~ (-4 t_3)$ & $-4 t_3 \tilde d_1^{ \prime\prime \prime} ~ (-4 t_3)$  & $$   \\[1mm]
 \hline
 $\rep{56_F}$ & $( \rep{1}\,,  \rep{1}\,, +1 )_{\rep{F}}^\prime$ & $( \repb{3 }\,, \rep{1}\,,  -\frac{2}{3 } )_{\rep{F}}^\prime$ &  $( \rep{3}\,,  \rep{2 }\,, +\frac{1 }{6 }  )_{\rep{F}}^\prime$  &  $$  \\[1mm]
 \hline
 $\Tc_3^{ \prime \prime \prime \prime}$  & $4 t_3 \tilde d_1^{ \prime\prime \prime} + \tilde d_2^{ \prime\prime \prime}~( 4 t_3 )$  &  $-\frac{4 }{3} t_3 \tilde d_1^{ \prime\prime \prime} -\frac{2}{3} \tilde d_2^{ \prime\prime \prime}~( -\frac{4 }{3} t_3 )$  &  $\frac{4 }{3} t_3 \tilde d_1^{ \prime\prime \prime} + \frac{1}{6} \tilde d_2^{ \prime\prime \prime} ~( \frac{4 }{3} t_3)$ &  $$   \\[1mm]
 \hline
 $\rep{56_F}$  & $( \rep{1}\,,  \rep{1}\,, +1 )_{\rep{F}}^{\prime \prime \prime}$ & $( \repb{3 }\,, \rep{1}\,,  -\frac{2}{3 } )_{\rep{F}}^{\prime \prime \prime}$ &  $( \rep{3}\,,  \rep{2 }\,, +\frac{1 }{6 }  )_{\rep{F}}^{\prime \prime\prime}$  &  $$   \\[1mm]
 \hline
$\Tc_3^{ \prime \prime \prime \prime}$  & $4 t_3 \tilde d_1^{ \prime\prime \prime} + \tilde d_2^{ \prime\prime \prime}~ (4 t_3)$  &  $-\frac{4 }{3 } t_3 \tilde d_1^{ \prime\prime \prime} -\frac{2}{3} \tilde d_2^{ \prime\prime \prime} ~ (-\frac{4}{3} t_3)$  &  $\frac{4 }{3} t_3 \tilde d_1^{ \prime\prime \prime} + \frac{ 1}{ 6} \tilde d_2^{ \prime\prime \prime} ~ (\frac{4}{3} t_3)$ &  $$   \\[1mm]
\hline
$\repb{28_H}_{\,, \dot \lambda}$ &  $( \rep{1}\,, \repb{2}\,,  - \frac{1}{2} )_{\rep{H}\,,\dot \lambda}^\prime$  &  $( \rep{1}\,, \rep{2}\,,  - \frac{1}{2} )_{\rep{H}\,,\dot \lambda}$ & $( \rep{1}\,, \repb{2}\,,  - \frac{1}{2} )_{\rep{H}\,,\dot \lambda}$  & $$   \\[1mm]
\hline
$\Tc_3^{ \prime \prime \prime \prime }$  &  $ - \frac{1}{2} \tilde d_2^{ \prime\prime \prime}~(0)$  &  $ - \frac{1}{2} \tilde d_2^{ \prime\prime \prime}~(0)$   & $ - \frac{1}{2} \tilde d_2^{ \prime\prime \prime}~(0)$  & $$   \\[1mm]
\hline
$\rep{56_H}$ &  $( \rep{1}\,, \repb{2}\,,  +\frac{1}{2} )_{\rep{H} }$  &  $( \rep{1}\,, \rep{2}\,,  +\frac{1}{2} )_{\rep{H} }^{ \prime \prime}$ &  $( \rep{1}\,, \repb{2}\,,  +\frac{1}{2} )_{\rep{H} }^{ \prime \prime}$  &    \\[1mm]
\hline
$\Tc_3^{ \prime \prime \prime \prime }$  &  $2t_3\tilde d_1^{ \prime\prime \prime} + \frac{1}{2} \tilde d_2^{ \prime\prime \prime} ~(2 t_3)$  & $2t_3\tilde d_1^{ \prime\prime \prime} + \frac{1}{2} \tilde d_2^{ \prime\prime \prime}~(2 t_3)$  & $2t_3\tilde d_1^{ \prime\prime \prime} + \frac{1}{2} \tilde d_2^{ \prime\prime \prime}~(2 t_3)$  & $$   \\[1mm]
 \hline\hline
\end{tabular}
\end{center}
\caption{The $\widetilde{ {\rm U}}(1)_{T_2^{\prime \prime \prime \prime } } \otimes \widetilde{ {\rm U}}(1)_{T_3^{\prime \prime \prime \prime} }$ charges for massless fermions and possible symmetry-breaking Higgs components in the SM.
The $\widetilde{ {\rm U}}(1)_{B-L } $ charges are displayed in the parentheses.
}
\label{tab:GSM_Tcharges}
\end{table}%

\para
Thus, we define the $\widetilde{ {\rm U}}(1)_{T_2^{\prime \prime \prime \prime } } \otimes \widetilde{ {\rm U}}(1)_{T_3^{ \prime \prime \prime \prime } }$ charges after this symmetry-breaking stage as
\beqn\label{eq:GSM_U1charges}
&& \Tc_2^{\prime \prime \prime \prime} \equiv \tilde c_1^{\prime \prime \prime } \Tc_2^{ \prime \prime \prime } + \tilde c_2^{ \prime \prime \prime} \Yc \,,\quad \Tc_3^{ \prime \prime \prime \prime }\equiv \tilde d_1^{ \prime \prime \prime } \Tc_3^{\prime \prime \prime } + \tilde d_2^{ \prime \prime \prime} \Yc \,,
\eeqn
and they are explicitly listed in Tab.~\ref{tab:GSM_Tcharges} for massless fermions and possible symmetry-breaking Higgs components.
Accordingly, we find the following global anomalies of
\beqn\label{eq:GSM_U1anoms}
&& \[ {\rm grav} \]^2 \cdot \widetilde{ {\rm U} }(1)_{ T_2^{\prime \prime \prime \prime} } = -40 t_2 \tilde c_1^{\prime \prime \prime} \,,\quad \[ \widetilde{ {\rm U} }(1)_{ T_2^{ \prime\prime \prime \prime} } \]^3 = -640 ( t_2 \tilde c_1^{ \prime \prime \prime } )^3 \,,\non
&& \[ {\rm grav} \]^2 \cdot \widetilde{ {\rm U} }(1)_{ T_3^{\prime \prime \prime \prime} } = -64 t_3 \tilde d_1^{ \prime \prime\prime } \,,\quad \[ \widetilde{ {\rm U} }(1)_{ T_3^{\prime \prime \prime \prime } } \]^3 = -1024 ( t_3 \tilde d_1^{ \prime \prime \prime } )^3 \,.
\eeqn
By matching with the global anomalies in Eq.~\eqref{eq:G331_U1anoms}, we find that
\beqn
&& \tilde c_1^{\prime \prime \prime} = \tilde d_1^{ \prime \prime \prime } = 1\,.
\eeqn
By identifying both $\widetilde{ {\rm U}}(1)_{T_2^{\prime \prime \prime \prime } } \otimes \widetilde{ {\rm U}}(1)_{T_3^{ \prime \prime \prime \prime } }$ to be the global $\widetilde{ {\rm U}}(1)_{B-L }$ symmetry in the SM, we also arrive at
\beqn
&& \tilde c_2^{ \prime\prime \prime} =  \tilde d_2^{ \prime\prime \prime}  =0 \,.
\eeqn
With a well-defined global $\widetilde{ {\rm U}}(1)_{B-L }$ symmetry after the $\Gc_{331}$ symmetry-breaking stage, one finds that all possible symmetry-breaking components from the $\rep{56_H}$ are charged under the global $\widetilde{ {\rm U}}(1)_{B-L }$.
According to the generalized neutrality condition, the Higgs field of $\rep{56_H}$ cannot develop any VEV through the symmetry-breaking pattern described in Eq.~\eqref{eq:Pattern}.
Thus, the Higgs field of $\rep{56_H}$ and the possible Yukawa coupling of $\rep{28_F} \rep{56_F} \rep{56_H} + H.c.$ should be absent.

%%%%%%%%%%%%%%%%%%%%%%%%%%%%
\subsection{The EWSB stage}
%%%%%%%%%%%%%%%%%%%%%%%%%%%%

\para
As we have previously mentioned, the EWSB is expected to be achieved by two components of $( \rep{1}\,, \repb{2}\,,  + \frac{1}{2} )_{\rep{H}}^{ \prime\prime \prime} \oplus ( \rep{1}\,, \rep{2}\,,  - \frac{1}{2} )_{\rep{H}}^{ \prime\prime \prime} \subset \rep{70_H}$ purely from the decomposition of the Higgs irreps in Eq.~\eqref{eq:SU8_Higgs_Br05}.
From their distinctive $\Tc_2^{ \prime\prime\prime \prime}$ charges in Tab.~\ref{tab:GSM_Tcharges}, only one of them can actually develop the EWSB VEV with the $\widetilde{ {\rm U}}(1)_{B-L}$-neutrality condition.
If one assumes that only the component of $( \rep{1}\,, \repb{2}\,,  + \frac{1}{2} )_{\rep{H}}^{ \prime\prime \prime}$ develop the EWSB VEV, one has
\beqn
&& \tilde c_2^{ \prime\prime \prime}=0\,.
\eeqn
This is consistent with the requirement of a well-defined global $\widetilde{ {\rm U}}(1)_{B-L }$ symmetry for the third-generational SM fermions.
The corresponding Yukawa coupling reads
\beqn\label{eq:top_Yukawa}
&&Y_\Tc \rep{28_F} \rep{28_F} \rep{70_H}  + H.c. \non
&\supset& Y_\Tc ( \rep{6}\,, \rep{1}\,, -\frac{1}{2} )_{ \mathbf{F}} \otimes  ( \rep{4}\,, \rep{4}\,, 0 )_{ \mathbf{F}}  \otimes ( \rep{4}\,, \repb{4}\,, +\frac{1}{2} )_{ \mathbf{H}} + H.c.  \non
&\supset& Y_\Tc ( \repb{3}\,, \rep{1}\,, -\frac{2}{3} )_{ \mathbf{F}} \otimes  ( \rep{3}\,, \rep{4}\,, -\frac{1}{12} )_{ \mathbf{F}}  \otimes ( \rep{1}\,, \repb{4}\,, +\frac{3}{4} )_{ \mathbf{H}}^\prime  + H.c.\non
&\supset& Y_\Tc ( \repb{3}\,, \rep{1}\,, -\frac{2}{3} )_{ \mathbf{F}} \otimes  ( \rep{3}\,, \rep{3}\,, 0 )_{ \mathbf{F}}  \otimes ( \rep{1}\,, \repb{3}\,, +\frac{2}{3} )_{ \mathbf{H}}^{ \prime \prime \prime }  + H.c.\non
&\supset& Y_\Tc ( \repb{3}\,, \rep{1}\,, -\frac{2}{3} )_{ \mathbf{F}} \otimes  ( \rep{3}\,, \rep{2}\,, +\frac{1}{6} )_{ \mathbf{F}}  \otimes ( \rep{1}\,, \repb{2}\,, +\frac{1}{2} )_{ \mathbf{H}}^{ \prime\prime \prime } + H.c. \non
&\Rightarrow&\frac{1}{\sqrt{2}} Y_\Tc  t_L {t_R}^c v_t + H.c. \,,
\eeqn
which precisely gives rise to the top quark mass with the natural Yukawa coupling of $Y_\Tc\sim \Oc(1)$ at the tree level.
The same result was also obtained in the two-generational ${\rm SU}(7)$ toy model~\cite{Chen:2022xge}.
Based on this fact, we recover the observation that the $\rep{28_F}$ only contains the $\rep{10_F}$ for the third-generational SM fermions, while the $\rep{56_F}$ only contains the $\rep{10_F}$'s for the first- and second-generational SM fermions~\cite{Barr:2008pn}.

\para
Alternatively, let us consider the hypothetical situation where only the component of $( \rep{1}\,, \rep{2}\,,  - \frac{1}{2} )_{\rep{H}}^{ \prime\prime \prime}$ developed the EWSB VEV, one has
\beqn
&& \tilde c_2^{ \prime\prime \prime}= -16 t_2\,.
\eeqn
The corresponding Yukawa coupling of
\beqn\label{eq:neutrino_Yukawa}
%%%%%%%%%%%%%%%%%%%%%%%%%%%%%%%%%%%%%%%%%%%%%
&& Y_\Tc \rep{28_F} \rep{28_F} \rep{70_H}  + H.c. \non
&\supset& Y_\Tc ( \rep{1}\,, \rep{6}\,, +\frac{1}{2} )_{ \mathbf{F}} \otimes  ( \rep{4}\,, \rep{4}\,, 0 )_{ \mathbf{F}}  \otimes ( \repb{4}\,, \rep{4}\,, -\frac{1}{2} )_{ \mathbf{H}}   + H.c.  \non
&\supset&  Y_\Tc( \rep{1}\,, \rep{6}\,, +\frac{1}{2} )_{ \mathbf{F}} \otimes  ( \rep{1}\,, \rep{4}\,, +\frac{1}{4} )_{ \mathbf{F}}  \otimes ( \rep{1}\,, \rep{4}\,, -\frac{3}{4} )_{ \mathbf{H}}^\prime  + H.c.  \non
&\supset& Y_\Tc \Big[ ( \rep{1}\,, \rep{3}\,, +\frac{1}{3} )_{ \mathbf{F}} \otimes  ( \rep{1}\,, \rep{3}\,, +\frac{1}{3} )_{ \mathbf{F}} \oplus ( \rep{1}\,, \repb{3}\,, +\frac{2}{3} )_{ \mathbf{F}} \otimes  ( \rep{1}\,, \rep{1}\,, 0 )_{ \mathbf{F}}  \Big]  \otimes ( \rep{1}\,, \rep{3}\,, -\frac{2}{3} )_{ \mathbf{H}}^{ \prime\prime\prime }  + H.c.  \non
&\supset& Y_\Tc \Big[ ( \rep{1}\,, \rep{2}\,, +\frac{1}{2} )_{ \mathbf{F}} \otimes  ( \rep{1}\,, \rep{1}\,, 0 )_{ \mathbf{F}}^\prime \oplus ( \rep{1}\,, \rep{2}\,, +\frac{1}{2} )_{ \mathbf{F}}^{\prime\prime } \otimes  ( \rep{1}\,, \rep{1}\,, 0 )_{ \mathbf{F}}  \non
&\oplus & ( \rep{1}\,, \repb{2}\,, +\frac{1}{2} )_{ \mathbf{F}}^\prime \otimes  ( \rep{1}\,, \rep{1}\,, 0 )_{ \mathbf{F}}^{\prime\prime }  \Big] \otimes ( \rep{1}\,, \rep{2}\,, -\frac{1}{2} )_{ \mathbf{H}}^{\prime\prime \prime  } + H.c. \non
&\Rightarrow& \frac{1}{\sqrt{2}} Y_\Tc  \Big(-  \nG_R^c \check \nG_R^{ \prime\, c} -  \nG_R^{\prime\prime \, c} \check \nG_R^{ c} +  \nG_R^{\prime \, c} \check \nG_R^{\prime\prime\, c}   \Big) u_{ \rep{2}}^{\prime\prime \prime  } + H.c.  \,,
\eeqn
generated the mass mixing terms between the right-handed heavy neutrinos.
Two consequences can be found with the choice of $\tilde c_2^{ \prime\prime \prime}= -16 t_2$:
\begin{enumerate}

\item the $\widetilde{{\rm U}}(1)_{ T_2^{ \prime\prime\prime \prime}}$ charges of the third-generational SM fermions became
\beqn
&& \Tc_2^{ \prime\prime\prime \prime} ( ( \repb{3}\,, \rep{1}\,, +\frac{1}{3} )_{ \rep{F}}^\lambda ) = -\frac{20}{3} t_2 \,,\quad  \Tc_2^{ \prime\prime\prime \prime} ( ( \rep{1}\,, \repb{2}\,,  -\frac{1}{2 } )_{ \rep{F}}^\lambda ) = 4 t_2 \,,\non
&& \Tc_2^{ \prime\prime\prime \prime} ( ( \repb{3}\,, \rep{1}\,,   -\frac{2}{3 } )_{ \rep{F}} ) =\frac{ 28}{3} t_2 \,,\quad \Tc_2^{ \prime\prime\prime \prime} ( ( \rep{1}\,, \rep{1}\,,   +1 )_{ \rep{F}} ) = -12 t_2 \,,\non
&&\Tc_2^{ \prime\prime\prime \prime} ( ( \rep{3}\,, \rep{2}\,,   +\frac{1}{6 } )_{ \rep{F}} ) = - \frac{4 }{3} t_2 \,,
\eeqn
which could not lead to a well-defined $\widetilde{{\rm U}}(1)_{ B-L}$ symmetry.

\item the Higgs field of $( \rep{1}\,, \repb{2}\,,  + \frac{1}{2} )_{\rep{H}}^{ \prime\prime \prime}$ could no longer develop the EWSB VEV, and hence the top quark was massless.

\end{enumerate}
Therefore, this hypothetical situation is impossible.
Consequently, the Higgs component of $( \repb{4} \,, \rep{4} \,, -\frac{1}{2} )_{\mathbf{H}} \subset \rep{70_H}$ in Eq.~\eqref{eq:SU8_Higgs_Br05} should be massive at the GUT scale.

%###################################################################
\section{Generalization to the ${\rm SU}(N)$ theories}
\label{section:BminusL_SUN}
%###################################################################

\para
In this section, we generalize the previous results to an arbitrary ${\rm SU}(N)$ theory.
We will show that a global $\widetilde{{\rm U}}(1)_{B-L}$ symmetry can always be defined as long as the ${\rm SU}(N)$ is spontaneously broken into the $\Gc_{\rm SM}$.
We consider the following symmetry-breaking pattern of
\beqn\label{eq:Pattern_SUN}
&& {\rm SU}(N) \, \[  \widetilde{ {\rm U}}(1)_T \] \xrightarrow{ v_U }  {\rm SU}(N_s )_s \otimes  {\rm SU}(N_W)_W  \otimes  {\rm U}(1)_{X_0} \, \[  \widetilde{ {\rm U}}(1)_{T^\prime} \]   \,,  \quad N_s \simeq N_W = \[  \frac{N}{2} \]  \non
&\rightarrow&{\rm SU}(N_s-1 )_s \otimes  {\rm SU}(N_W)_W  \otimes  {\rm U}(1)_{X} \, \[  \widetilde{ {\rm U}}(1)_{T^{\prime \prime} } \] ... \non
&\rightarrow& {\rm SU}(3 )_c \otimes  {\rm SU}(N_W)_W  \otimes  {\rm U}(1)_{X^\prime }  \, \[  \widetilde{ {\rm U}}(1)_{T^{\prime \prime \prime } } \] ... \non
&\rightarrow& {\rm SU}(3 )_c \otimes  {\rm SU}(N_W-1 )_W  \otimes  {\rm U}(1)_{X^{\prime \prime} }  \, \[  \widetilde{ {\rm U}}(1)_{T^{\prime \prime \prime \prime } } \]  ... \non
&\rightarrow&  {\rm SU}(3 )_c \otimes  {\rm SU}(2 )_W  \otimes  {\rm U}(1)_{Y }  \, \[  \widetilde{ {\rm U}}(1)_{B-L } \] \,,
\eeqn
where the generalized global $\widetilde{{\rm U}}(1)_{T}$ symmetries at each stages are labeled in the square brackets.
The pattern of the ${\rm SU}(N_s \geq 3)$ and ${\rm SU}(N_W \geq 2)$ breaking to the ${\rm SU}(3)_c$ and the ${\rm SU}(2)_W$ symmetries is outlined purely for defining the global $\widetilde{{\rm U}}(1)_{B-L}$ symmetry, while it does not represent the necessary sequence when other physical considerations are taken into account.
We decompose the ${\rm SU}(N)$ fundamental irrep as follows
\beqn
&& \rep{N} = ( \rep{N}_s \,, \rep{1}\,, - \frac{1}{ N_s } ) \oplus ( \rep{1}\,, \rep{N}_W \,,  + \frac{1}{ N_W } ) \,,
\eeqn
for the zeroth stage.

\para
Before the detailed analyses, some general results from the ${\rm SU}(8)$  theory are useful.
Through the 't Hooft anomaly matching condition at different symmetry-breaking stages of the ${\rm SU}(8)$ theory, we have found that the generalized global $\widetilde{ {\rm U}}(1)_T$ charges always read
\beqn
&& \Tc_2^\prime = \Tc_2 + \tilde c_2 \Xc_0 \,, \quad  \Tc_3^\prime = \Tc_3 + \tilde d_2 \Xc_0 \,,
\eeqn
from Eq.~\eqref{eq:G441_match} and etc.
This is obvious, since the contributions to the $\[ {\rm grav} \]^2 \cdot \widetilde{ {\rm U}}(1)_T$ and the $\[ \widetilde{ {\rm U}}(1)_T \]^3$ anomalies from the gauged ${\rm U}(1)_{X}$ charges must be vanishing.
The matching of these global anomalies must lead to $\tilde c_1 = \tilde d_1 =1$ in Eq.~\eqref{eq:G441_U1anoms} and other sequential stages.
Though our main conjecture of a flavor-unified theory involves several distinctive rank-$k$ IAFFSs in general, it is sufficient to discuss the global $\widetilde{ {\rm U}}(1)_{B-L}$ symmetry in the rank-$2$ IAFFS of an ${\rm SU}(N)$ theory.
There are two arguments for this simplification.
\begin{itemize}

\item The Higgs fields from the rank-$2$ IAFFS will be responsible for each symmetry-breaking stage after the zeroth stage.
Therefore, the generalized $\widetilde{ {\rm U}}(1)_T$ symmetries can be defined accordingly.
Meanwhile, the Higgs fields from the rank-$3$ IAFFS of the ${\rm SU}(8)$ cannot be responsible for all symmetry-breaking stages.

\item As long as the global $B-L$ symmetry can be defined in the rank-$2$ IAFFS (presumably for the third-generational SM fermions), the same $\widetilde{ {\rm U}}(1)_{B-L}$ charges in the other rank-$k$ IAFFS must be defined in accordance with those defined in the rank-$2$ IAFFS. 
In other words, we require a universal global $\widetilde{ {\rm U}}(1)_{B-L}$ charges for three-generational SM fermions.

\end{itemize}
More specifically, it is sufficient to consider the global $\widetilde{ {\rm U}}(1)_T$ symmetry for the anti-fundamental fermions and Higgs fields in the rank-$2$ IAFFS.

\para
We consider the global $B-L$ symmetry in the ${\rm SU}(2m)$ theory, where the $\widetilde{ {\rm U}}(1)_T$ charges for the anti-fundamental fermions and Higgs fields are
\beqn
&& \Tc( \repb{N_F}) = - (2m -2 )t  \,,\quad  \Tc(\repb{N_H} ) = 2t \,,
\eeqn
The zeroth symmetry-breaking stage reads
\beqn
&& {\rm SU}(2m ) \xrightarrow{ v_U }  {\rm SU}(m )_s \otimes  {\rm SU}(m )_W  \otimes  {\rm U}(1)_{X_0}  \,.
\eeqn
The decomposition of the anti-fundamental fermions and Higgs fields are
\beqn
&& \repb{N}_{ \mathbf{F}/ \mathbf{H} } = ( \repb{m}\,, \rep{1}\,, + \frac{1 }{m } )_{ \mathbf{F}/ \mathbf{H} } \oplus ( \rep{1 }\,, \repb{m }\,, - \frac{1 }{m } )_{ \mathbf{F}/\mathbf{H}} \,.
\eeqn
Their global $\widetilde{ {\rm U}}(1)_{T^\prime}$ charges are given by
\beqn\label{eq:Tp_SU2m}
&& \Tc^\prime ( ( \repb{m}\,, \rep{1}\,, + \frac{1 }{m } )_{ \mathbf{F} } )=  - (2m -2 ) t + \frac{1 }{m }  \tilde c_2\,, \quad  \Tc^\prime ( ( \rep{1}\,, \repb{m} \,, - \frac{1 }{m } )_{ \mathbf{F} } )=  - (2m -2 ) t  -  \frac{1 }{m }  \tilde c_2 \,,\non
%%%%%%%%%%%%%%%%%%%%%%%%%%%%%%%%%%%%%%%%%%%%%
&& \Tc^\prime ( ( \repb{m}\,, \rep{1}\,, + \frac{1 }{m } )_{ \mathbf{H} } )=  2t +  \frac{1 }{m }  \tilde c_2 \,, \quad  \Tc^\prime ( ( \rep{1}\,, \repb{m} \,, - \frac{1 }{m } )_{ \mathbf{H} } )=  2t -  \frac{1 }{m }  \tilde c_2 \,.
\eeqn
According to Eq.~\eqref{eq:Pattern_SUN}, the first stage of ${\rm SU}(m )_s \otimes {\rm U}(1 )_{X_0} \to {\rm SU}(m-1 )_s \otimes {\rm U}(1 )_X$ can only be achieved by the $( \repb{m}\,, \rep{1}\,, + \frac{1 }{m } )_{ \mathbf{H} }$, since $( \repb{m}\,, \rep{1}\,, + \frac{1 }{m } )_{ \mathbf{H} } = ( \repb{m-1}\,, \rep{1}\,, + \frac{1 }{m-1 } )_{ \mathbf{H} } \oplus ( \rep{1}\,, \rep{1}\,, 0 )_{ \mathbf{H} }$.
The $\widetilde{ {\rm U}}(1)_{T^\prime}$-neutrality condition leads to the solution of $\tilde c_2 = -2 m t$.
In turn, the global $\widetilde{ {\rm U}}(1)_{T^\prime}$ charges in Eq.~\eqref{eq:Tp_SU2m} become
\beqn
&& \Tc^\prime ( ( \repb{m}\,, \rep{1}\,, + \frac{1 }{m } )_{ \mathbf{F} } )=  - 2m  t  \,, \quad  \Tc^\prime ( ( \rep{1}\,, \repb{m} \,, - \frac{1 }{m } )_{ \mathbf{F} } )=  - 2m  t  + 4t \,,\non
%%%%%%%%%%%%%%%%%%%%%%%%%%%%%%%%%%%%%%%%%%%%%
&& \Tc^\prime ( ( \repb{m}\,, \rep{1}\,, + \frac{1 }{m } )_{ \mathbf{H} } )=  0 \,, \quad  \Tc^\prime ( ( \rep{1}\,, \repb{m} \,, - \frac{1 }{m } )_{ \mathbf{H} } )=  4 t   \,.
\eeqn
The sequential symmetry-breaking stages of ${\rm SU}(m-1 )_s \otimes {\rm U}(1 )_X \to ... \to {\rm SU}(3 )_c \otimes {\rm U}(1 )_{X^\prime }$ can only be achieved by the $( \repb{m-1}\,, \rep{1}\,, + \frac{1 }{m-1 } )_{ \mathbf{H} }\,,...\,,( \repb{4}\,, \rep{1}\,, + \frac{1 }{4 } )_{ \mathbf{H} }$.
Thus, one must have
\beqn
&& \Tc^{ \prime \prime \prime }=...  = \Tc^{ \prime \prime } + 0 \cdot \Xc = \Tc^\prime + 0 \cdot \Xc_0 \,, 
\eeqn
which leads to 
\beqn
&& \Tc^{\prime \prime \prime } ( ( \repb{3}\,, \rep{1}\,, + \frac{1 }{3 } )_{ \mathbf{F} } )=  - 2m  t \,,\quad \Tc^{\prime \prime \prime } ( ( \rep{1}\,, \rep{1}\,, 0 )_{  \mathbf{\check F} } )=  - 2m  t   \,, \quad  \Tc^{\prime \prime \prime } ( ( \rep{1}\,, \repb{m} \,, - \frac{1 }{m } )_{ \mathbf{F} } )=  (- 2m  + 4) t \,,\non
%%%%%%%%%%%%%%%%%%%%%%%%%%%%%%%%%%%%%%%%%%%%%
&& \Tc^{\prime \prime \prime } ( ( \repb{3}\,, \rep{1}\,, + \frac{1 }{3 } )_{ \mathbf{H} } )=  0 \,, \quad  \Tc^{\prime \prime \prime } ( ( \rep{1}\,, \repb{m} \,, - \frac{1 }{m } )_{ \mathbf{H} } )=  4 t   \,.
\eeqn
Here, the $( \rep{1}\,, \rep{1}\,, 0 )_{  \mathbf{\check F} }$ represent all sterile neutrinos by decomposing the $( \repb{m}\,, \rep{1}\,, + \frac{1 }{m } )_{ \mathbf{F} }$ to the $( \repb{3}\,, \rep{1}\,, + \frac{1 }{3 } )_{ \mathbf{F} }$.

\para
The followup symmetry-breaking stage of ${\rm SU}(m )_W \otimes {\rm U}(1 )_{X^\prime } \to {\rm SU}(m-1 )_W \otimes {\rm U}(1 )_{X^{\prime \prime} }$ is achieved by the $( \rep{1}\,, \repb{m} \,, - \frac{1 }{m } )_{ \mathbf{H} }$.
Their global $\widetilde{ {\rm U}}(1)_{T^{ \prime\prime \prime \prime} }$ charges are given by
\beqn\label{eq:Tpppp_SU2m}
&& \Tc^{\prime \prime \prime \prime } ( ( \repb{3}\,, \rep{1}\,, + \frac{1 }{3 } )_{ \mathbf{F} } )=  - 2m  t + \frac{1}{3} \tilde c_2^{ \prime\prime \prime}  \,,\quad \Tc^{\prime \prime \prime \prime } ( ( \rep{1}\,, \rep{1}\,, 0 )_{  \mathbf{\check F} } )=  - 2m  t   \,, \quad  \Tc^{\prime \prime \prime \prime } ( ( \rep{1}\,, \repb{m} \,, - \frac{1 }{m } )_{ \mathbf{F} } )=  (- 2m  + 4) t - \frac{1}{m} \tilde c_2^{ \prime\prime \prime} \,,\non
%%%%%%%%%%%%%%%%%%%%%%%%%%%%%%%%%%%%%%%%%%%%%
&& \Tc^{\prime \prime \prime \prime } ( ( \repb{3}\,, \rep{1}\,, + \frac{1 }{3 } )_{ \mathbf{H} } )=  \frac{1}{3} \tilde c_2^{ \prime\prime \prime} \,, \quad  \Tc^{\prime \prime \prime \prime } ( ( \rep{1}\,, \repb{m} \,, - \frac{1 }{m } )_{ \mathbf{H} } )=  4 t - \frac{1}{m} \tilde c_2^{ \prime\prime \prime} \,.
\eeqn
The $\widetilde{ {\rm U}}(1)_{T^{ \prime\prime \prime \prime} }$-neutrality condition of the $( \rep{1}\,, \repb{m} \,, - \frac{1 }{m } )_{ \mathbf{H} }$ leads to the solution of $\tilde c_2^{\prime \prime \prime} = 4mt$.
In turn, the global $\widetilde{ {\rm U}}(1)_{T^{ \prime\prime \prime \prime} }$ charges in Eq.~\eqref{eq:Tpppp_SU2m} become
\beqn
&& \Tc^{\prime \prime \prime \prime } ( ( \repb{3}\,, \rep{1}\,, + \frac{1 }{3 } )_{ \mathbf{F} } )=  - \frac{2}{3} m  t  \,,\quad \Tc^{\prime \prime \prime \prime } ( ( \rep{1}\,, \rep{1}\,, 0 )_{  \mathbf{\check F} } )=  - 2m  t   \,, \quad  \Tc^{\prime \prime \prime \prime } ( ( \rep{1}\,, \repb{m} \,, - \frac{1 }{m } )_{ \mathbf{F} } )=  - 2m  t \,,\non
%%%%%%%%%%%%%%%%%%%%%%%%%%%%%%%%%%%%%%%%%%%%%
&& \Tc^{\prime \prime \prime \prime } ( ( \repb{3}\,, \rep{1}\,, + \frac{1 }{3 } )_{ \mathbf{H} } )=  \frac{4}{3} m t\,, \quad  \Tc^{\prime \prime \prime \prime } ( ( \rep{1}\,, \repb{m} \,, - \frac{1 }{m } )_{ \mathbf{H} } )= 0 \,.
\eeqn
With the same argument that the sequential symmetry-breaking stages of ${\rm SU}(m-1 )_W \otimes {\rm U}(1 )_{X^{ \prime\prime}} \to ... \to {\rm SU}(2 )_W \otimes {\rm U}(1 )_{Y }$ can only be achieved by the $( \rep{1}\,, \repb{m-1} \,, - \frac{1 }{m-1 } )_{ \mathbf{H} }\,, ...\,, ( \rep{1}\,, \repb{2} \,, - \frac{1 }{2 } )_{ \mathbf{H} }$, one must have
\beqn
&&  \Bc - \Lc =  ... =\Tc^{\prime \prime \prime \prime } \,,
\eeqn
and this leads to 
\beqn
&& (  \Bc - \Lc ) ( ( \repb{3}\,, \rep{1}\,, + \frac{1 }{3 } )_{ \mathbf{F} } )=  - \frac{2}{3} m  t  \,,\quad (  \Bc - \Lc ) ( ( \rep{1}\,, \rep{1}\,, 0 )_{  \mathbf{\check F} } )=  - 2m  t   \,, \quad  (  \Bc - \Lc ) ( ( \rep{1}\,, \repb{2} \,, - \frac{1 }{2 } )_{ \mathbf{F} } )=  - 2m  t \,,\non
%%%%%%%%%%%%%%%%%%%%%%%%%%%%%%%%%%%%%%%%%%%%%
&& (  \Bc - \Lc ) ( ( \repb{3}\,, \rep{1}\,, + \frac{1 }{3 } )_{ \mathbf{H} } )=  \frac{4}{3} m t\,, \quad  (  \Bc - \Lc ) ( ( \rep{1}\,, \repb{2} \,, - \frac{1 }{2 } )_{ \mathbf{H} } )= 0 \,.
\eeqn
By choosing $t= \frac{1 }{2m}$, one has $(  \Bc - \Lc ) ( ( \repb{3}\,, \rep{1}\,, + \frac{1 }{3 } )_{ \mathbf{F} } )= - \frac{1}{3}$ for the right-handed down-type anti-quarks, and $(  \Bc - \Lc ) ( ( \rep{1}\,, \rep{1}\,, 0 )_{  \mathbf{\check F} } ) = (  \Bc - \Lc ) ( ( \rep{1}\,, \repb{2} \,, - \frac{1 }{2 } )_{ \mathbf{F} } )= -1$ for both the left-handed sterile neutrinos and the left-handed EW lepton doublets.
The same procedure of deriving the global $B-L$ symmetry can be applied to the ${\rm SU}(2m+1)$ theory as well, with the similar symmetry-breaking pattern as in Eq.~\eqref{eq:Pattern_SUN}.

%###################################################################
\section{Conclusion}
\label{section:conclusion}
%###################################################################

%%%%%%%%%%%%%%%%%%%%%%%%%%%%
\subsection{Summary of results}
%%%%%%%%%%%%%%%%%%%%%%%%%%%%

\para
In this work, we study the origin of the global $\widetilde{ {\rm U}}(1)_{B-L}$ symmetry in the flavor-unified ${\rm SU}(N)$ theories, where three-generational SM fermions transform differently.
Specifically, we focus on the minimal ${\rm SU}(8)$ theory. 
Based on our analyses of the $\widetilde{ {\rm U}}(1)_{B-L}$ origin in this framework, we find that the non-anomalous global $\widetilde{ {\rm U}}(1)_{B-L}$ in the SM originate from the generalized global $\widetilde{ {\rm U}}(1)_{T}$ symmetries in the GUT beyond the minimal ${\rm SU}(5)$ theory. 
With the generalized $\widetilde{ {\rm U}}(1)_{T}$-neutrality conditions to Higgs components that can develop the VEVs for different stages of symmetry breaking, the seemingly possible Higgs field of $\rep{56_H}$ in Eq.~\eqref{eq:SU8_Yukawa} cannot develop any VEV from Tabs.~\ref{tab:G331_Tcharges} and \ref{tab:GSM_Tcharges}.
Thus, the minimal set of ${\rm SU}(8)$ Higgs fields are reduced to
\beqn\label{eq:SU8_Higgs_real}
 \{ H \}_{ {\rm SU}(8)}^{n_g=3} &=&  \repb{8_H}_{\,, \lambda}  \oplus \repb{28_H}_{\,, \dot \lambda }  \oplus \rep{70_H} \oplus \underline{ \rep{63_H} }  \,, \non
 && \lambda = ( 3\,, {\rm IV}\,, {\rm V}\,, {\rm VI}) \,, ~  \dot \lambda = (\dot 1\,, \dot 2\,, \dot {\rm VII}\,, \dot {\rm IIX}\,, \dot {\rm IX} ) \,.
\eeqn
According to the $\widetilde{ {\rm U}}(1)_{T_2} \otimes \widetilde{ {\rm U}}(1)_{T_3}$ charges assigned in Tab.~\ref{tab:SU8_Tcharges}, the renormalizable ${\rm SU}(8)$ Higgs potential cannot include terms such as $(\rep{70_H} )^2$, $(\rep{70_H} )^2 \cdot \rep{63_H}$, or $(\rep{70_H} )^4$.
The possible Higgs mixing terms can only appear as the non-renormalizable ones, such as the
\beqn
\Vc&\supset& \frac{1}{ M_{\rm pl} } \epsilon^{ \lambda_1 \, ... \lambda_4 } \repb{8_H}_{\,, \lambda_1 } \, ... \, \repb{8_H}_{\,, \lambda_4 } \cdot \rep{70_H} + H.c.  \,.
\eeqn
The renormalizable Yukawa couplings in Eq.~\eqref{eq:SU8_Yukawa} are reduced to 
\beqn\label{eq:SU8_Yukawa_real}
-\Lc_Y&=&  Y_\Bc \repb{8_F}^\lambda \rep{28_F}  \repb{8_{H}}_{\,,\lambda } +  Y_\Dc \repb{8_F}^{\dot \lambda} \rep{56_F}  \repb{28_{H}}_{\,,\dot \lambda } +  Y_\Tc \rep{28_F} \rep{28_F} \rep{70_H} + H.c.\,.
\eeqn
Accordingly, one can find the one-loop $\beta$ coefficient of the minimal ${\rm SU}(8)$ theory to be $b_1=-9 <0$.
Hence, the minimal flavor-unified ${\rm SU}(8)$ theory is asymptotically free.
Furthermore, the Higgs field of $\rep{70_H}$ contains two EWSB components from Eq.~\eqref{eq:SU8_Higgs_Br05}.
By applying the generalized $\widetilde{ {\rm U}}(1)_{T}$-neutrality condition, we find only one of them can develop the EWSB VEV, which also leads to the top quark Yukawa coupling term in Eq.~\eqref{eq:top_Yukawa}.
Apparently, the results we obtained in the ${\rm SU}(8)$ Higgs sector will be crucial in the future analyses of: (i) the origin of the observed SM quark/lepton mass hierarchies as well as the Cabibbo-Kobayashi-Maskawa mixing pattern~\cite{Chen:2024cht}, and (ii) the gauge coupling unification with multiple effects that may modify the running behaviors~\footnote{A previous study of the gauge coupling runnings in the supersymmetric ${\rm SU}(5)$ was made by Ref.~\cite{Langacker:1992rq}.}.
Both topics in the ${\rm SU}(8)$ theory will be studied elsewhere.

\para
The existence of massless sterile neutrinos in the flavor-unified theories with gauge groups of ${\rm SU}(N \geq 6)$ is a generic feature.
On the other hand, it is widely known that very small neutrino masses can originate from various seesaw mechanisms~\cite{Yanagida:1979as,Gell-Mann:1979vob,Mohapatra:1980yp,Lazarides:1980nt,Wetterich:1981bx,Schechter:1981cv,Foot:1988aq} as well as the $d=5$ Weinberg operator~\cite{Weinberg:1979sa}.
The 't Hooft anomaly matching condition to the generalized global $\widetilde{ {\rm U}}(1)_{T}$ symmetries at different stages is useful to count the numbers of massless sterile neutrinos precisely, without knowing the possible mechanism of generating their masses.
The ${\rm SU}(8)$ fermion sector contains twenty-seven left-handed sterile neutrinos from nine $\repb{8_F}^\Lambda$, while only four of them obtain Dirac masses with their right-handed mates.
The remaining twenty-three left-handed sterile neutrinos cannot obtain masses above the EW scale through the 't Hooft anomaly matching condition.
With this fact, one can expect three-body decay modes of massive vectorlike fermions with two SM fermions plus one sterile neutrinos as their final states.
Some of the relevant flavor-changing charged and neutral currents in the $\Gc_{331}$ theory have been obtained in the two-generational ${\rm SU}(7)$ toy model, which can be automatically generalized to the three-generational case~\footnote{In the flavor-unified theories, the effective $\Gc_{331}$ theory can only exhibit the flavor universality among three-generational SM fermions.}.

%%%%%%%%%%%%%%%%%%%%%%%%%%%%
\subsection{Discussions}
%%%%%%%%%%%%%%%%%%%%%%%%%%%%

\para
Our current discussions focus on the minimal flavor-unified ${\rm SU}(8)$ theory.
The generalization to larger ${\rm SU}(N)$ theories can always lead to a well-defined global $B-L$ symmetry.
A next possible ${\rm SU}(9)$ theory with fermions of
\beqn\label{eq:SU9_Frampton}
\{ f_L \}_{ {\rm SU}(9)}^{n_g=3}&=& \repb{9_F}^\Lambda \oplus \rep{84_F} \,,~  \Lambda = ( 1\,, 2\,, 3\,, {\rm IV}\,, ...\,, {\rm IX})\,,~ {\rm dim}_{ \mathbf{F}}= 165 \,,  
\eeqn
was suggested long ago~\cite{Frampton:1979tj,Frampton:2009ce}.
A renormalizable Yukawa coupling of $\rep{84_F} \rep{84_F} \rep{84_H} + H.c.$ is vanishing due to the rank-$3$ anti-symmetric property of $\rep{84_F}$.
Similar to Eq.~\eqref{eq:Yukawa_341_03}, one can only formulate $d=5$ non-renormalizable Yukawa coupling such as $ \frac{1}{ M_{\rm pl} } \rep{84_F} \rep{84_F} \rep{80_H} \rep{84_H}  + H.c.$ with the adjoint Higgs field of $\rep{80_H}$.
Since all three-generational left-handed quark doublets and the right-handed up-type quarks are packed in the rank-$3$ anti-symmetric $\rep{84_F}$, one can thus expect a suppression factor of $\Oc(v_U/ M_{\rm pl})$ for the top quark mass.
This is unnatural compared with the $\Oc(1)$ Yukawa coupling for the top quark mass in Eq.~\eqref{eq:top_Yukawa} of the ${\rm SU}(8)$ theory.

\para
Thus, the next-minimal theory should be an ${\rm SU}(9)$ theory with the following rank-$2$ IAFFS and rank-$4$ IAFFS
\beqn\label{eq:SU9_3gen_fermions}
\{ f_L \}_{ {\rm SU}(9)}^{n_g=3}&=& \Big[  \repb{9_F}^\lambda \oplus \rep{36_F} \Big] \bigoplus \Big[  \repb{9_F}^{ \dot \lambda } \oplus \rep{126_F} \Big] \,,~ {\rm dim}_{ \mathbf{F}}= 252\,, \non
&& \lambda = ( 3\,, {\rm IV}\,, {\rm V}\,, {\rm VI}\,, {\rm VII}) \,, ~  \dot \lambda = (\dot 1\,, \dot 2 \,, \dot {\rm IIX}\,, \dot {\rm IX} \,, \dot {\rm X} )  \,.
\eeqn
The corresponding global non-anomalous DRS symmetries at the GUT scale read
\beqn\label{eq:SU9_DRS}
\widetilde{ \Gc}_{\rm DRS}&=&  \[ \widetilde{ {\rm SU} }(5)_{ \lambda }   \otimes \widetilde{ {\rm U} }(1)_{T_2 } \] \bigotimes   \[ \widetilde{ {\rm SU} }(5)_{\dot \lambda }  \otimes \widetilde{ {\rm U} }(1)_{T_4 } \] \,.
\eeqn
Accordingly, one can find the following Yukawa couplings of
\beqn\label{eq:SU9_Yukawa}
-\Lc_Y&=& Y_\Bc  \repb{9_F}^\lambda \rep{ 36_F}  \repb{9_{H}}_{\,,\lambda } +  Y_\Tc \rep{36_F} \rep{36_F} \repb{126_H}  +  Y_\Dc  \repb{9_F}^{\dot \lambda} \rep{126_F}  \repb{84_{H}}_{\,,\dot \lambda}  \non
&+&   \frac{1 }{ M_{\rm pl} } \rep{126_F} \rep{126_F} \repb{9_H}_{\,, \lambda}^\dag \rep{80_H}  + H.c.\,.
\eeqn
Altogether, the ${\rm SU}(9)$ Higgs fields are the following
\beqn\label{eq:SU9_Higgs}
 \{ H \}_{ {\rm SU}(9)}^{n_g=3} &=&  \repb{9_H}_{\,, \lambda}  \oplus \repb{84_H}_{\,, \dot \lambda }   \oplus \repb{126_H} \oplus \underline{ \rep{80_H} }  \,, \non
 && \lambda = ( 3\,, {\rm IV}\,, {\rm V}\,, {\rm VI}\,, {\rm VII}) \,, ~  \dot \lambda = (\dot 1\,, \dot 2 \,, \dot {\rm IIX}\,, \dot {\rm IX} \,, \dot {\rm X} )  \,, 
\eeqn
where the adjoint Higgs field of $\rep{80_H}$ is real and responsible for the GUT scale symmetry breaking of ${\rm SU}(9) \to {\rm SU}(5) \otimes {\rm SU}(4) \otimes {\rm U}(1)$.
Their non-anomalous $\widetilde{ {\rm U}}(1)_{T_2} \otimes \widetilde{ {\rm U}}(1)_{T_4}$ charges are given in Tab.~\ref{tab:SU9_Tcharges}.

\begin{table}[htp]
\begin{center}
\begin{tabular}{c|ccccc}
\hline\hline
 & $\repb{9_F}^\lambda$  &  $\rep{36_F}$  &  $\repb{9_H}_{\,,\lambda}$ & $\repb{126_H}$  &  $$  \\
\hline
$\Tc_2$  &  $-7t_2$  &  $+5t_2$  & $+2t_2$  &  $ -10 t_2$  &  $$  \\[1mm]
\hline\hline
 & $\repb{9_F}^{\dot \lambda}$  &  $\rep{126_F}$  &  $\repb{84_H}_{\,,\dot \lambda}$ &   &    \\
 \hline
 $\Tc_4$  &  $ -7t_4$  &  $+t_4$  & $+6 t_4$  &  &  \\[1mm]
 \hline\hline
\end{tabular}
\end{center}
\caption{The $\widetilde{ {\rm U}}(1)_{T_2} \otimes \widetilde{ {\rm U}}(1)_{T_4}$ charges for all massless fermions and Higgs fields in the $\gSU(9)$ theory.}
\label{tab:SU9_Tcharges}
\end{table}%

\para
One can immediately find the one-loop $\beta$ coefficient of the next-minimal ${\rm SU}(9)$ theory to be $b_1= \frac{61}{6}>0$.
This means that the minimal ${\rm SU}(8)$ theory described here is the only asymptotically free flavor-unified theory in the deep UV regime.
Other flavor-unified theories of this class become strongly interacting in the deep UV regime, and the issue of the asymptotic safety will be relevant therein.

%###################################################################
\section*{Acknowledgements}
%###################################################################
%
%
\para
We would like to thank Tianjun Li, Kaiwen Sun, Yuan Sun, Jian Tang, and Wenbin Yan for very enlightening discussions in preparing this work. 
N.C. would like to thank Shanxi University and Yantai University for hospitality when preparing this work.
N.C. is partially supported by the National Natural Science Foundation of China (under Grants No. 12035008 and No. 12275140) and Nankai University.
Y.N.M. is partially supported by the National Natural Science Foundation of China (under Grant No. 12205227), the Fundamental Research Funds for the Central Universities (WUT: 2022IVA052), and Wuhan University of Technology.

\appendix

%###################################################################
\section{Decomposition rules and charge quantizations in the ${\rm SU}(8)$ theory}
\label{section:Br}
%###################################################################

%

\para
%We list the branching rules of the Higgs fields and to look for the SM-singlet directions for each of them.
We define the decomposition rules in the ${\rm SU}(8)$ theory according to the symmetry-breaking pattern in Eq.~\eqref{eq:Pattern}.
After the GUT-scale symmetry breaking, we define the ${\rm U}(1)_{X_0}$ charges as follows
\beqn\label{eq:X0charge}
\hat X_0( \rep{8} ) &\equiv& {\rm diag} ( \underbrace{ - \frac{1}{4}  \,, - \frac{1}{4}\,, - \frac{1}{4}\,, -\frac{1}{4} }_{ \rep{4_s} }\,, \underbrace{ +\frac{1}{4} \,, +\frac{1}{4}\,, +\frac{1}{4}\,, +\frac{1}{4} }_{ \rep{4_W} } )\,.
\eeqn
Sequentially, the ${\rm U}(1)_{X_1}$, ${\rm U}(1)_{X_2}$, and ${\rm U}(1)_{Y}$ charges are defined according to the ${\rm SU}(4)_s$, ${\rm SU}(4)_W$, and ${\rm SU}(3)_W$ fundamental representations as follows
\beqs
\beqn
\hat X_1(\rep{4_s}) &\equiv&  {\rm diag} \, \Big( \underbrace{  (- \frac{1}{12}+ \Xc_0 ) \mathbb{I}_{3\times 3} }_{ \rep{3_c} } \,, \frac{1}{4}+ \Xc_0 \Big) \,,\label{eq:X1charge_4sfund}\\[1mm]
%%%%%%%%%%%%%%%%%%%%%%%%%%%%%%%%%%%%%%%%%%%%%
\hat X_2 ( \rep{4_W} )&\equiv& {\rm diag} \, \Big( \underbrace{  ( \frac{1}{12} + \Xc_1 ) \mathbb{I}_{3\times 3} }_{ \rep{3_W} } \,, -\frac{1}{4} + \Xc_1 \Big)  \,, \label{eq:X2charge_4Wfund} \\[1mm]
%%%%%%%%%%%%%%%%%%%%%%%%%%%%%%%%%%%%%%%%%%%%%
\hat Y ( \rep{4_W} )&\equiv&  {\rm diag} \, \Big(  ( \frac{1}{6}+ \Xc_2 ) \mathbb{I}_{2\times 2} \,,- \frac{1}{3}+ \Xc_2 \,, \Xc_2 \Big) \non
&=& {\rm diag} \, \Big(  \underbrace{ ( \frac{1}{4} + \Xc_1 ) \mathbb{I}_{2\times 2} }_{ \rep{2_W} } \,,  ( - \frac{1}{4} + \Xc_1  ) \mathbb{I}_{2\times 2} \Big) \,, \label{eq:Ycharge_4Wfund} \\[1mm]
%%%%%%%%%%%%%%%%%%%%%%%%%%%%%%%%%%%%%%%%%%%%%
\hat Q_e ( \rep{4_W} )&\equiv& T_{ {\rm SU}(4) }^3 +  \hat Y ( \rep{4_W} ) = {\rm diag} \, \Big( \frac{3}{4} + \Xc_1  \,, ( - \frac{1}{4} + \Xc_1  ) \mathbb{I}_{3\times 3}  \Big) \,. \label{eq:Qcharge_4Wfund}
\eeqn
\eeqs
Based on the above definitions for the fundamental representations, the rules for other higher rank anti-symmetric representations can be derived by tensor productions.

%\bibliographystyle{utphys.bst}
%\bibliography{references}

\providecommand{\href}[2]{#2}\begingroup\raggedright\endgroup

\end{document}